\documentclass[12pt,preprint]{aastex}
\def\degpoint{\ifmmode ^{\rm{o}}\!. \else $^{\rm{o}}\!.$\fi}
\newcommand{\degrees}{$^{\rm{o}}$}
\newcommand{\ms}{\mbox{m\ s$^{-1}$}}
\newcommand{\kms}{\mbox{km \ s$^{-1}$}}
\newcommand{\Msun}{\mbox{M$_{\odot}$}}
\newcommand{\Rsun}{\mbox{R$_{\odot}$}}
\newcommand{\Mjup}{\mbox{M$_{\rm Jup}$}}

\newcommand{\Mearth}{\mbox{M$_{\oplus}$}}

\newcommand{\ltsimeq}{\raisebox{-0.6ex}{$\,\stackrel
         {\raisebox{-.2ex}{$\textstyle <$}}{\sim}\,$}}
\newcommand{\gtsimeq}{\raisebox{-0.6ex}{$\,\stackrel
         {\raisebox{-.2ex}{$\textstyle >$}}{\sim}\,$}}

\begin{document}

\title{A Search for Multi-Planet Systems Using the Hobby-Eberly 
Telescope\footnotemark[1] }

\author{Robert A.~Wittenmyer\altaffilmark{2,3}, Michael 
Endl\altaffilmark{2}, William D.~Cochran\altaffilmark{2}, Harold 
F.~Levison\altaffilmark{4}, Gregory W.~Henry\altaffilmark{5}}
\altaffiltext{2}{McDonald Observatory, University of Texas at Austin, 
Austin, TX 78712}
\altaffiltext{3}{Department of Astrophysics, School of Physics, 
University of NSW, 2052, Australia}
\altaffiltext{4}{Department of Space Studies, Southwest Research 
Institute, Boulder, CO 80302 }
\altaffiltext{5}{Center of Excellence in Information Systems, Tennessee 
State University, 3500 John A.\ Merritt Blvd., Box 9501, Nashville, TN 
37209, USA}
\email{
rob@phys.unsw.edu.au}

\shorttitle{A Search for Multi-Planet Systems}
\shortauthors{Wittenmyer et al.}
\footnotetext[1]{Based on observations obtained with the Hobby-Eberly 
Telescope, which is a joint project of the University of Texas at 
Austin, the Pennsylvania State University, Stanford University, 
Ludwig-Maximilians-Universit{\" a}t M{\" u}nchen, and 
Georg-August-Universit{\" a}t G{\" o}ttingen.}

\begin{abstract}

\noindent Extrasolar multiple-planet systems provide valuable 
opportunities for testing theories of planet formation and evolution.  
The architectures of the known multiple-planet systems demonstrate a 
fascinating level of diversity, which motivates the search for 
additional examples of such systems in order to better constrain their 
formation and dynamical histories.  Here we describe a comprehensive 
investigation of 22 planetary systems in an effort to answer three 
questions: 1) Are there additional planets? 2) Where could additional 
planets reside in stable orbits? and 3) What limits can these 
observations place on such objects?  We find no evidence for additional 
bodies in any of these systems; indeed, these new data do not support 
three previously announced planets (HD~20367b: Udry et al.~2003, 
HD~74156d: Bean et al.~2008, and 47~UMa~c: Fischer et al.~2002).  The 
dynamical simulations show that nearly all of the 22 systems have large 
regions in which additional planets could exist in stable orbits.  The 
detection-limit computations indicate that this study is sensitive to 
close-in Neptune-mass planets for most of the systems targeted.  We 
conclude with a discussion on the implications of these non-detections.

\end{abstract}


\section{Introduction}

About 12\% ($N=31$) of known planetary systems contain more than one 
planet.  Now that radial-velocity precision at the 1-2 \ms\ level is 
being achieved by several planet search programs \citep{butler06, 
lovis06}, Neptune-mass planets are becoming detectable.  Recent 
discoveries of ``super-Earths'' (m~sin~$i$ \ltsimeq 10 $M_\oplus$) by 
the High-Accuracy Radial Velocity Planet Search (HARPS) instrument 
\citep{bouchy08, mayor08, udry07, bonfils07} suggest that super-Earths 
may be common.

The presence of close-in giant planets (``hot Jupiters'') inferred by 
precision radial-velocity surveys has emphasized the importance of 
post-formational dynamical evolution processes such as planetary 
migration.  The core-accretion model of planetary formation 
\citep{lissauer95, pollack96} posits that rocky cores form in the outer 
regions of the protoplanetary disk and experience runaway gas accretion 
once they reach a mass of $\sim$10 Earth masses.  These giant planets 
then migrate inward to become hot Jupiters \citep{bodenheimer00}. 
Alternatively, the disk-instability model suggests that such planets 
form by direct gravitational collapse of the protoplanetary disk 
\citep{boss95, boss98}.  Multi-planet systems can be formed by this 
method \citep{boss03}, though subsequent evolution can easily eject 
planets, resulting in a wide variety of system end-states 
\citep{levison98}.  The discovery of additional multi-planet systems 
will provide valuable added constraints to these two models of planet 
formation. \citet{trilling98} have proposed that gas giant planets 
migrating inward can overflow their Roche lobes and be stripped of their 
gaseous envelopes.  Under the core-accretion model of planet formation, 
a Neptune-mass rocky core would then remain in a close orbit, and the 
detection of such objects would lend support to that theory. 
Alternatively, the nondetection of close-in, low-mass ($m_p < 15$ 
\Mearth) planets would tend to favor the disk-instability model, in 
which gas giant planets have no solid cores.  Hence, an intensive effort 
to characterize the population of detectable planets around nearby stars 
will be extremely valuable for understanding the processes of planet 
formation and evolution.

The architectures of multi-planet systems can shed light on their 
formation and dynamical history.  \citet{chatterjee07} performed 
simulations of systems with three giant planets and found that at least 
one planet would be ejected before the system stabilised.  When two 
planets remained (80\% of cases), their median eccentricities were 
$e\sim$0.4.  Similarly, randomly generated planetary systems simulated 
by \citet{juric07} typically retained 2-3 giant planets after $10^8$~yr.  
That all five planets \citep{fischer08} in the 55~Cancri system have 
relatively low eccentricities ($e<0.2$) suggests that systems with 
inactive dynamical histories (i.e.~free of major perturbation events) 
may be able to retain several giant planets in nearly circular orbits.

The final configuration of a planetary system is dependent on the 
post-formation migration and dynamical interaction processes.  
\citet{mandell07} showed that the migration of a Jupiter-mass planet 
through a disk of planetesimals can result in the formation of an 
interior terrestrial-mass planet.  Simulations of known multi-planet 
systems by \citet{bq04} and \citet{barnes04} suggest that planetary 
systems are ``packed'' -- that is, they contain the maximum number of 
planets that is dynamically possible.  \citet{barnes04} investigated the 
dynamically stable regions of the HD~74156 system.  Those authors used 
the results to predict that an additional planet, between planets b and 
c, could be present.  The detection by \citet{bean08} of such an object 
lends support to the ``packed planetary systems'' hypothesis 
\citep{barnes08}, which would imply that multiple-planet systems are 
common.  However, our own results (see \S~3) do not support this 
hypothesis.

A series of papers by Ida \& Lin \citep{idalin04a, idalin04b} predicts a 
paucity of planets of 10-100 Earth masses within $\sim 1$~AU (the 
``planet desert'').  Their core-accretion simulations also predict an 
abundance of close-in ($a\ltsimeq$0.1~AU) planets with masses below 
about 10 $M_\oplus$.  Ida and Lin further suggest that the distribution 
of planetary mass vs. semimajor axis will constrain the dominant 
formation processes of planets.  In a subsequent paper, 
\citet{idalin08a} show that the frequency of giant planets depends 
sensitively on the Type~I migration rate, which must be slowed by a 
factor $C_1\sim$0.03-0.1 in order to reproduce the distribution of 
detected planets.


In this work, we describe an intensive three-year radial-velocity 
campaign to search for additional planets in known planetary systems 
(\S~2).  Section~3 gives the results of the orbit fits and the search 
for new planets, with discussion about a few of the interesting systems.  
Section~4 describes the dynamical simulations used to determine the 
regions in each system where additional planets could reside in stable 
orbits. The detection limits, which determine the sensitivity of this 
survey, are presented in \S~5.  Finally, Section~6 assesses the impact 
of these new data and analyses on the theories of planet formation and 
the population-level statistics of extrasolar planets.  This work thus 
presents a three-fold approach to the question of planetary system 
architecture: 1) Are additional planets present in these known planetary 
systems? 2) Where could additional objects reside in stable orbits? 3) 
What limits can be placed on such objects?

\section{Observational Data}

Twenty-two targets were chosen for this project from the list of 
$\sim$150 planet hosts known in 2004~September.  A majority of the 
observational data were obtained at McDonald Observatory with the 9.2~m 
Hobby-Eberly Telescope (HET: Ramsey et al.~1998) using its High 
Resolution Spectrograph (HRS) \citep{tull98}.  The targets were selected 
according to the following criteria: 1) HET observability, with 
declination between -11\degrees~and +72\degrees, and 2) Either a 
long-period ($P\gtsimeq$1 yr) planet such that inner planets may be 
dynamically stable, or a very short-period ($P\ltsimeq$10 days) hot 
Jupiter which would allow for previously undetected outer planets, and 
3) The orbital solution for the known planet in each system has RV 
residuals of 10-20 m s$^{-1}$, so that an additional planet may be 
present but undetected.  The targets and their stellar parameters are 
listed in Table~\ref{stellarparms}.  Except where noted, masses are 
obtained from \citet{takeda07}, [Fe/H], $T_{eff}$, and V~sin~$i$ from 
\citet{vf05}, and the chromospheric emission ratio log $R'_{HK}$ 
\citep{noyes84} computed from measurements of the Ca~II S-index obtained 
with the 2.7m telescope using the techniques developed by 
\citet{paulson02}.  The uncertainties on the stellar masses given in 
\citet{takeda07} are asymmetric about the central value; for the 
purposes of Table~\ref{stellarparms} and the determination of planetary 
parameters, the adopted stellar mass uncertainty was taken to be the 
larger of the two.

All of the HET observations for this program were performed at a 
spectral resolution of 60,000, with the 316 gr/mm cross-disperser and a 
central wavelength of 5936\AA.  An iodine cell temperature-controlled at 
70\degr~C was used as the velocity metric \citep{marcy92}.  This setup, 
identical to that used for the ongoing planet search program 
\citep{cochran04, endl08}, places the iodine region ($\sim$5000-6000 
\AA) almost entirely onto the blue CCD, which is cosmetically superior 
to the red CCD.  For each target, an iodine-free template spectrum was 
obtained near the beginning of the first season in which is was 
observable.  We determined precise radial velocities following the 
general recipe outlined by \citet{butler96}, using an advanced version 
of our own code ``Austral'' \citep{endl00}.

We observed each target with the HET in queue mode using a random 
observing interval of 2-10 days between visits.  Each visit consisted of 
one spectrum, except for seven bright targets (HD~3651, HD~19994, 
HD~38529, HD~74156, 47~UMa, HD~128311, HD~136118) for which 3 
consecutive spectra were obtained in each visit.  HET data consisting of 
multiple exposures per visit were binned using the weighted mean value 
of the velocities in each visit.  We adopted the quadrature sum of the 
rms about the mean and the mean internal error as the the error bar of 
each binned point.  This procedure was done for HD~3651, HD~19994, 
HD~38529, HD~74156, 47~UMa, HD~128311, and HD~136118.  Targets were 
observed with the HET from 2004~December through 2007~November.  During 
the three years of this study, supplemental observations were also made 
using the 2.7m Harlan J.~Smith telescope at McDonald Observatory.  All 
available published radial-velocity data were also gathered from the 
literature for the purpose of fitting orbits to the known planets.  
Those data are summarized in Table~\ref{rvdata1}.  All radial-velocity 
data obtained from McDonald Observatory are given in 
Tables~\ref{3651vels}-\ref{hjs190228}.

\section{Refined Planetary System Parameters}

\subsection{Orbit Fitting Results }

Available published data were combined with velocities from the HET and 
the 2.7m to fit Keplerian orbits using GaussFit \citep{jefferys87}, 
which is a generalized least-squares program used here to solve a 
Keplerian radial-velocity orbit model.  The GaussFit model has the 
ability to allow the offsets between data sets to be a free parameter.  
This is important because the radial velocities cited in published 
works, and those computed from HET and 2.7m data, are not absolute 
radial velocities, but rather are measured relative to the iodine-free 
stellar template.  The Geneva planet-search group, however, makes use of 
a simultaneous thorium-argon calibration rather than an iodine 
absorption cell \citep{baranne96}.  Each data set thus has an arbitrary 
zero-point offset which must be accounted for in the orbit-fitting 
procedure.

The best-fit Keplerian orbital solutions and planetary parameters are 
shown in Table~\ref{planetparms}.  A summary of the fit results for each 
individual data set is given in Table~\ref{datasummary}.  In computing 
the planetary minimum mass M~sin~$i$ and semimajor axis $a$, the stellar 
masses listed in Table~\ref{stellarparms} were used. The addition of a 
large amount of new data and the use of multiple independent data sets 
in fitting Keplerian orbits have generally improved the precision of the 
derived planetary parameters by a factor of 2-4 over the published 
results summarized in the Catalog of Nearby Exoplanets \citep{butler06}.  
In particular, the precision of the orbital periods have been improved 
by the addition of new data, due to the increased number of orbits now 
observed.  Our parameters generally agree within 2$\sigma$ of previously 
published estimates.  In this section, we highlight interesting results 
from the combined fits.

For each object, we searched for periodic signals in the residuals to 
the known planet's orbit using a Lomb-Scargle periodogram \citep{lomb76, 
scargle82}.  To assess the statistical significance of those periods, 
the false alarm probabilities (FAP) were calculated using the bootstrap 
randomization method detailed by \citet{kurster97}.  The bootstrap 
method randomly shuffles the velocity observations while keeping the 
times of observation fixed. The periodogram of this shuffled data set is 
then computed and its highest peak recorded.  In this way, we can 
determine the probability that a periodogram peak of a given power level 
will arise by chance, without making any assumptions about the error 
distribution of the data.  All bootstrap FAP estimates result from 10000 
such realizations.  Those results are shown in Table~\ref{faps}.

\textit{HD 20367}.  A planet orbiting HD~20367 was first announced in a 
conference proceedings \citep{udry03}, but has not yet appeared in a 
refereed journal.  The Geneva planet search group 
website\footnote{http://obswww.unige.ch/$\sim$udry/planet/hd20367.html} 
lists the planet's period as 469.5~days, with an eccentricity of 0.32 
and M~sin~$i$=1.17 \Mjup.  Eighty-one observations of HD~20367 were 
obtained with the HET over three observing seasons, as well as 19 
observations from the 2.7m, but period searches of these data give no 
indication of such a signal.

\begin{figure}
\plottwo{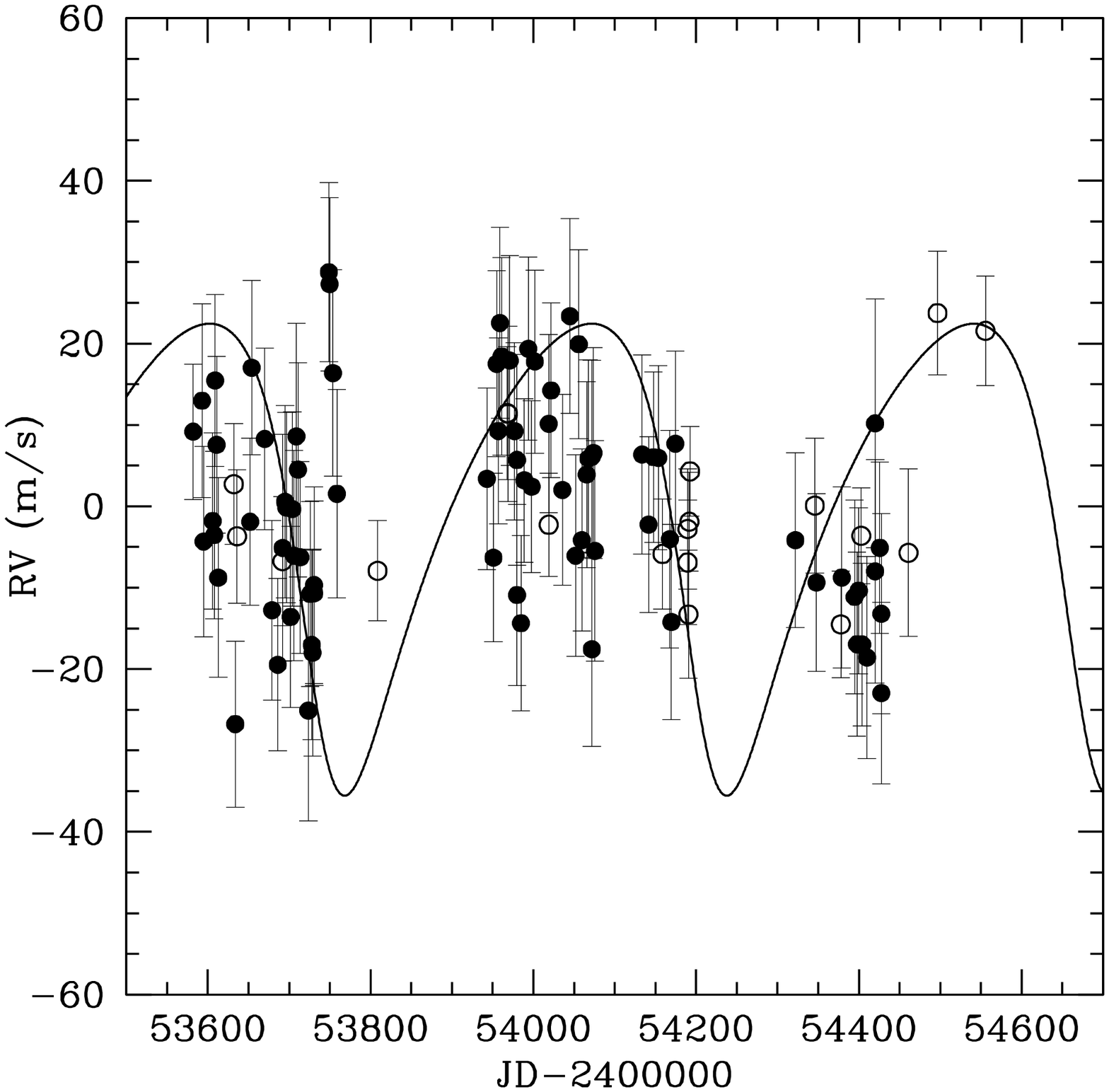}{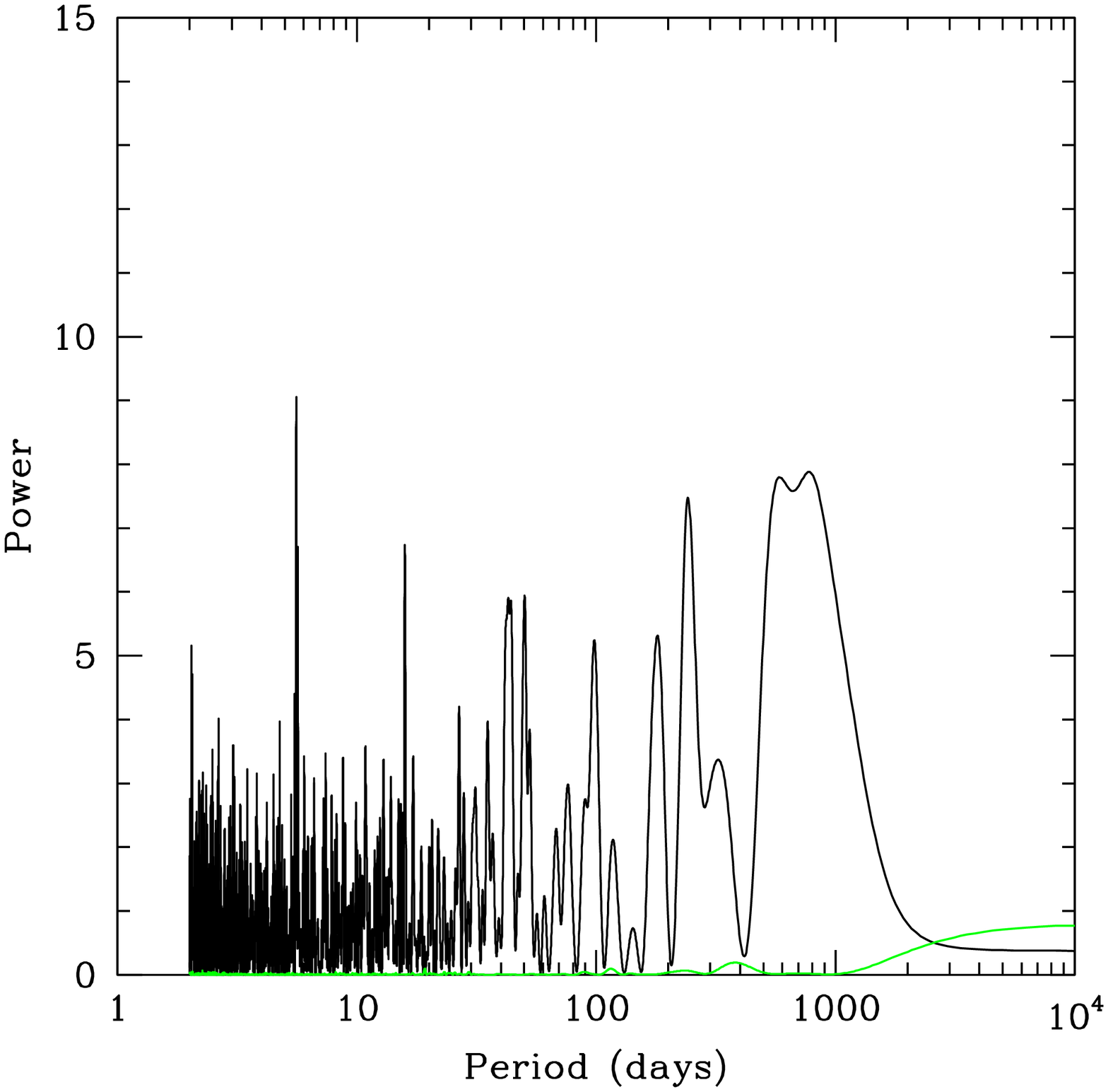}
\caption{\label{20367pgrams} Left panel: Radial-velocity data for 
HD~20367.  Filled circles: HET, open circles: 2.7m.  The Geneva group's 
orbital solution for the proposed planet is shown as a solid line.  
Right panel: Lomb-Scargle periodogram of the velocities.  The 5.5-day 
stellar rotation period is evident, but no other periodocities are 
significant.}
\end{figure}

\begin{figure}
\plotone{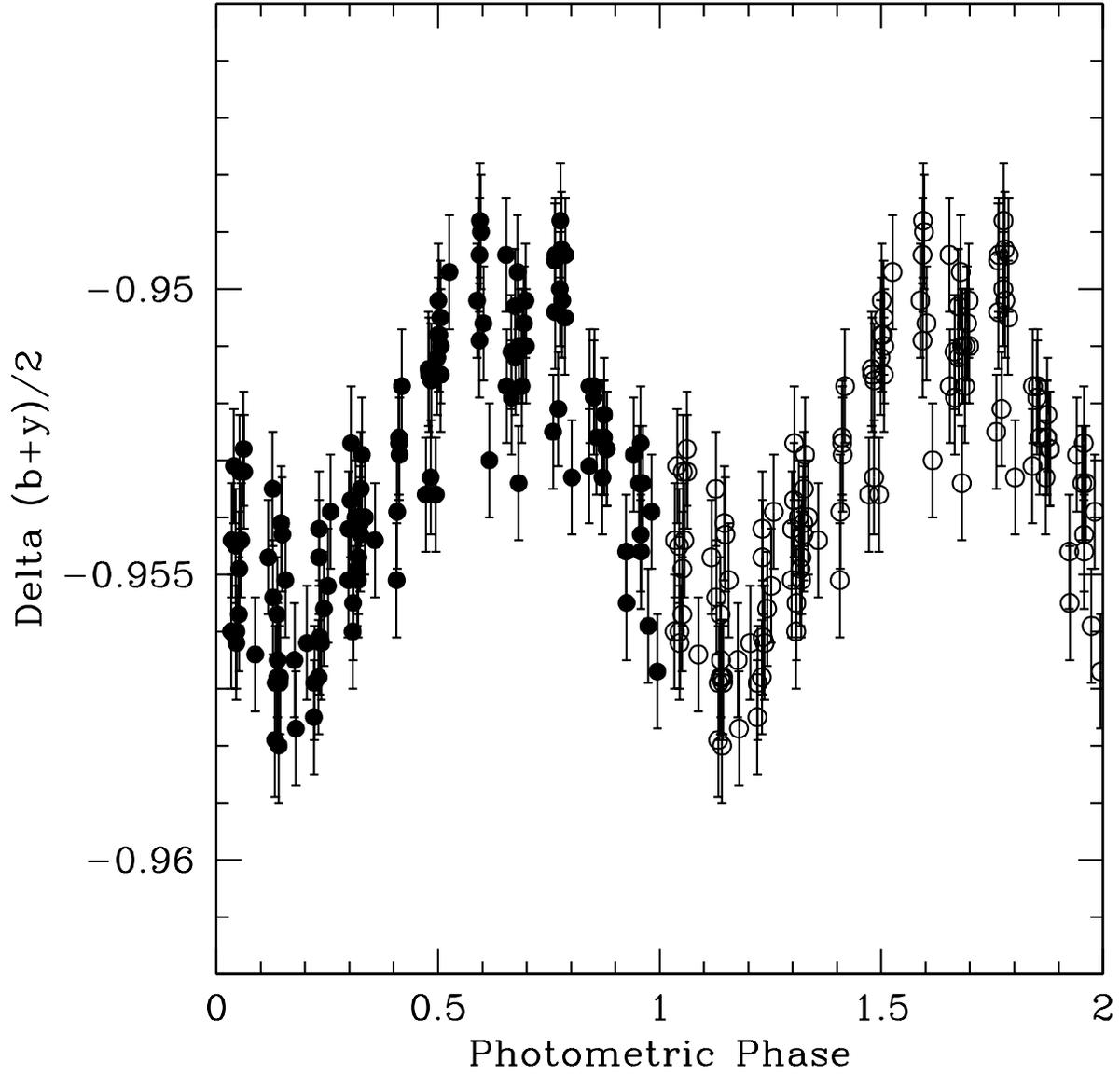}
\caption[HD 20367 rotation period]{\label{20367phot} Photometric 
observations of HD~20367 phased to the stellar rotation period of 5.50 
days.  Two cycles are shown for clarity. }
\end{figure}

Figure~\ref{20367pgrams} shows the radial-velocity data from HET and the 
2.7m telescopes, and the periodogram of those data.  The Geneva group's 
solution has been overplotted.  The highest peak, at 5.58 days, has a 
bootstrap FAP of 8.5\%.  The dominant periodicity of 5.58 days, which 
was evident early in the observation campaign, prompted a photometric 
investigation to search for transits and to rule out stellar rotation.  
We obtained 132 observations of HD~20367 from 2006 September to 2007 
January with the T10 0.8m automated photometric telescope (APT) at 
Fairborn Observatory in southern Arizona.  The T10 APT and its precision 
photometer are very similar to the T8 APT described in \citet{h99}.  The 
precision of a single observation is typically around 0.001 mag.  The 
results indicate a stellar rotation period of 5.50$\pm$0.02 days, with a 
photometric amplitude of 0.0055$\pm$0.0003 mag (Figure~\ref{20367phot}).  
From these observations, we conclude that the 5.6-day radial-velocity 
periodicity is caused by starspots rotating into and out of view.  This 
is consistent with the estimate of $P_{rot}=$6 days reported by 
\citet{wright04}, and the high level of chromospheric activity for this 
star (log $R'_{HK}=-4.50$).  The literature contains conflicting age 
estimates for HD~20367: \citet{holmberg07} estimate an age of 
4.4$^{+1.6}_{-2.1}$~Gyr, whereas \citet{wright04} report an age of 
0.9~Gyr.  Based on the rapid rotation rate, and the high level of 
chromospheric emission, the younger age estimate is favored.

The lack of any Keplerian signal in the 100 observations presented here 
leads us to conclude that there is not convincing evidence for the 
existence of HD~20367b.

\textit{HD 74156}.  For HD~74156, we fit the two planets at 51 and 2473 
days using ELODIE and CORALIE data from \citet{naef04}, and 82 
independent HET visits.  This system warrants closer scrutiny in light 
of the report by \citet{bean08} of a third planet, with a period 
of 346~days and a radial-velocity semi-amplitude $K=10.5$~\ms.  That 
result was obtained using the same HET spectra as considered in this 
work, but velocities were derived using an independent method described 
in \citet{bean33636}.  Here, we further investigate the possibility of 
an additional planet in the HD~74156 system.  Applying our orbit-fitting 
methods as described above to the velocities for HD~74156 given in 
\citet{bean08}, a periodogram peak is evident near 346 days, and we 
obtain a three-planet Keplerian orbit fit which is consistent with that 
of \citet{bean08}.  This indicates that the fitting method used here is 
not responsible for our non-detection of HD~74156d.  

It is possible that the HET velocities derived by \citet{bean08} are of 
superior quality to those presented here.  However, the rms of the HET 
data about a two-planet fit reported by \citet{bean08} is 8.5 \ms, 
whereas we obtain an rms of 8.3 \ms\ for those data.  These results 
suggest that there is no significant difference in quality between the 
two extant sets of HET velocities for HD~74156.  The uncertainties 
quoted by \citet{bean08} are generally smaller than ours by a factor of 
2-3.  We repeated the fitting procedure, reducing the HET uncertainties 
by a factor of 2 and 3, but there was no significant change in the 
residuals: no signal is evident at periods near 346 days.

Since the total rms scatter about our two-planet fit is 11.5 \ms, and 
the semi-amplitude of planet~d is $K=10.5$\ms, it is possible that a 
third planetary signal may have been lost in the noise.  To test this 
possibility, we performed the following Monte Carlo simulations.  From 
each of the two data sets considered in the fits described here, we 
generated 1000 simulated sets of velocities consisting of three 
Keplerian signals plus a Gaussian noise term.  This noise was equivalent 
to the mean uncertainty of each data set (ELODIE+CORALIE: 10.8 \ms, HET: 
8.3 \ms) added in quadrature to a stellar jitter of 4 \ms\ (the jitter 
estimate used in Bean et al.~2008).  The parameters of the three 
simulated planets were those from \citet{bean08}.  These simulated 
datasets retained the times of observation and the error bars of the 
originals.  We then fit the simulated data with a two-planet model 
exactly as described above, and examined the residuals of the two-planet 
fit by the periodogram method, to determine whether the signal of 
planet~d was recovered.  The criteria for recovery were that the period 
of the second planet had to be detected correctly and with a FAP of less 
than 0.1\%.  This FAP was computed using the analytic FAP formula of 
\citet{hb86}.  Of the 1000 trials, only 11 did not result in a 
successful recovery of the signal of the second planet.  The correct 
period was recovered 995 times, and the FAP exceeded 0.1\% only 6 times; 
the worst FAP was 0.3\%.  These results indicate that our method should 
have been able to detect the signal of HD~74156d, had it been present 
with the parameters given by \citet{bean08}.

In \citet{bean08}, the iodine-free stellar template spectrum was 
obtained at a resolving power of $R=$120,000, rather than the $R=$60,000 
which is standard for targets in this paper.  We obtained an $R=$120,000 
template spectrum on 2007~Nov~12, but the velocities computed using this 
template resulted in a 2-planet fit with a slightly higher rms (HET: 8.9 
\ms) than the original $R=$60,000 template (HET: 8.3 \ms).  All analysis 
for HD~74156 in this paper refers to velocities obtained using the 
$R=$60,000 template.

\begin{figure}
\plottwo{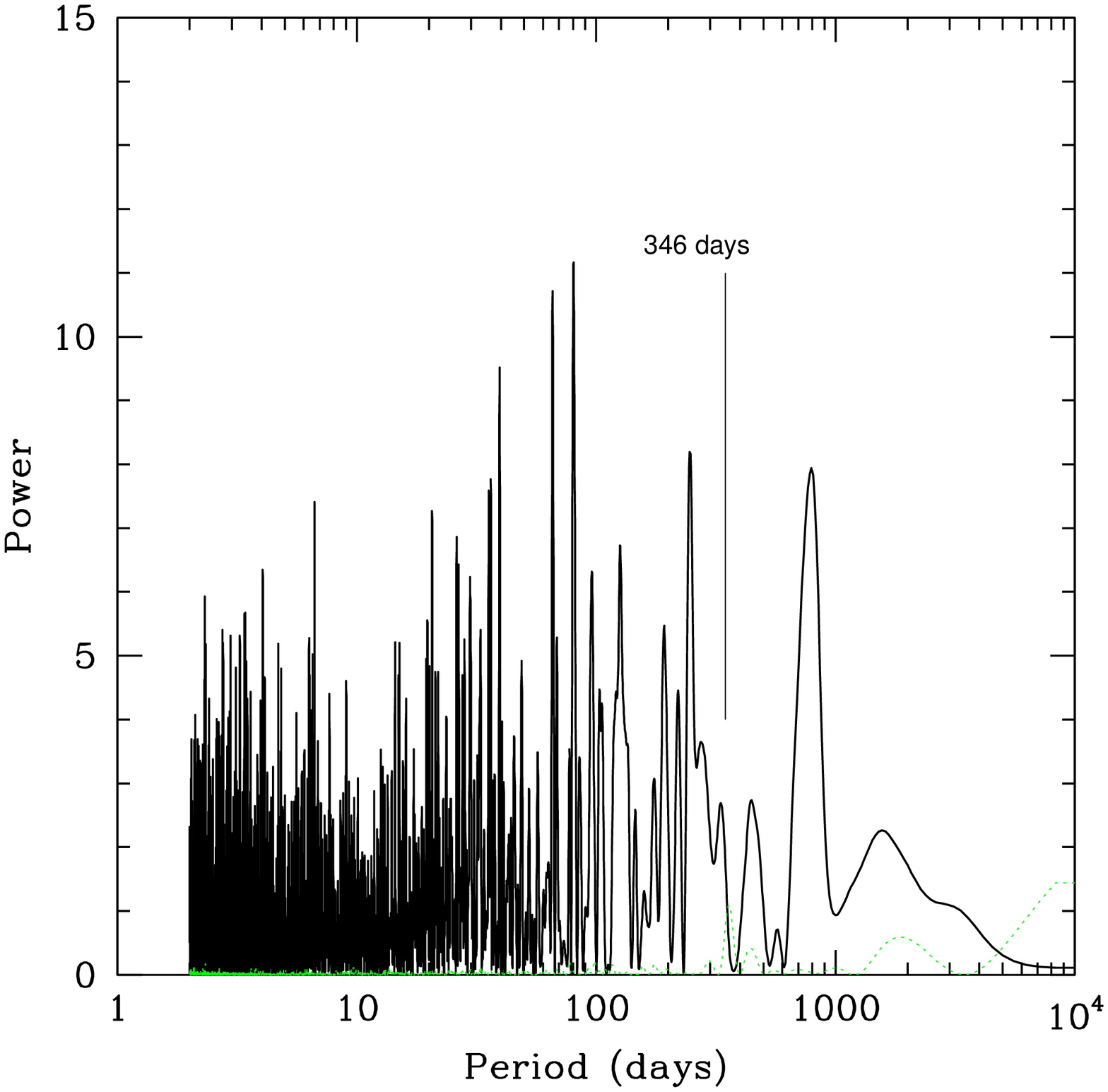}{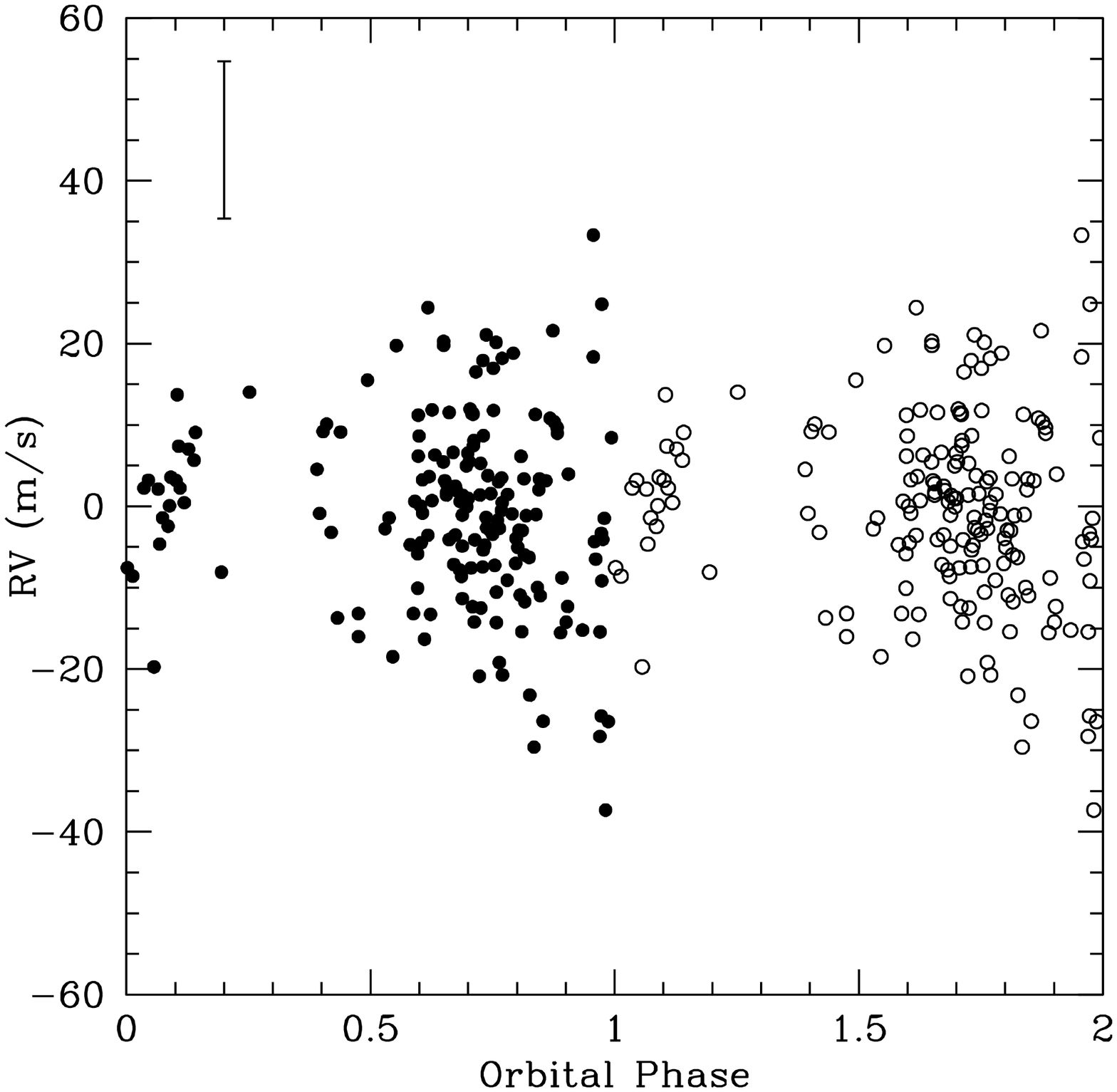}
\caption{\label{346phase} Left panel: Periodogram of the residuals of a 
2-planet fit for the HD~74156 system.  The window function is shown as a 
grey dotted line, and the 346-day period of planet~d is marked.  Right 
panel: The residuals to the 2-planet fit, phased to a period of 346.6 
days \citep{bean08}.  For clarity, two cycles are shown, and the error 
bars have been omitted.  A reference error bar representing the mean 
uncertainty of 9.65 \ms\ is shown. }
\end{figure}

A periodogram of the residuals to our 2-planet fit is shown in the left 
panel of Figure~\ref{346phase}, and those residuals are phased to the 
346.6~day period in the right panel.  The window function (grey dotted 
line) has a broad peak near 346 days due to the 1-year observing window.  
The phase gaps (right panel) are expected since the trial period is 
close to 1~year.  No clear Keplerian signal is evident despite the large 
number of data points ($N=177$).  We conclude from these data that there 
is not sufficient evidence for a third planet in the HD~74156 system.


\textit{47 UMa (=HD~95128)}.  In \citet{destructor}, we performed these 
fits to an earlier set of data from McDonald Observatory.  Those results 
did not provide convincing evidence for the outer planet reported by 
\citet{fischer02} at P$\sim$2594 days; rather, we obtained a best-fit 
2-planet model with $P_2\sim$6900 days.  Here we include an additional 
14 epochs from HET, and the best-fit 2-planet model now calls for $P_2 
\sim$9660 days.  As in previous attempts to fit a second planet, the 
parameters $e_2$ and $\omega_2$ needed to be held fixed, at the values 
proposed by \citet{fischer02}: $e_2 = 0.005$ and $\omega_2 
=$127\degrees.  The rms about a single-planet model is 10.2\ms, compared 
to 8.6\ms\ when a second planet is included.  Considering the continued 
ambiguity in the parameters for a second planet, and the 
ever-lengthening period of such an object, we use the one-planet fit for 
all further analysis in this work.

\textit{HD 114783}.  \citet{vogt02} reported the planet orbiting 
HD~114783, and recently, \citet{wright07} proposed an outer companion 
with a period of at least 8~yr.  Here, we combine the Keck data given in 
\citet{butler06} with HET observations.  A single-planet fit has a total 
rms of 6.25 \ms\ and $\chi^{2}_{\nu}$=4.91, whereas a two-planet fit 
reduces the rms to 4.42 \ms\ and the $\chi^{2}_{\nu}$ to 1.81.  The data 
considered in \citet{wright07} were of insufficient duration to 
establish a solution for the outer planet, but the combination of data 
allows for a Keplerian fit to converge.  Although a 2-Keplerian model 
can be fit to these data, it is of limited utility: the outer planet has 
a 50\% uncertainty in period ($P_2=5098\pm 2576$ days).  Our results 
support those of \citet{wright07}, that a second object is likely 
present, although there is not yet a sufficient time baseline of 
observations to establish its nature.  The 1-planet fit was used to 
derive the parameters given in Table~\ref{planetparms}, and was also 
used for the detection-limits determination in \S~5.

\textit{HD 128311}.  The inner planet ($P\sim$450 days) in the HD~128311 
system was first discovered by \citet{butler03}, who noted a linear 
trend in the residuals to the fit, as well as the extremely high 
activity level.  Those authors estimated the stellar jitter at 30 \ms, 
and expressed concern that the planetary signal may have its origin in 
the stellar velocity jitter.  Additional data proved that the inner 
planet was indeed real, and \citet{vogt05} reported a second planet at 
the 2:1 mean-motion resonance (MMR).  They published a solution 
consisting of two superposed Keplerian orbits, noting that preliminary 
dynamical tests showed the system to be unstable, and that the system 
was likely in a protected 2:1 resonance.  \citet{gk06}, in their 
dynamical analysis of available radial-velocity data, suggested that the 
observed signal could be attributed to a 1:1 resonance, i.e.~a pair of 
Trojan planets.  In this work, we fit a double Keplerian model to the 
combined Keck and HET data.  Convergence is achieved, with a total rms 
of 16.9 \ms\ about both data sets (Keck--15.8 \ms, HET--17.9 \ms).  The 
residuals show a strong periodicity near 11.5~days, with bootstrap FAP 
less than 0.01\%.  Photometry of HD~128311 by G.~Henry in \citet{vogt05} 
indicates a stellar rotation period of 11.53~days with a photometric 
amplitude of 0.03~mag.  Hence, it is quite clear that the residual 
signal is caused by stellar rotation in this highly active star.

\textit{HD 130322}.  HD~130322 is host to a hot Jupiter in a 10.7-day 
period, discovered with the CORALIE observations of \citet{udry00}.  
Four data sets are available for this object: CORALIE \citep{udry00}, 
Keck \citep{butler06}, HET, and 2.7m.  Fitting all four sets together 
results in a total rms of 14.8 \ms, but removing the CORALIE data drops 
the rms to 9.3 \ms.  In addition to the large scatter about the fit, a 
highly significant periodicity remains at 35 days (FAP$<$0.01\%), which 
vanishes when the CORALIE data are removed.  Due to these 
irregularities, we elect to exclude those data from the fits. The 
precision of the derived orbital parameters is not significantly 
affected by this removal, since the CORALIE data span only 167 days.  
For all further analysis in this work, we refer to the fit which 
excluded the CORALIE data.  As given in Table~\ref{faps}, a residual 
period is present at $P\sim$438 days (FAP=0.16\%).  However, the HET 
velocities obtained using a second iodine-free template spectrum show no 
such periodicity.  Those results show a residual period at 2.518 days, 
with a bootstrap FAP of 0.35\%.  A second planet can be fitted at this 
shorter period, and preliminary dynamical tests show that it would 
remain stable for at least $10^7$~yr; however, the disagreement between 
the two templates makes it imprudent for us to claim a detection at this 
time.



\section{Dynamical Mapping}

With the increasing availability of computing power and planetary 
systems, many investigators have undertaken N-body simulations of known 
planetary systems in an effort to characterise regions in which 
additional bodies could be found.  \citet{menou03} performed a 
comprehensive test-particle analysis of 85 systems to determine the 
extent of the habitable zones in the presence of the known planet(s).  
Due to disruptions from the known giant planet's ``zone of influence,'' 
they found that only one-fourth of the systems had dynamical 
habitability comparable to our own Solar system.  In addition to test 
particles, massive ``test planets'' have also been used to test 
observational claims for new planets and to probe known multiple-planet 
systems for additional regions of stability \citep{rivera00, rivera01, 
raymond05, rivera07}.  Likewise, in this section we perform 
test-particle and massive-body simulations on the systems targeted by 
the intensive radial-velocity monitoring described in \S~3.

\subsection{Test Particle Simulations}

We performed test particle simulations using SWIFT\footnote{SWIFT is 
publicly available at http://www.boulder.swri.edu/$\sim$hal/swift.html.} 
\citep{levison94} to investigate the dynamical possibility of additional 
low-mass planets in each of the systems considered here.  SWIFT is a 
numerical integration package which is designed to solve the equations 
of motion for gravitational interactions between massive bodies (star, 
planets) and massless test particles.  Neptune-mass planets can be 
treated as test particles (1 Neptune mass = 0.054 \Mjup) since the 
exchange of angular momentum with jovian planets is small.  We chose the 
regularized mixed-variable symplectic integrator (RMVS3) version of 
SWIFT for its ability to handle close approaches between massless, 
non-interacting test particles and planets.  This version is most 
efficient when the gravitational interactions are dominated by a single 
body (the central star).  A symplectic integrator has the advantage that 
errors in energy and angular momentum do not accumulate.  Particles are 
removed if they are (1) closer than 1 Hill radius to the planet, (2) 
closer than 0.01~AU to the star, or (3) farther than 10~AU from the 
star.  A planetary-mass object passing within 1 Hill radius of another 
planet, or within 0.01~AU (2 \Rsun) of the star's barycenter, is 
unlikely to survive the encounter.  Since the purpose of these 
simulations is to determine the regions in which additional planets 
could remain in stable orbits, we set the outer boundary at 10~AU 
because the current repository of radial-velocity data cannot detect 
objects at such distances.

For each planetary system, 390 test particles were placed in initially 
circular orbits, spaced every 0.005~AU in the region between 
0.05-2.0~AU.  We have chosen to focus on this region because the 
duration of our high-precision HET data is currently 3-4 years for the 
objects in this study.  The test particles were coplanar with the 
existing planet, which had the effect of confining the simulation to two 
dimensions.  The initial orbital positions of the particles were 
randomly distributed in orbital phase with respect to the existing 
planets.  The method used here are the same as \citet{swiftpaper}, in 
which we performed test-particle simulations for six highly eccentric 
planetary systems.  Input physical parameters (Table~\ref{planetparms}) 
for the known planet in each system were obtained from our Keplerian 
orbit fits combining published velocity data and new observations from 
McDonald Observatory.  The planetary masses were taken to be their 
minimum values (sin~$i=1$).  By choosing the minimum mass for the 
planets, the regions of dynamical stability shown by the test-particle 
results are larger.  Since the system inclinations are almost certainly 
not edge-on, and hence the true planetary masses are higher, we expect 
the actual regions of stability to be smaller than shown here.  The 
systems were integrated for $10^7$ yr, following \citet{barnes04} and 
allowing completion of the computations in a reasonable time.  We 
observed that nearly all of the test-particle removals occurred within 
the first $10^6$ yr; after this time, the simulations had essentially 
stabilized to their final configurations.

\subsection{Test Particle Results}

\begin{figure}
\plottwo{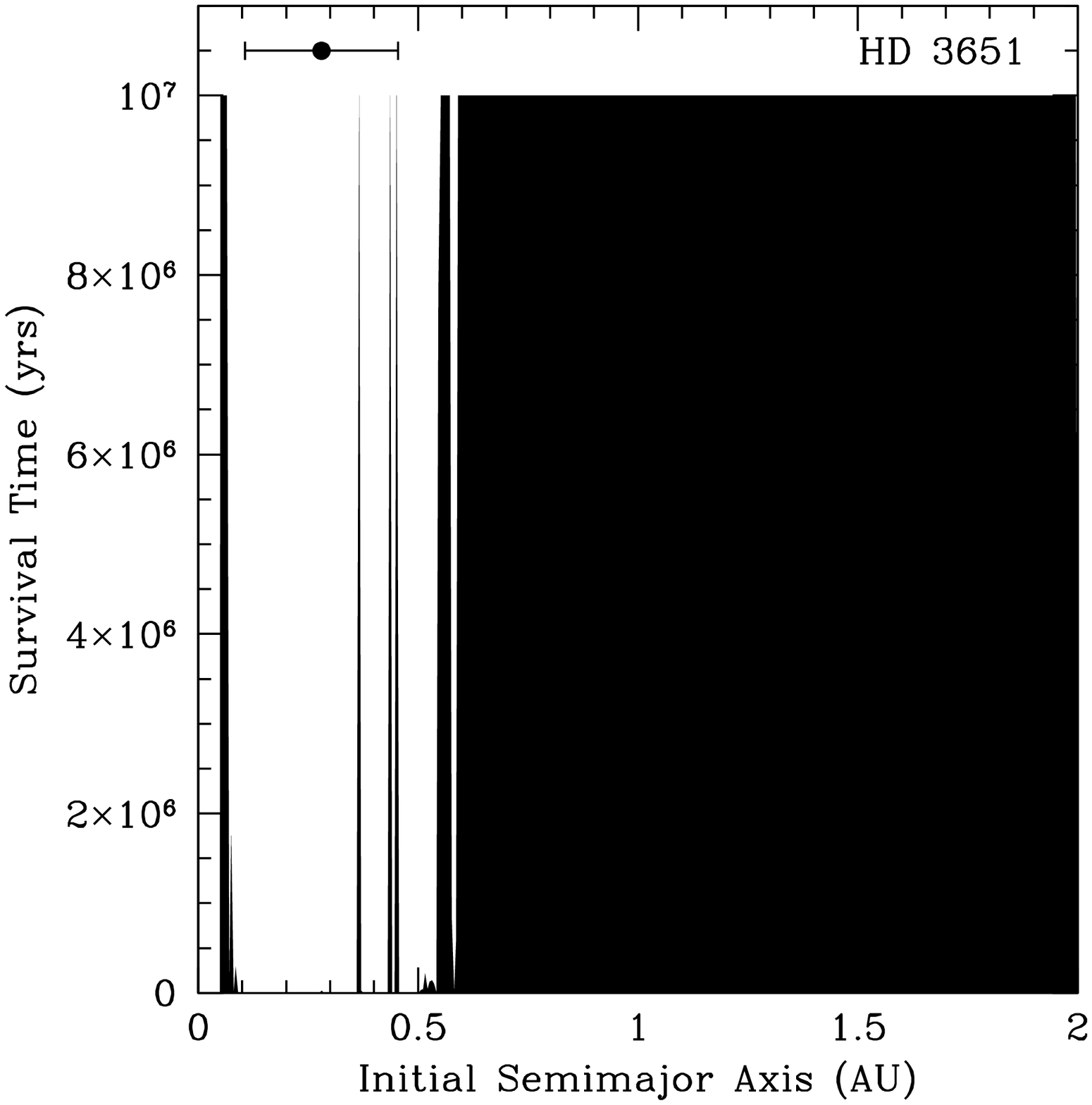}{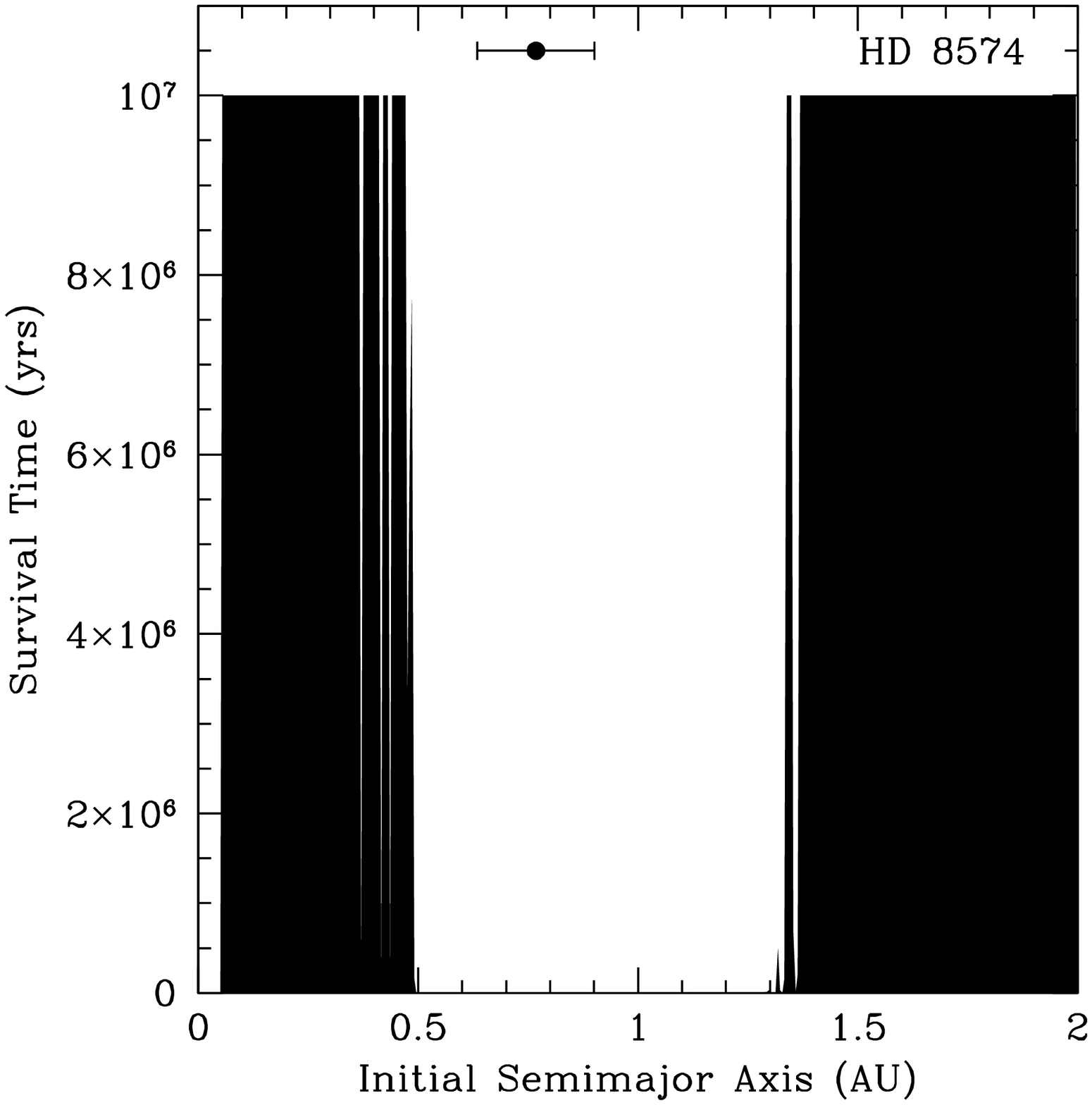}
\caption[Test particle results for HD~3651 and HD~8574]{Left panel: 
Survival time as a function of initial semimajor axis
for test particles in the HD~3651 system after $10^7$ yr.  The filled
regions indicate test particles which survived.  The orbital excursion of
HD~3561b is indicated by the horizontal error bars at the top.  Particles
were placed on initially circular orbits with $0.05<a<2.00$~AU.  Right
panel: Same, but for the HD~8574 system. } \label{tp1}
\end{figure}

\begin{figure}
\plottwo{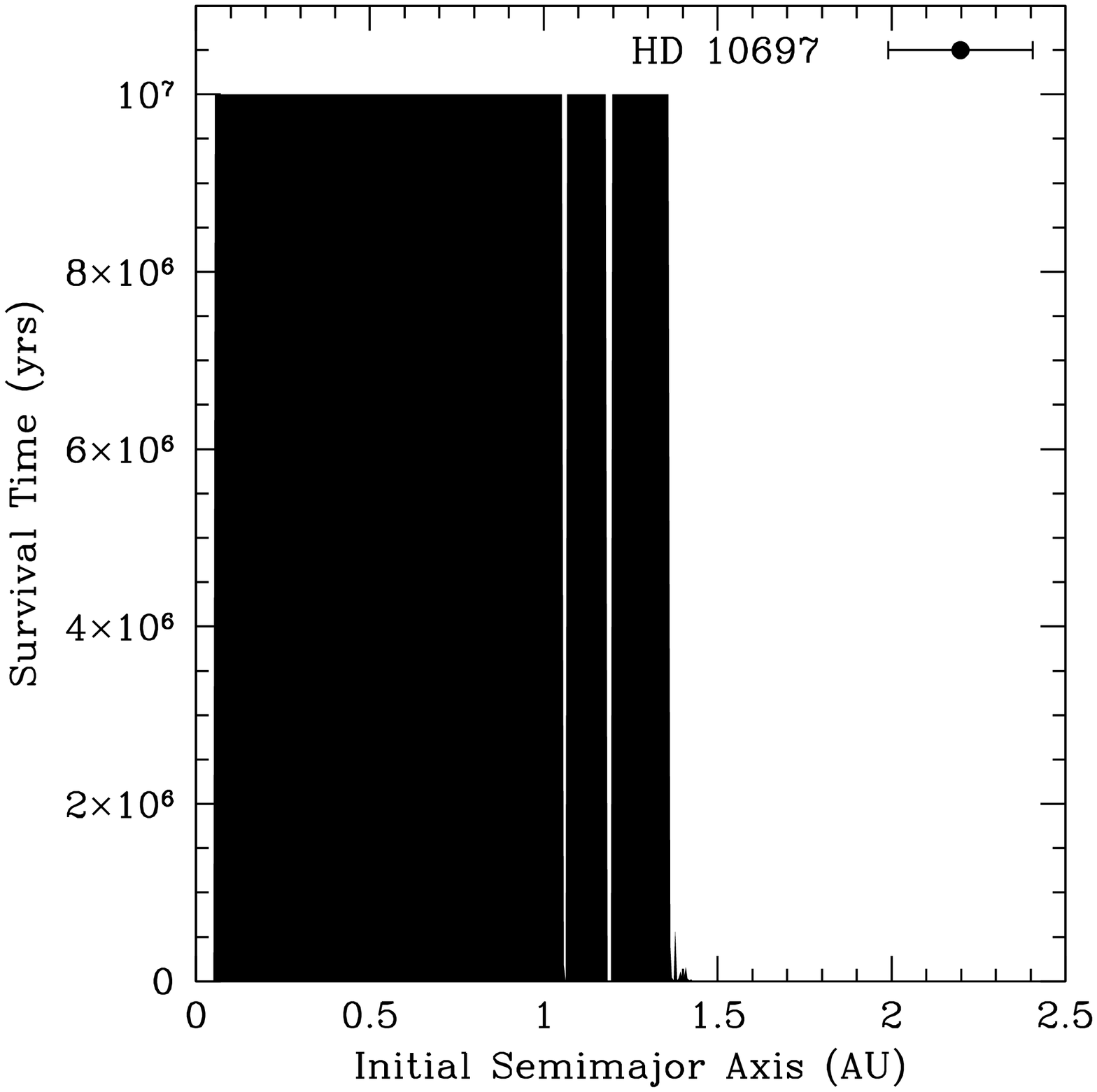}{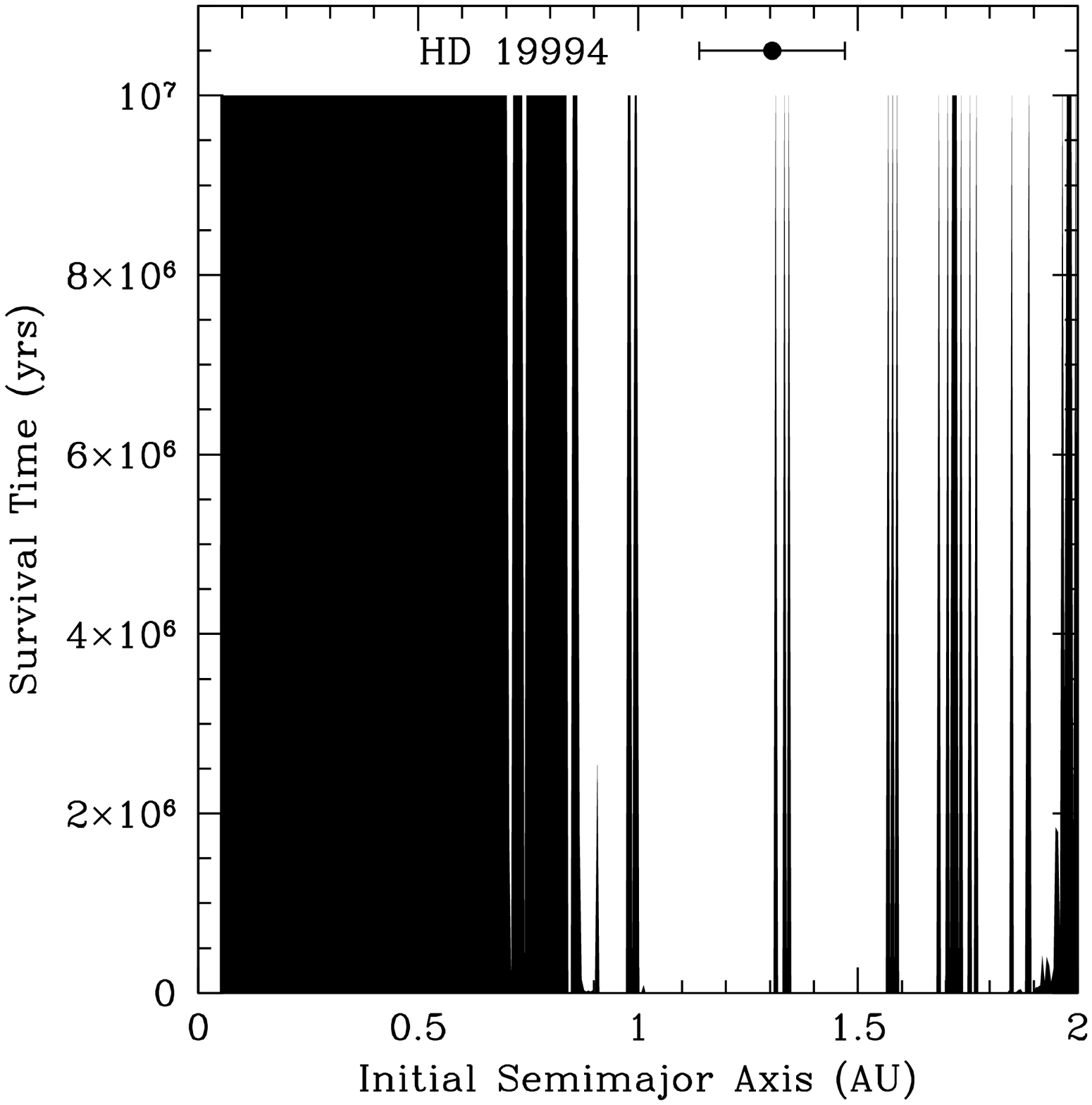}
\caption[Test particle results for HD~10697 and HD~19994]{Same as 
Fig.~\ref{tp1}, but for the HD~10697 (left) and HD~19994
(right) systems. } \label{tp2}
\end{figure}

\begin{figure}
\plottwo{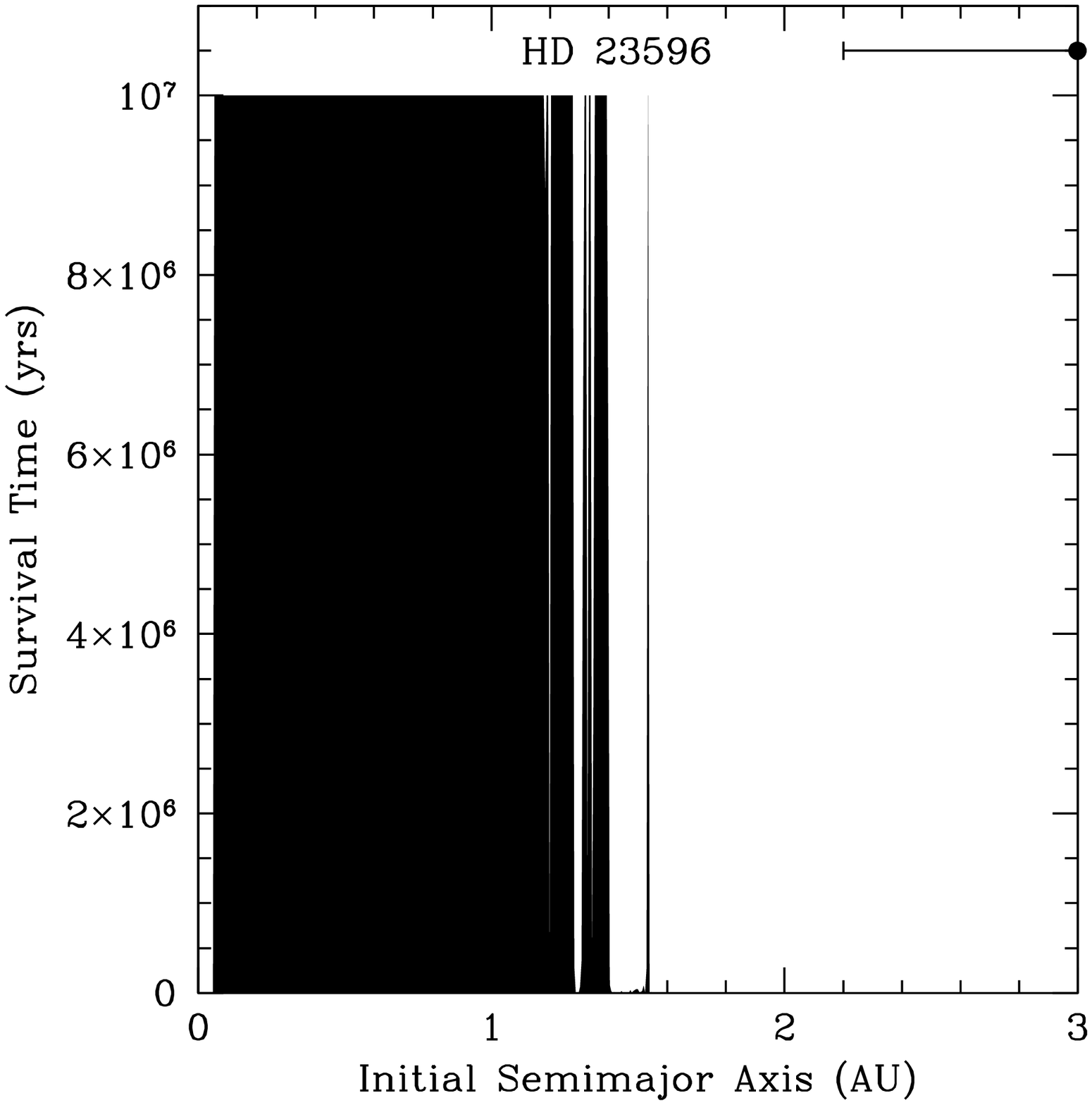}{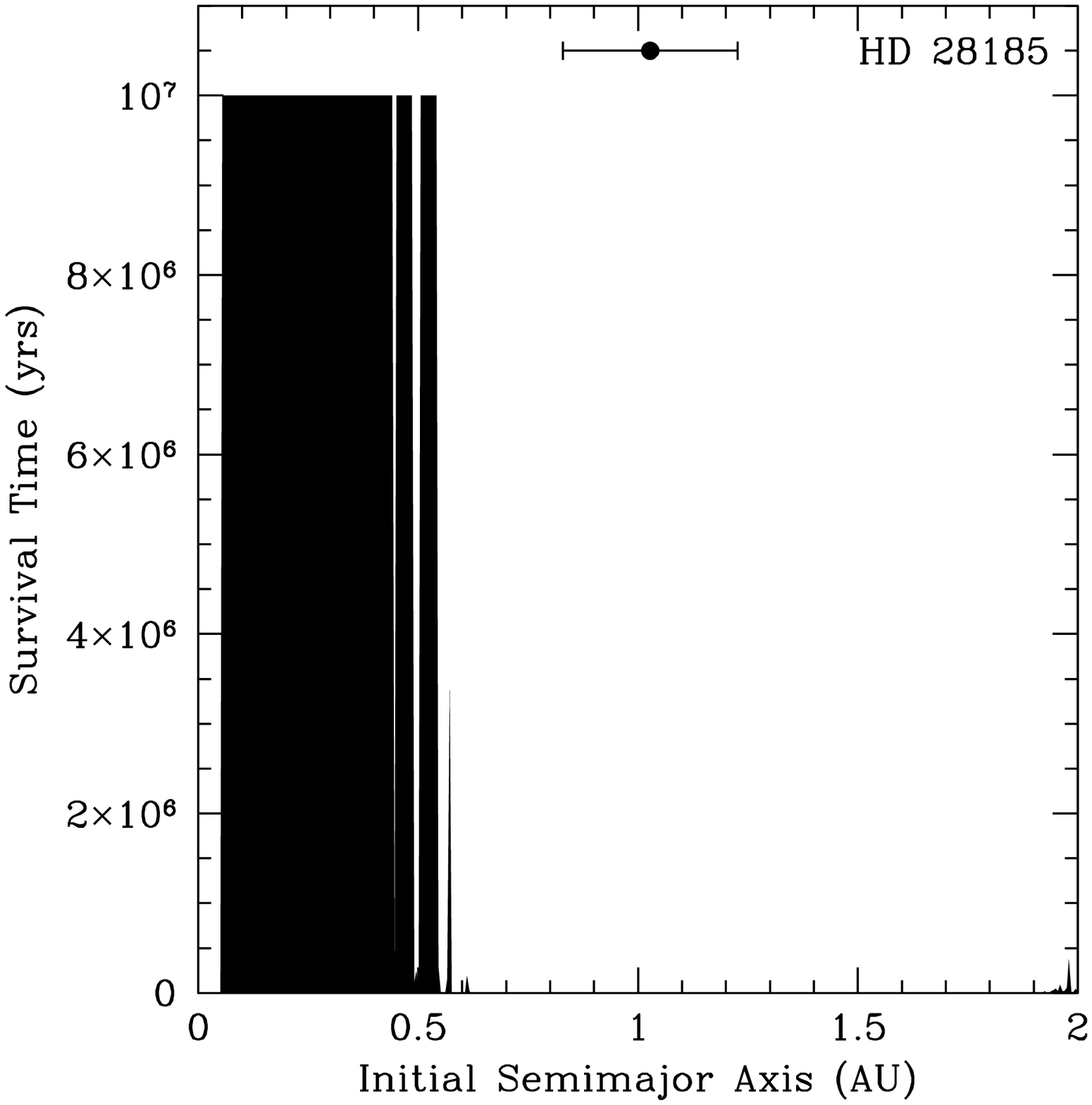}
\caption[Test particle results for HD~23596 and HD~28185]{Same as 
Fig.~\ref{tp1}, but for the HD~23596 (left) and HD~28185
(right) systems. } \label{tp3}
\end{figure}

\begin{figure}
\plottwo{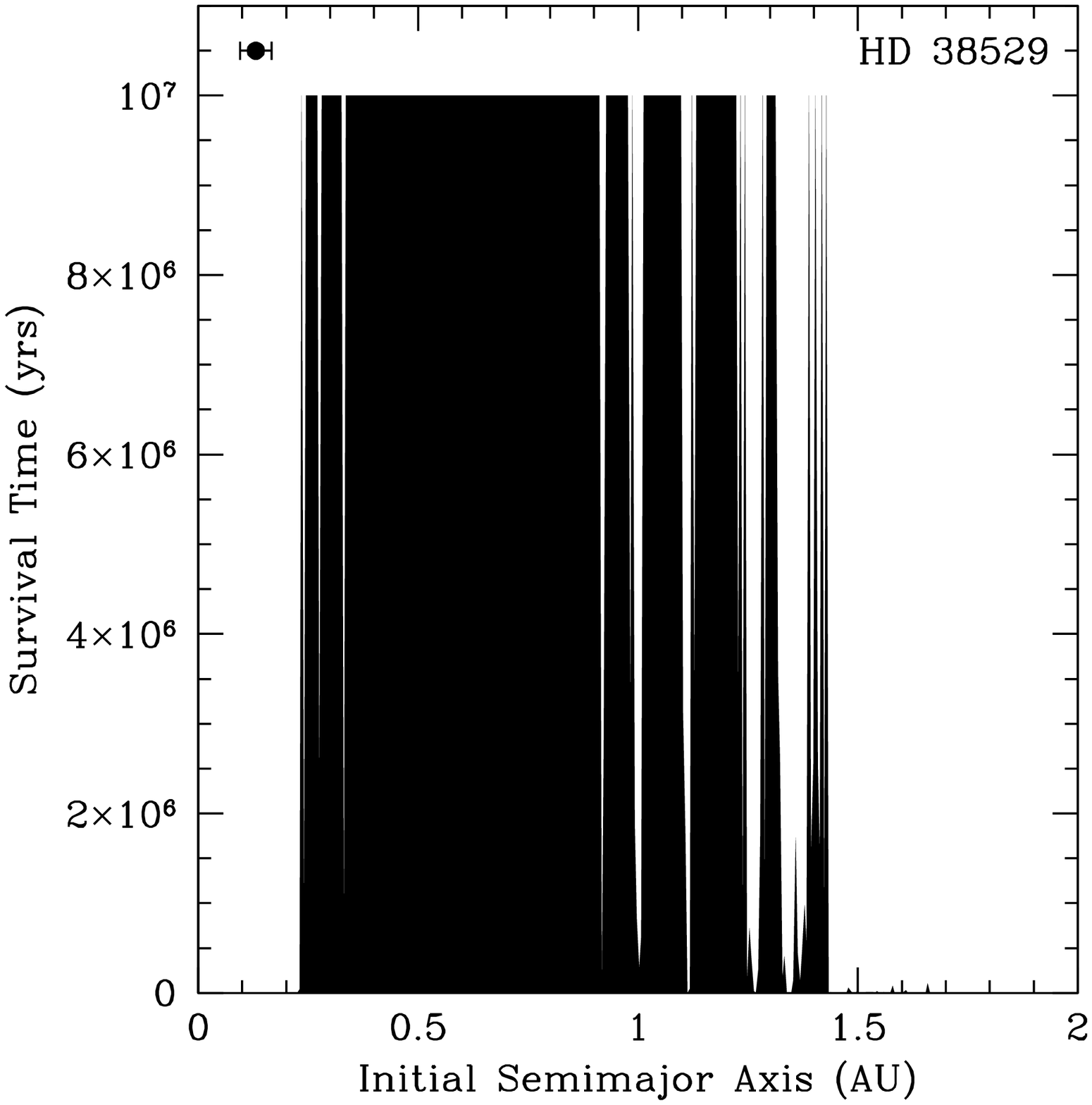}{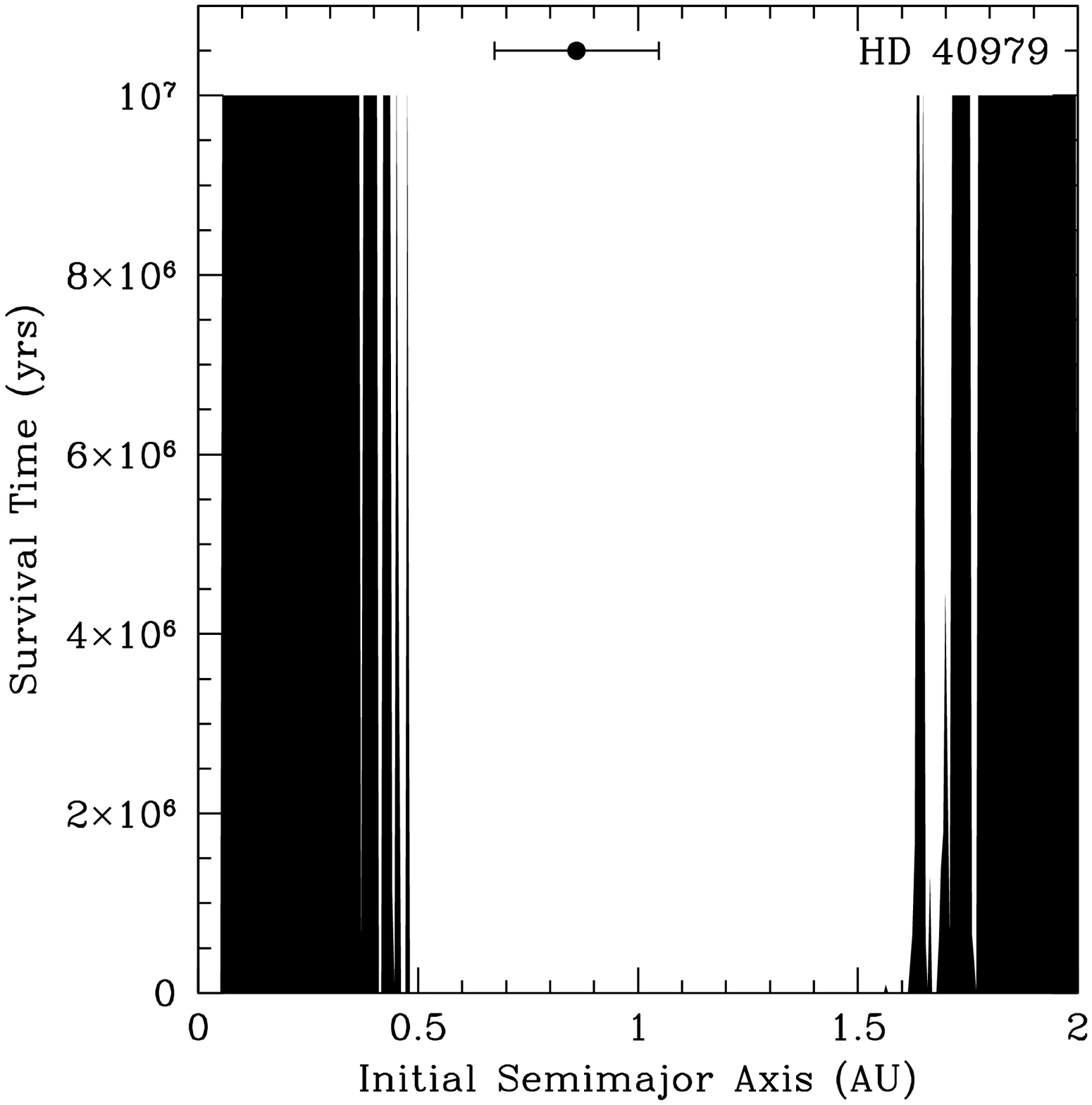}
\caption[Test particle results for HD~38529 and HD~40979]{Same as 
Fig.~\ref{tp1}, but for the HD~38529 (left) and HD~40979 
(right) systems. } \label{tp4}
\end{figure}

\begin{figure}
\plottwo{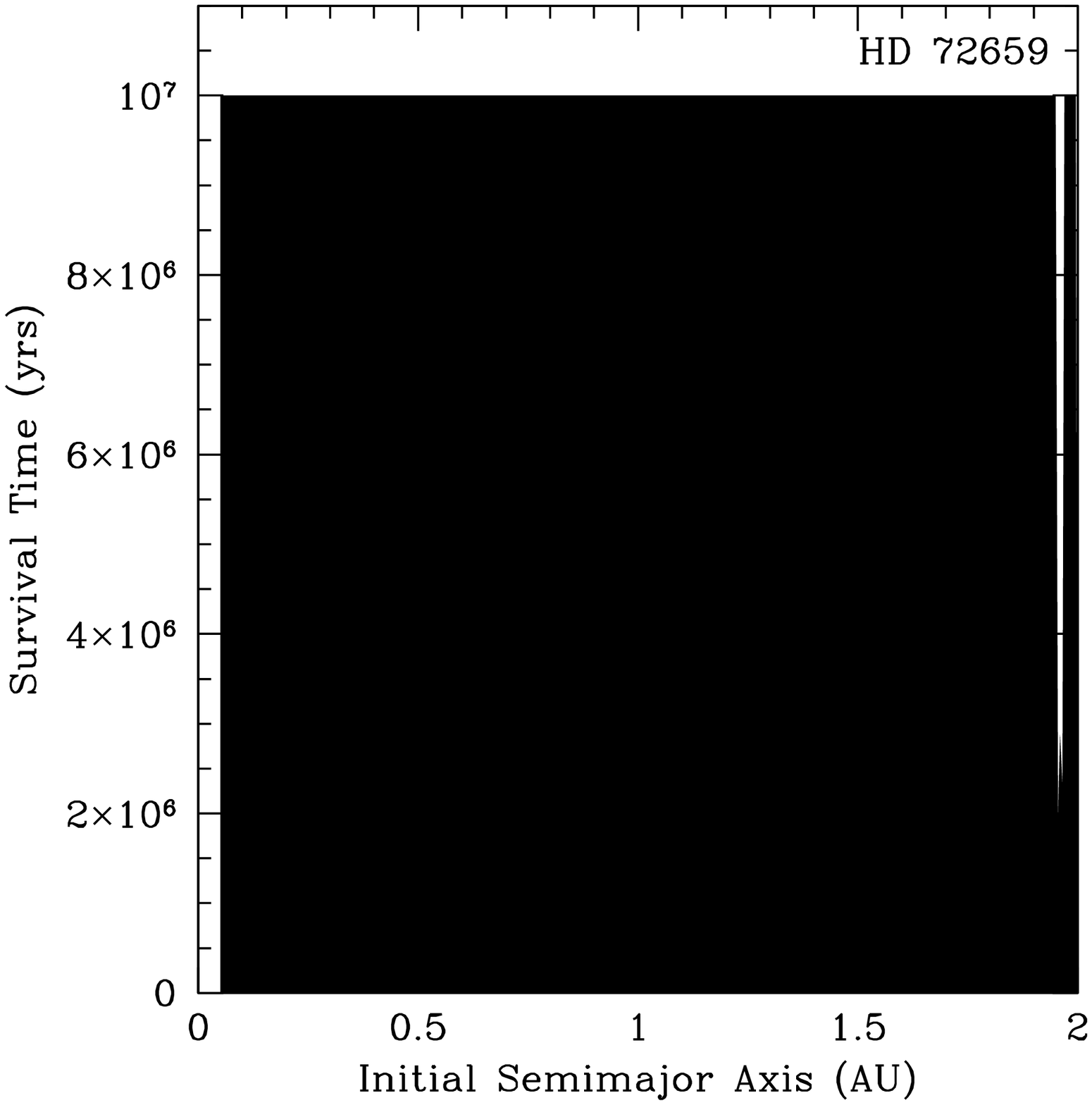}{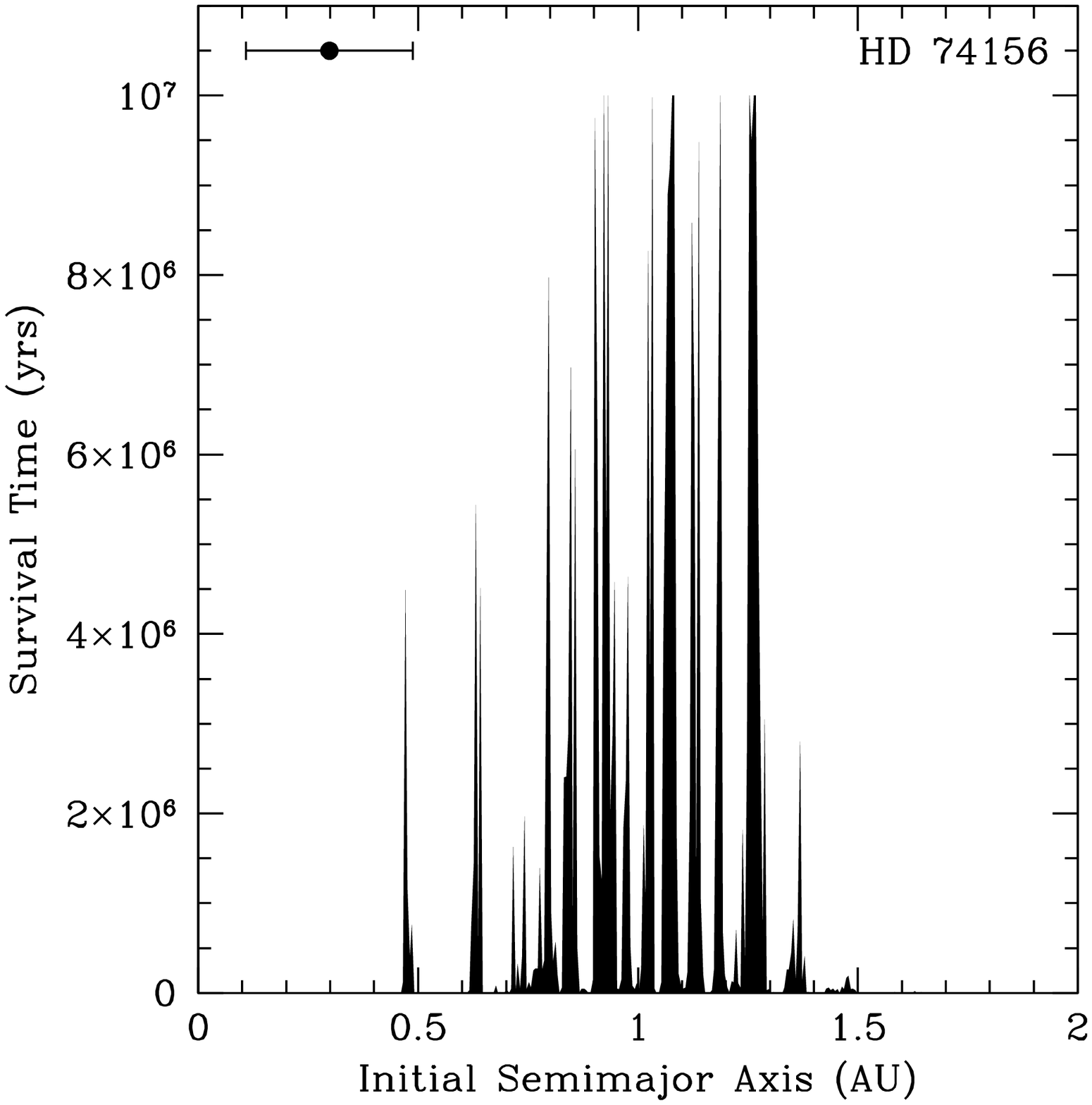}
\caption[Test particle results for HD~72659 and HD~74156]{Same as 
Fig.~\ref{tp1}, but for the HD~72659 (left) and HD~74156
(right) systems.  HD~72659b, with an orbital excursion of 3.48-6.48~AU, is
off the plot.  The recently-announced planet HD~74156d, between planets b
and c, was not included in the simulation, but would reside in the narrow
stable strip. } \label{tp5}
\end{figure}

\begin{figure}
\plottwo{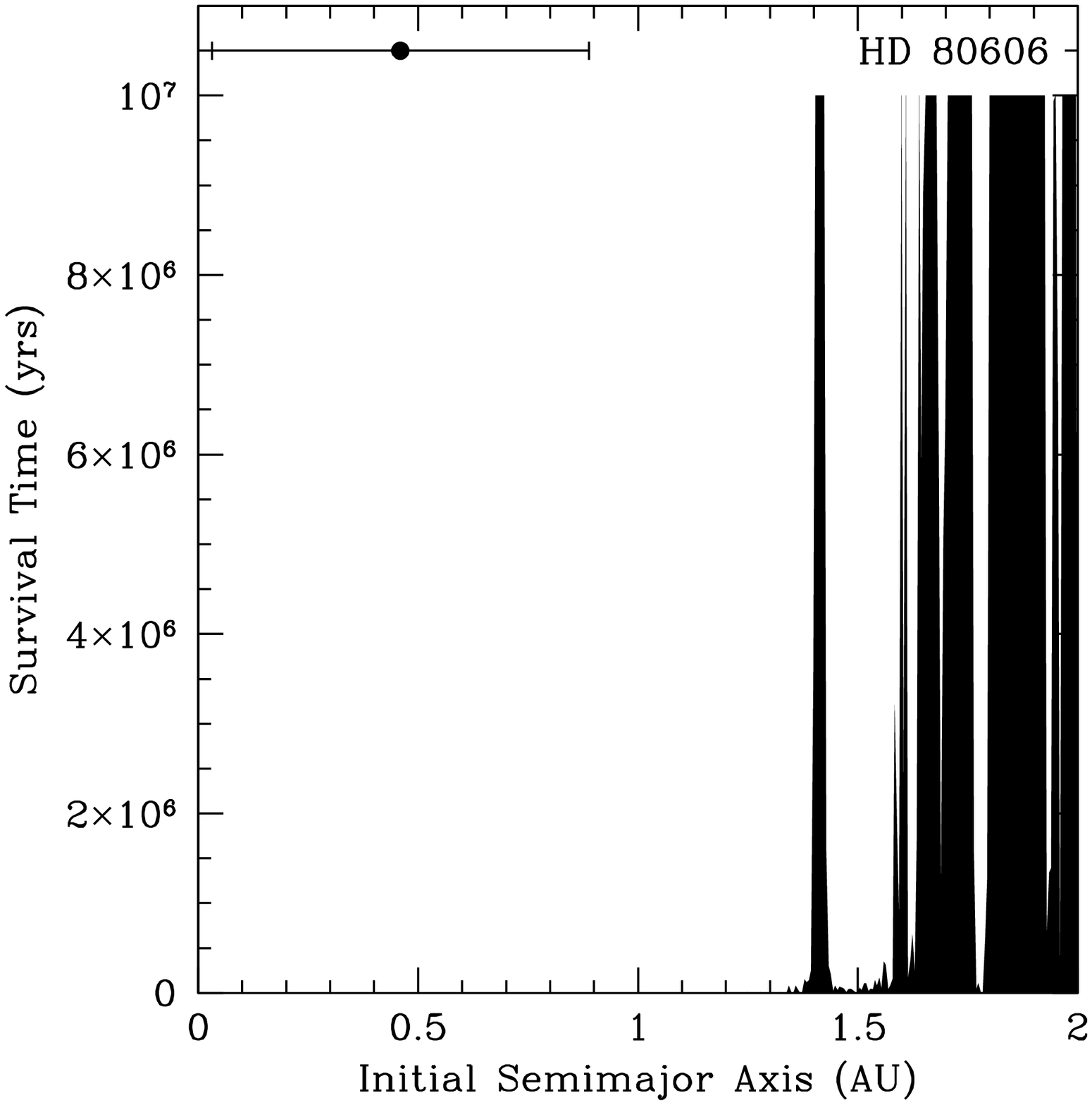}{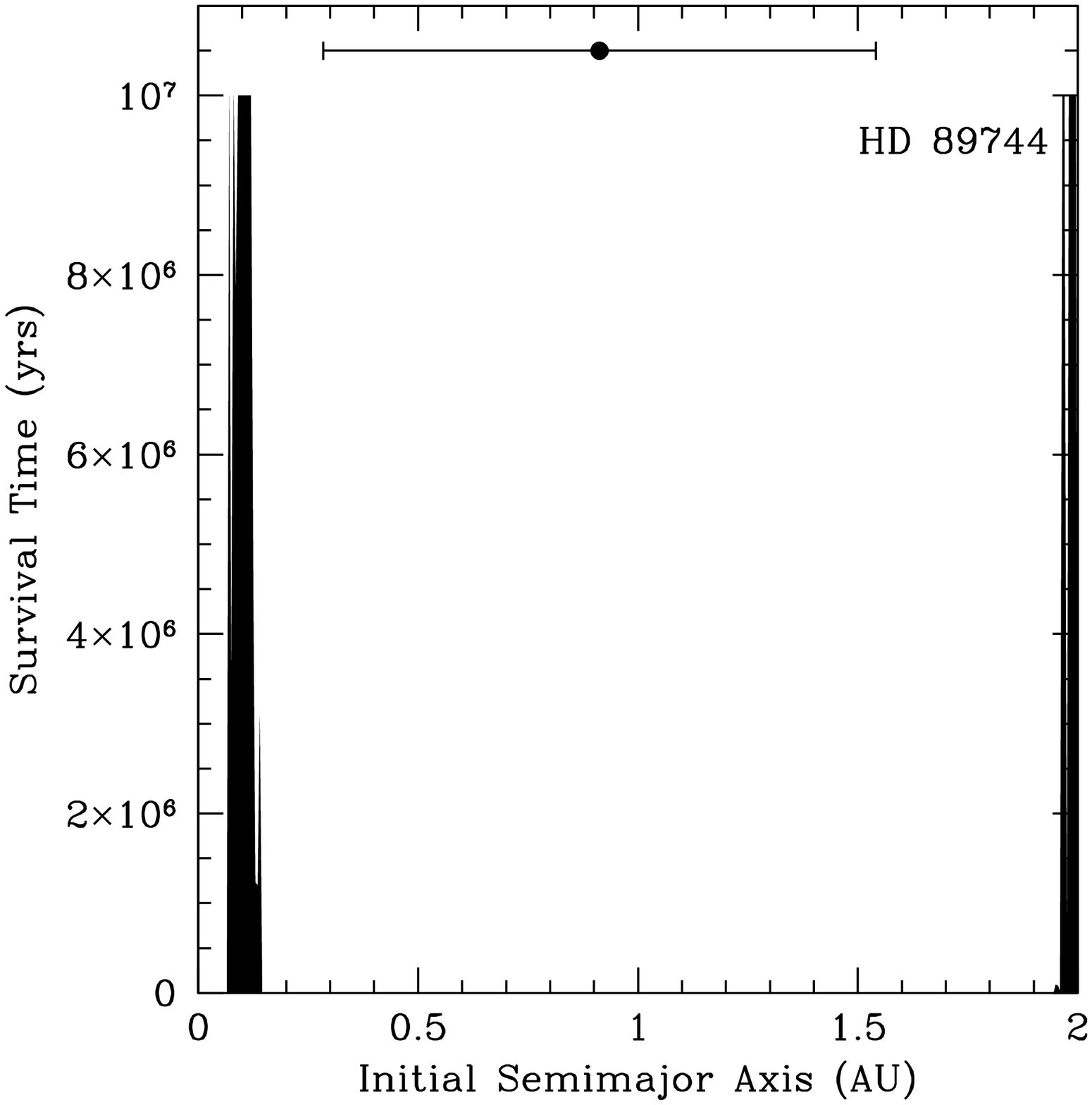}
\caption[Test particle results for HD~80606 and HD~89744]{Same as
Fig.~\ref{tp1}, but for the HD~80606 (left) and HD~89744
(right) systems. } \label{tp6}
\end{figure}

\begin{figure}
\plottwo{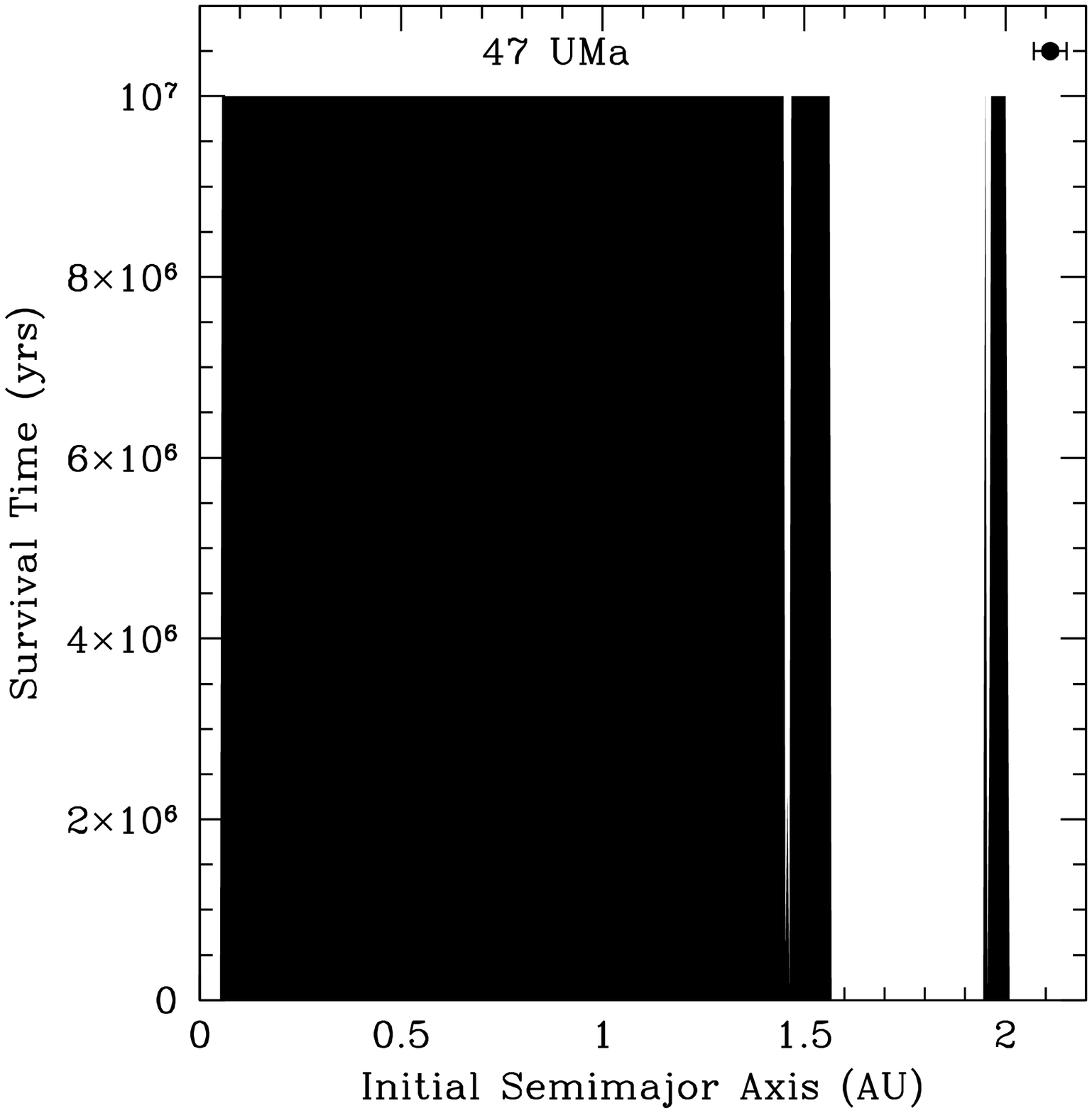}{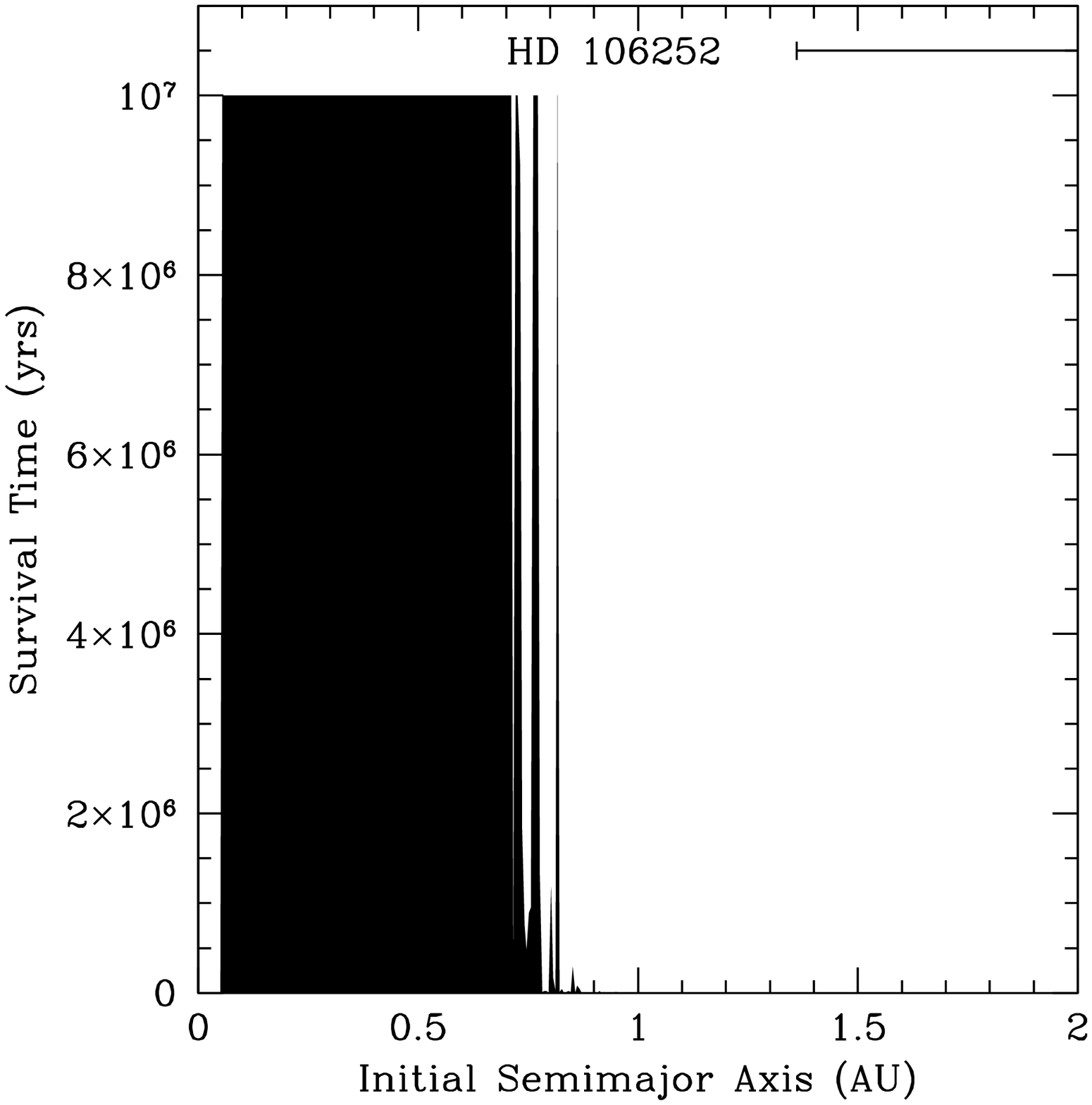}
\caption[Test particle results for 47~UMa and HD~106252]{Same as 
Fig.~\ref{tp1}, but for the 47~UMa (left) and HD~106252 (right) systems. 
Only 47~UMa~b was considered in the simulations.  An outer body would be 
too distant to affect the region under consideration.} \label{tp7}
\end{figure}

\begin{figure}
\plottwo{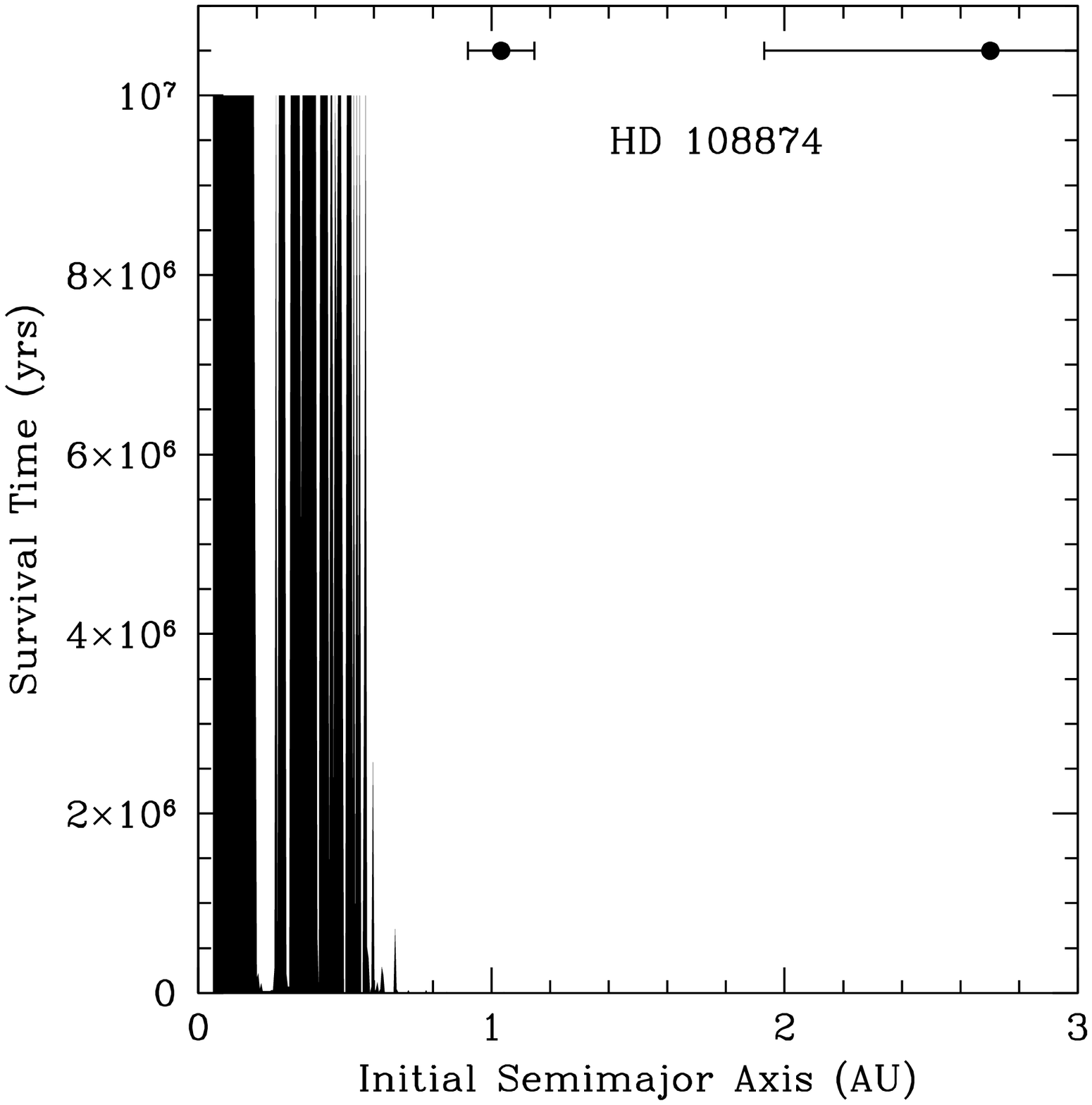}{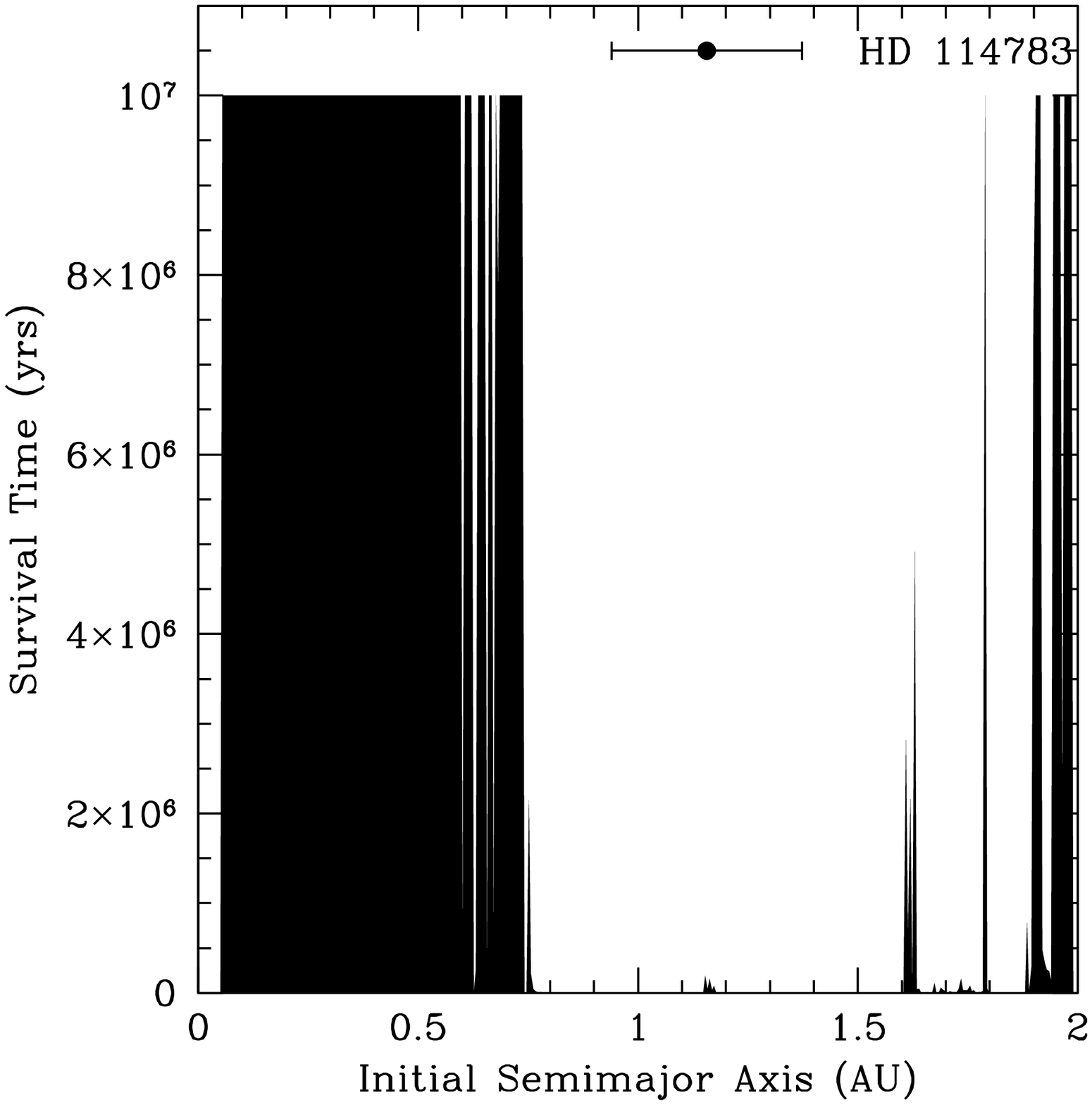}
\caption[Test particle results for HD~108874 and HD~114783]{Same as 
Fig.~\ref{tp1}, but for the HD~108874 (left) and HD~114783 (right) 
systems. } \label{tp8}
\end{figure}

\begin{figure}
\plottwo{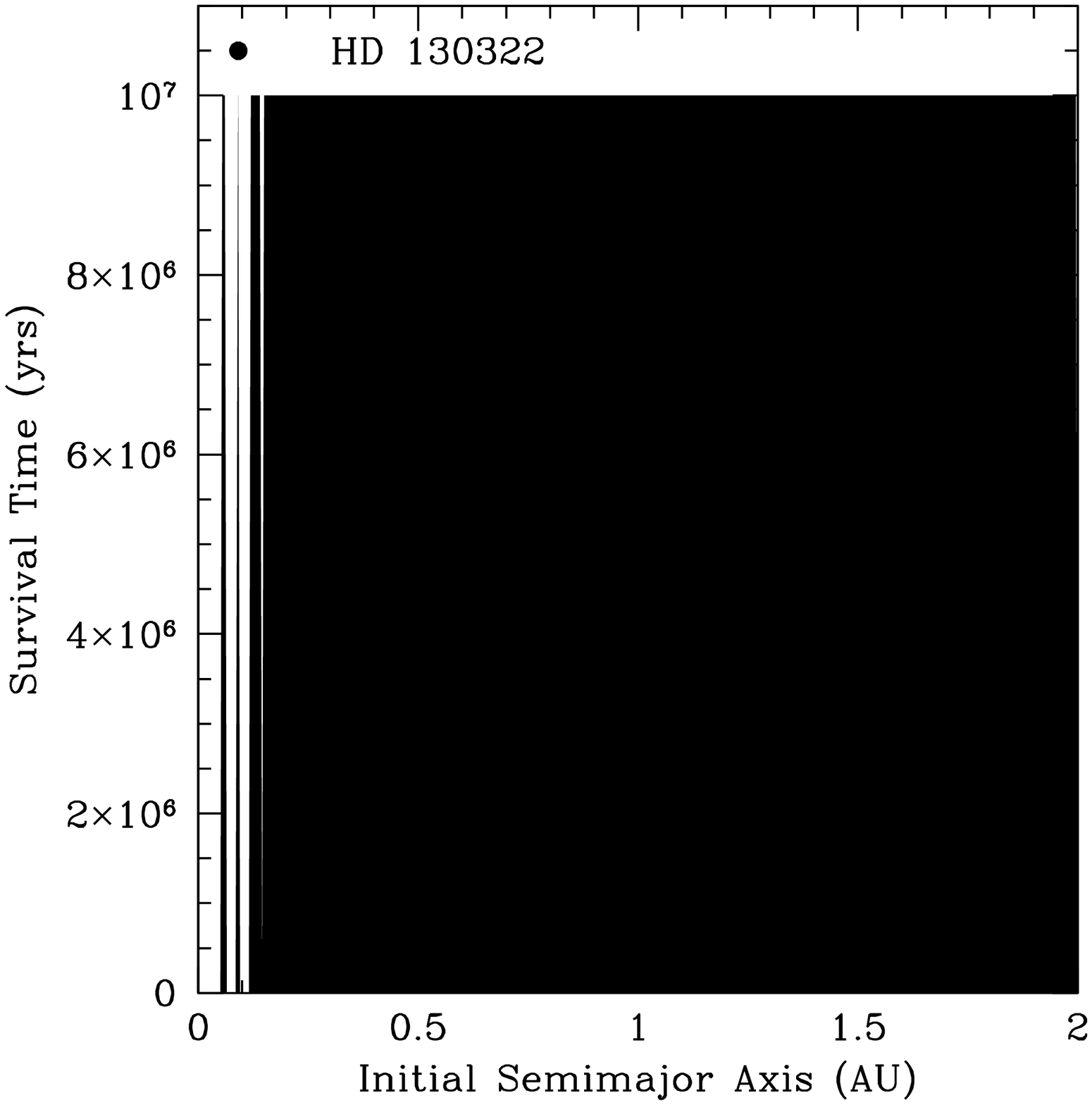}{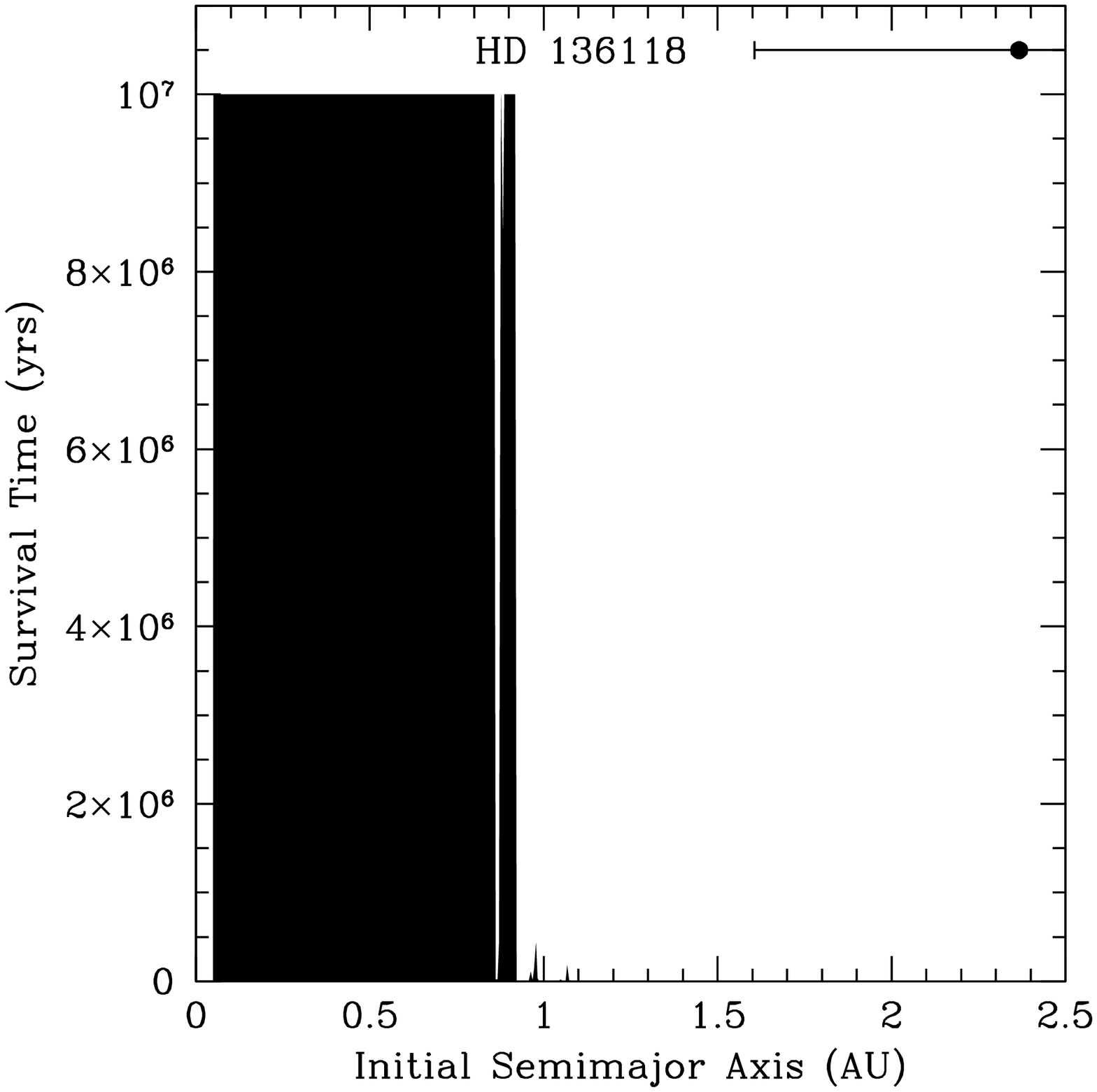}
\caption[Test particle results for HD~130322 and HD~136118]{Same as 
Fig.~\ref{tp1}, but for the HD~130322 (left) and 
HD~136118
(right) systems. } \label{tp9}
\end{figure}

\begin{figure}
\plottwo{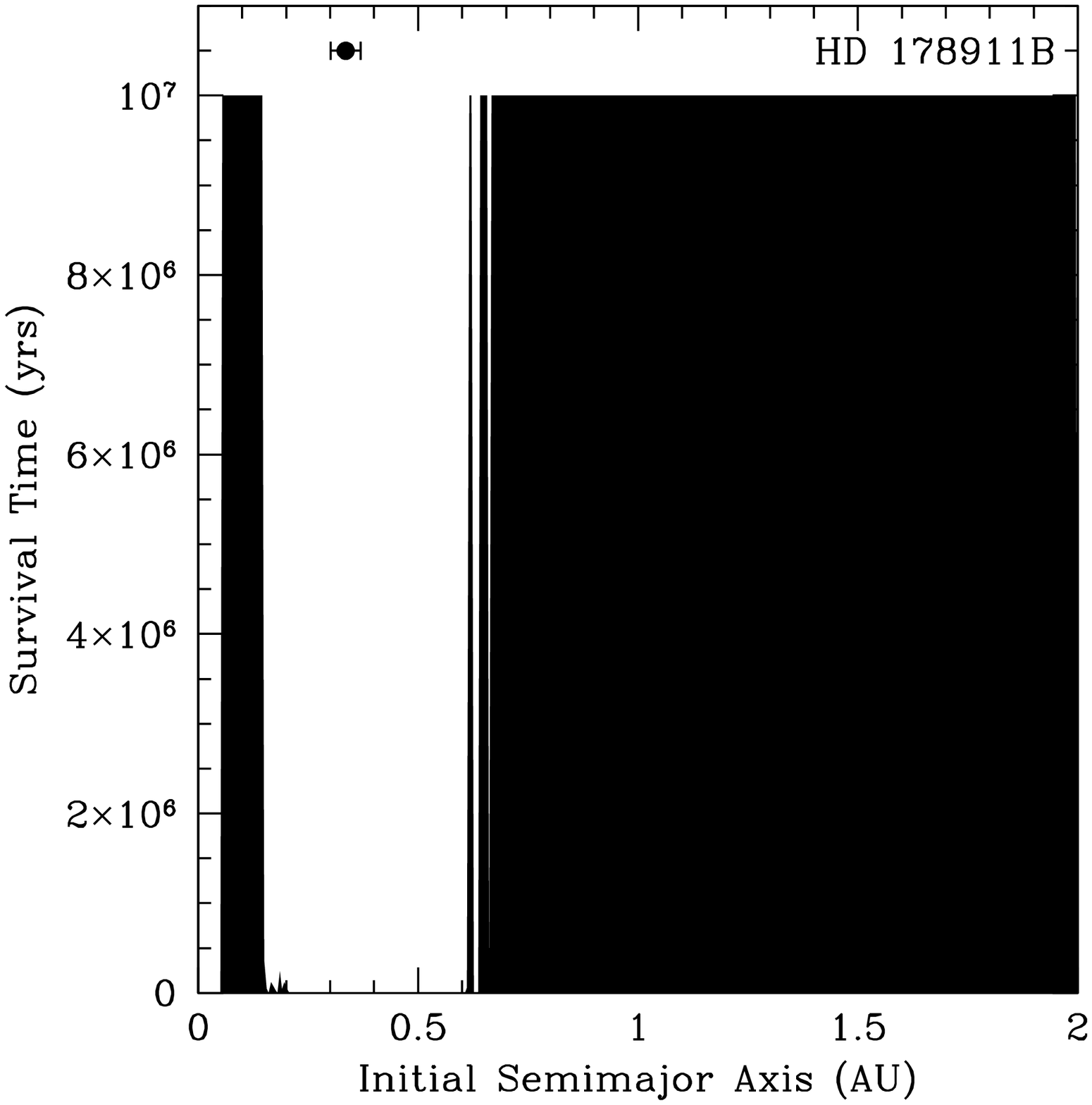}{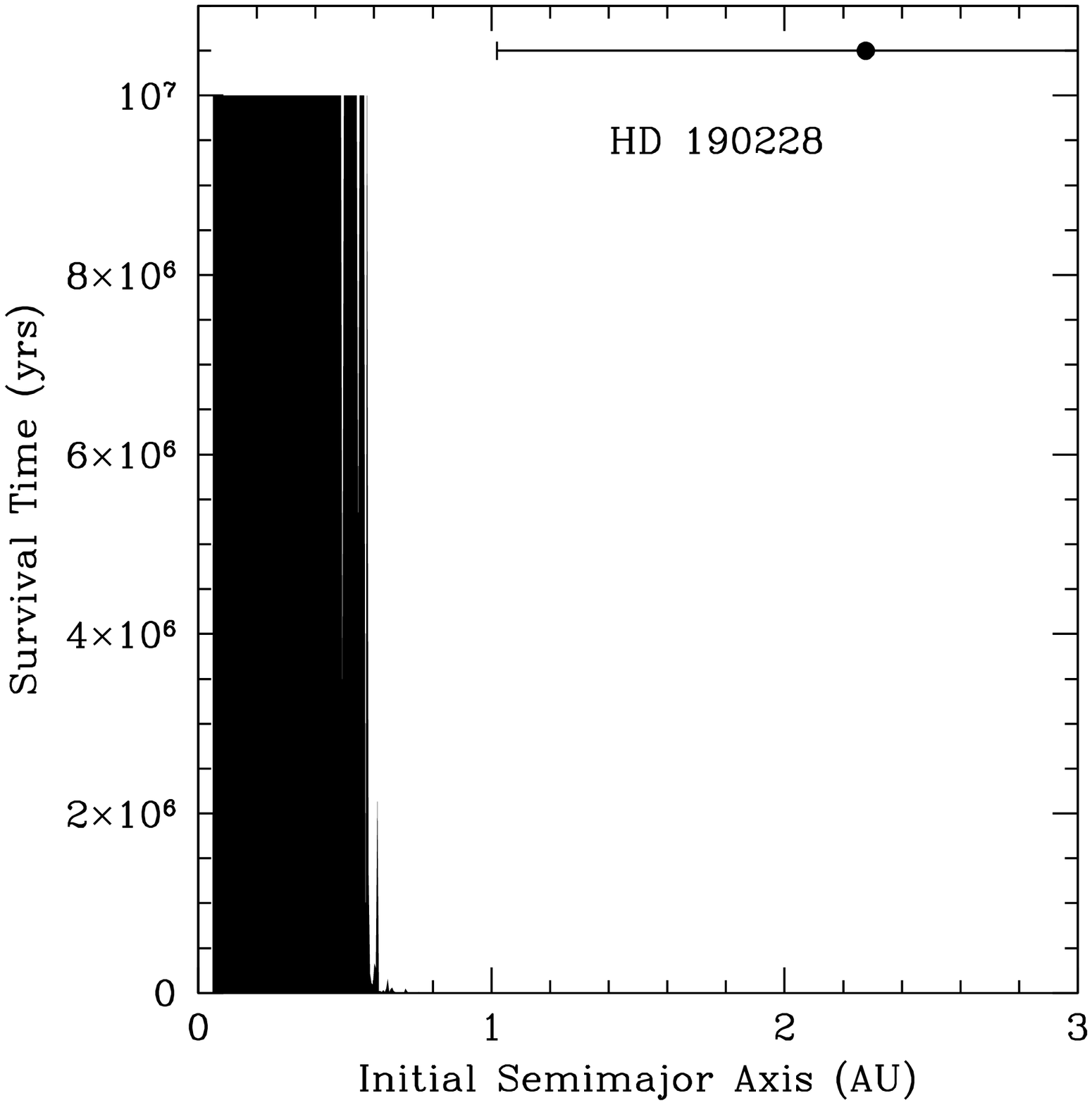}
\caption[Test particle results for HD~178911B and HD~190228]{Same as 
Fig.~\ref{tp1}, but for the HD~178911B (left) and 
HD~190228 (right) systems. } \label{tp10}
\end{figure}

The results of the test-particle simulations are shown in 
Figures~\ref{tp1}-\ref{tp10}.  The survival time of the test particles 
is plotted against their initial semimajor axis.  Two systems targeted 
by the radial-velocity observations were not included in these 
simulations: HD~20367, because there is no evidence for a planet, and 
HD~128311, since the Keplerian orbit solution obtained in \S~3.1 results 
in an unstable system.  As shown in Figure~\ref{tp1}, the short-period 
planet HD~3651b sweeps clean the region inside of about 0.5~AU.  
However, a small number of test particles remained in low-eccentricity 
orbits near the 1:3 and 2:1 mean-motion resonances (MMR).  Since these 
regions lie within the orbital excursion of HD~3651b, these appear to be 
protected resonances.  The eccentricity of the test particles in the 
region of the 1:3 MMR oscillated between 0.00 and 0.31 with a 
periodicity of about $1.2\times~10^5$~yr, while those in the 2:1 
resonance remained at $e\ltsimeq$0.07 throughout the simulation.  All 
particles beyond about 0.6~AU also remained in stable orbits, which is 
not surprising given the low mass of the planet.  In simulations by 
\citet{mandell07} and \citet{raymond06}, a migrating Jupiter-mass planet 
captured planetesimals into low-order resonances, and these accreted 
into terrestrial planets during the 200~Myr run.  The architecture of 
the HD~3651 system, with a 0.2 \Mjup\ planet at 0.3~AU, is similar to 
the configuration modeled by \citet{mandell07}.  Given the stable 
regions evident near the 1:3 and 2:1 resonances for HD~3651b, it is 
possible that terrestrial-mass planets were captured into these regions 
during the migration process.  The detection limits for HD~3651 (\S~5) 
complement the dynamics well, and the current data can place upper 
limits of 1-2 Neptune masses (17-34 Earth masses) on such objects.

For most of the systems, the test-particle results give few surprises.  
Broad stable regions exist interior and exterior to HD~8574b, with the 
inner 0.47~AU retaining 100\% of particles.  For HD~10697 and HD~23596, 
particles remained in the inner 1.35~AU and 1.4~AU, respectively.  The 
HD~19994 system, shown in Figure~\ref{tp2}, proved to be quite 
interesting.  One would expect any particles in orbits which cross that 
of the planet to be removed straightaway, but a few particles remained 
near the 1:1 resonance with the planet, in the range 1.29-1.33~AU.  
\citet{laughlin02} investigated the possibility of planets in a 1:1 
resonance, and concluded that such configurations are indeed possible, 
In the ``eccentric resonance,'' one 1:1 configuration described by 
\citet{laughlin02}, one planet is in a nearly circular orbit while the 
other is in a highly eccentric orbit.  Though the orbits cross, the 
longitudes of pericenter are sufficiently different to avoid close 
encounters.

In the HD~28185 system (Fig.~\ref{tp3}), no stable regions exist 
exterior to the planet out to the maximum separation tested 
($a=2.0$~AU).  Figure~\ref{tp4} shows the results for the HD~38529 and 
HD~40979 systems.  There is a broad region of stability between the 
widely-separated planets HD~38529b and c, consistent with the results of 
\citet{barnes04}.  The outer planet does not fall within the range of 
Fig.~\ref{tp4}, but has an orbital excursion of 2.43--4.99~AU.  For 
HD~74156, the recently-announced planet~d \citep{bean08} in a 346-day 
period between planets~b and c, was not included in the simulation.  
Only those particles in a narrow strip near 1.25~AU survived the full 
10~Myr; planet~d would fall within the stable region.  

The 47~UMa system (Figure~\ref{tp7}) included only the inner planet 
($a=2.11$~AU) for this experiment.  The parameters of an outer planet 
are highly uncertain \citep{destructor, naef04}, and such an object 
would be too distant to affect the inner 2~AU explored here.  A large 
region interior to the planet is stable for the full duration, including 
the habitable zone.  This result is consistent with that of 
\citet{jones01}, who also found the 47~UMa habitable zone to be stable 
for an Earth-mass planet at 1~AU.  With an M~sin~$i$ of 6.9 \Mjup, 
HD~106252b clears out all particles outside of $a\sim$0.7~AU.  For the 
HD~108874 system, no test particles survive between the two planets 
(Figure~\ref{tp8}), but those in the innermost 0.3~AU remain stable.  
Particles interior to HD~114783b were stable to about $a\sim$0.7~AU.  As 
expected for the HD~130322 hot-Jupiter system, all particles with 
$a>0.15$~AU survived (Fig.~\ref{tp9}).  In the HD~178911B system 
(Fig.~\ref{tp10}, some particles remained in the inner 0.1~AU despite 
the large mass (M~sin~$i$=6.95 \Mjup) and relative proximity 
($a=0.34$~AU) of the planet.

\subsection{Massive Body Simulations}

Regions stable for massless test particles may not be stable for massive 
bodies.  Alternatively, regions unstable for test particles may be able 
to host a massive planet.  In the latter case, the existing planet(s) 
may adjust their orbits in response to the perturbation induced by the 
introduced planet.  For these reasons, it is important to also consider 
the effect of massive ``test planets'' in order to obtain a more 
complete dynamical picture of the systems under consideration.  In this 
section, we explore the effect of inserting massive bodies into a known 
planetary system.

\begin{figure}
\plotone{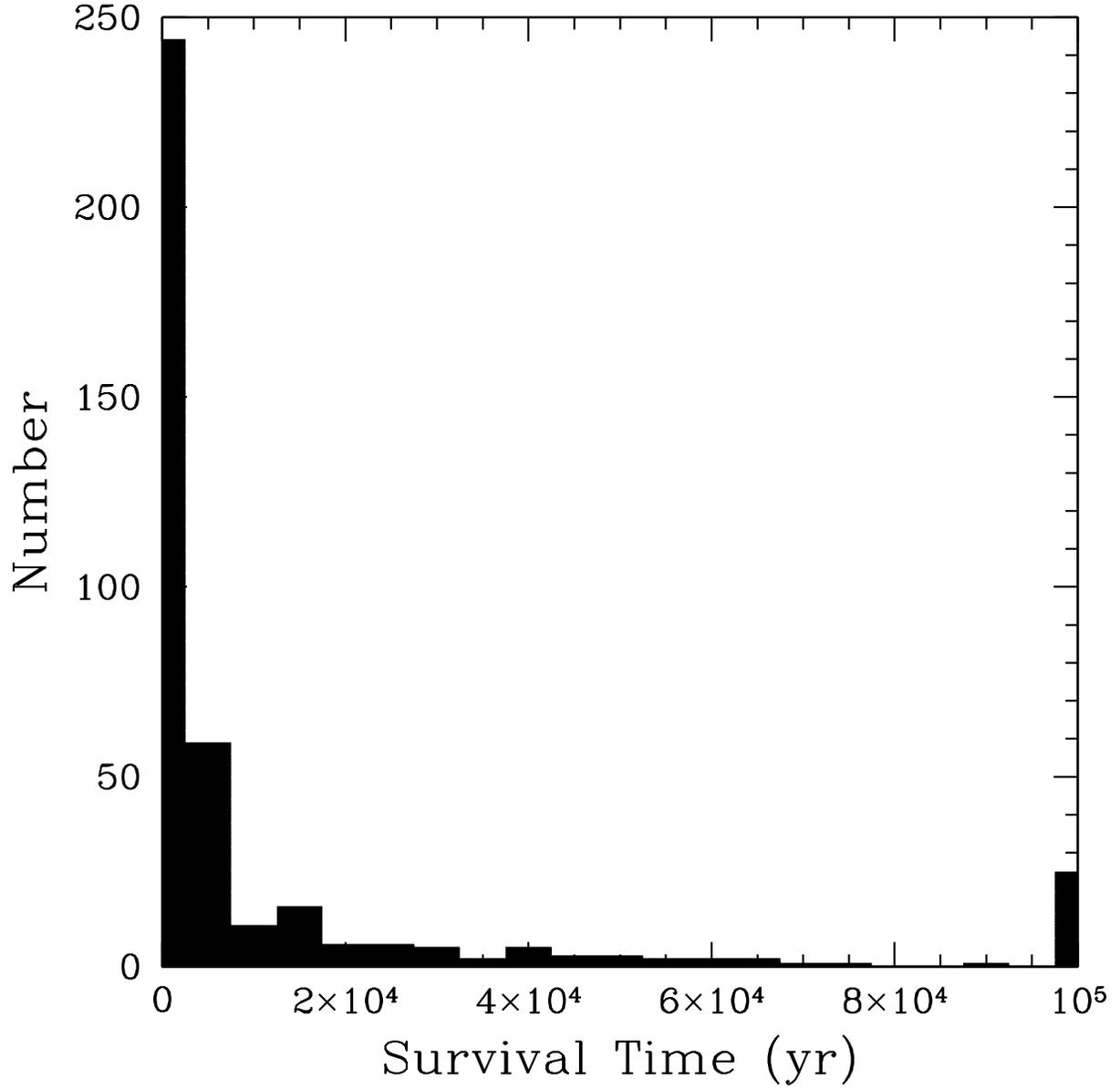}
\caption{Histogram of the survival times for the unstable test 
configurations ($N=352$).  Twenty realizations survived longer than 
$10^5$~yr. }
\label{smashed}
\end{figure}

SWIFT's RMVS3 integrator cannot handle close encounters, when massive 
bodies are closer to each other than 3 Hill radii.  For the massive-body 
simulations, we use the Mercury orbital integrator \citep{chambers99}, 
which has a hybrid feature that switches from an MVS integration to a 
Bulirsch-Stoer method when objects are within 3 Hill radii of each 
other.  General relativistic effects have not been included.  For these 
tests, fictitious planets were placed in each system on initially 
circular orbits at 0.05~AU intervals from 0.05-2.00~AU.  The masses of 
the bodies were set at a Saturn mass (=0.3 \Mjup); this is comparable to 
the mass detectable by the radial-velocity survey, and is the mass used 
by \citet{raymond05} in a similar investigation.  These simulations ran 
for $10^6$~yr, and we observed that unstable configurations usually 
resulted in system destruction within $10^5$~yr.  Figure~\ref{smashed} 
shows a histogram of the survival times for the unstable trials.

\subsection{Massive Body Results}

\begin{figure}
\plottwo{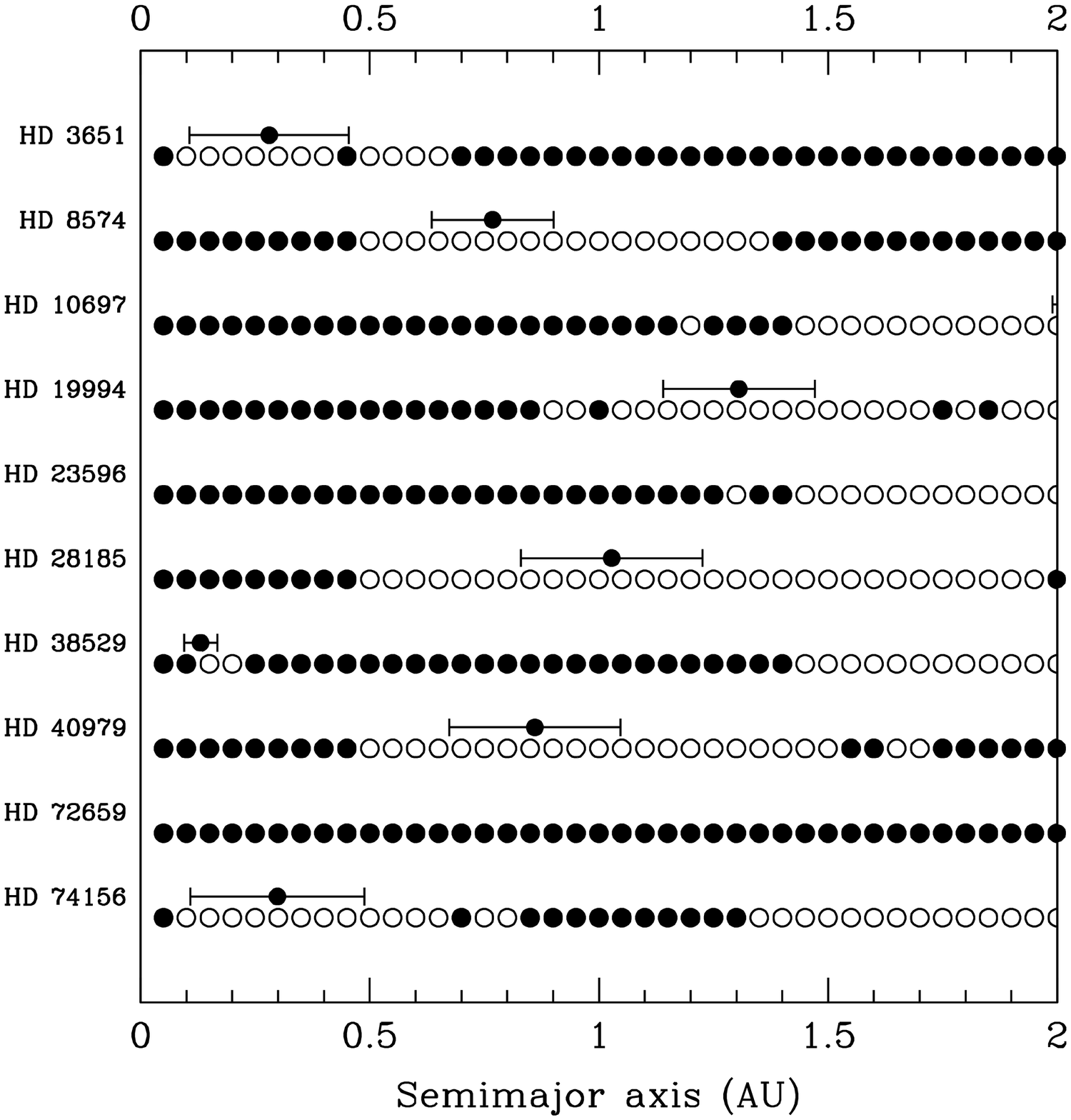}{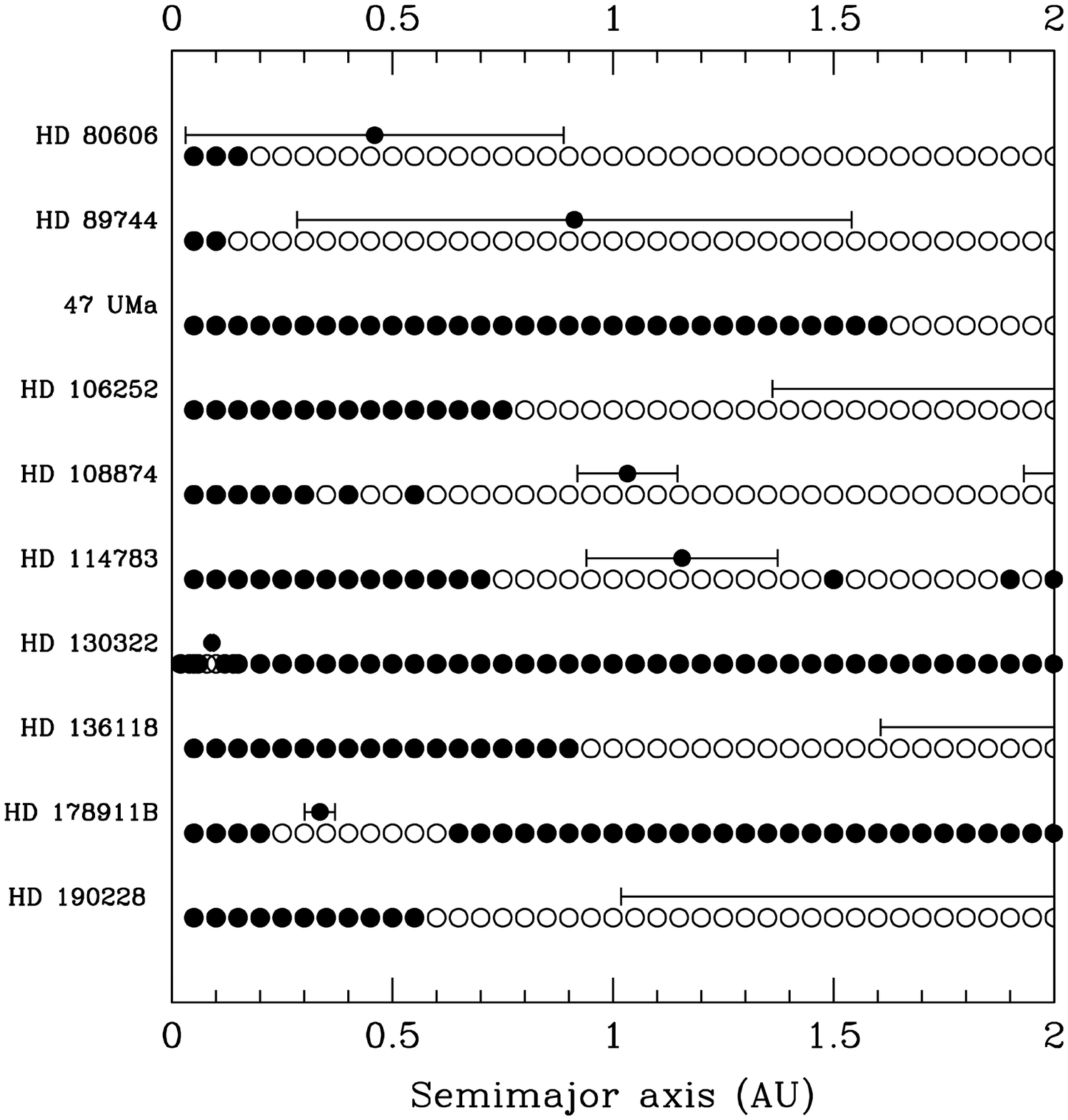}
\caption{Survival of Saturn-mass planets for $10^6$~yr on initially circular 
orbits in 20 planetary systems.  The orbital excursions of the existing 
planets are indicated by the horizontal error bars.  Open circles 
represent unstable locations, filled circles were stable for $10^6$~yr.}
\label{massivebodies}
\end{figure}

\begin{figure}
\plottwo{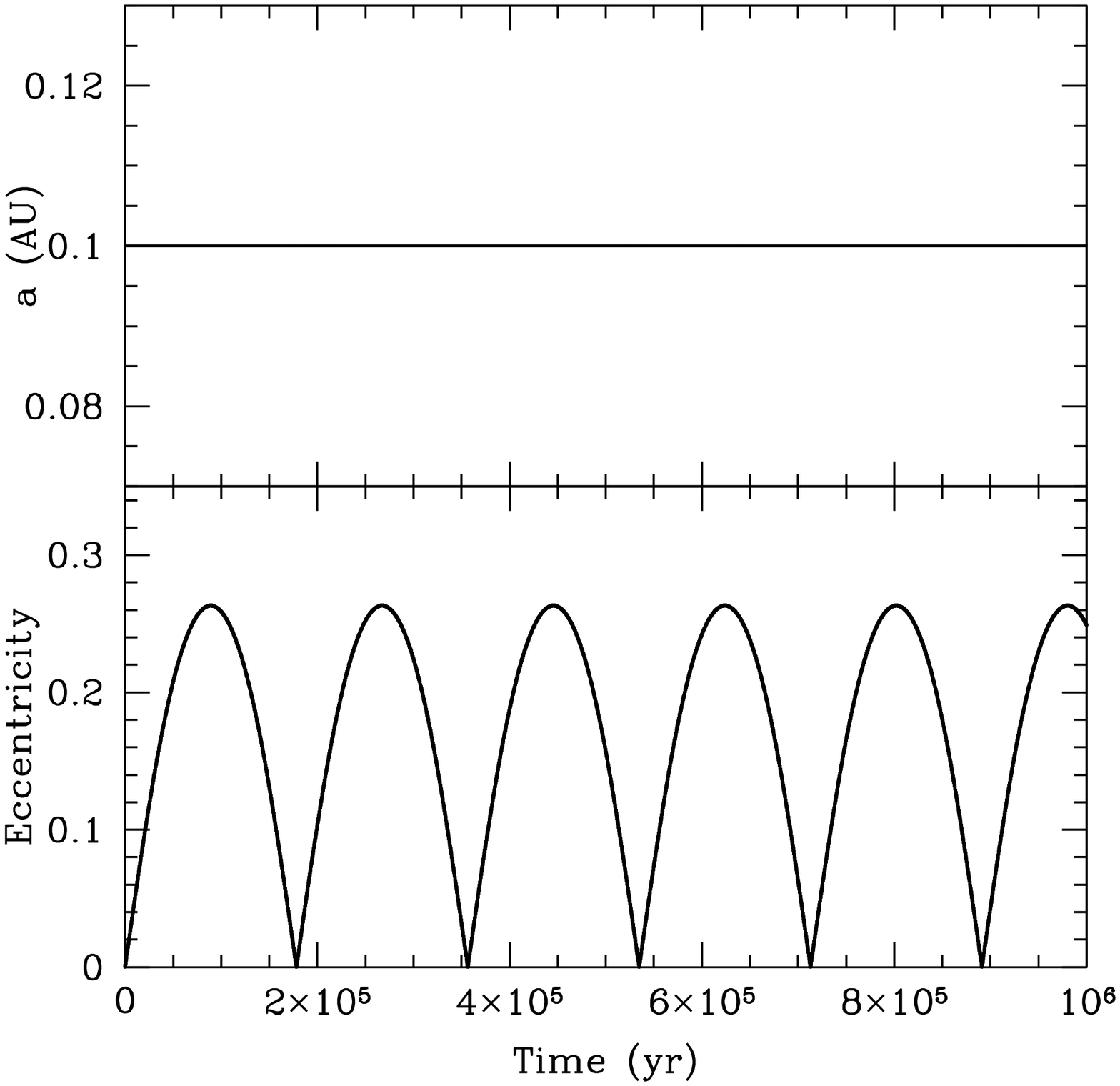}{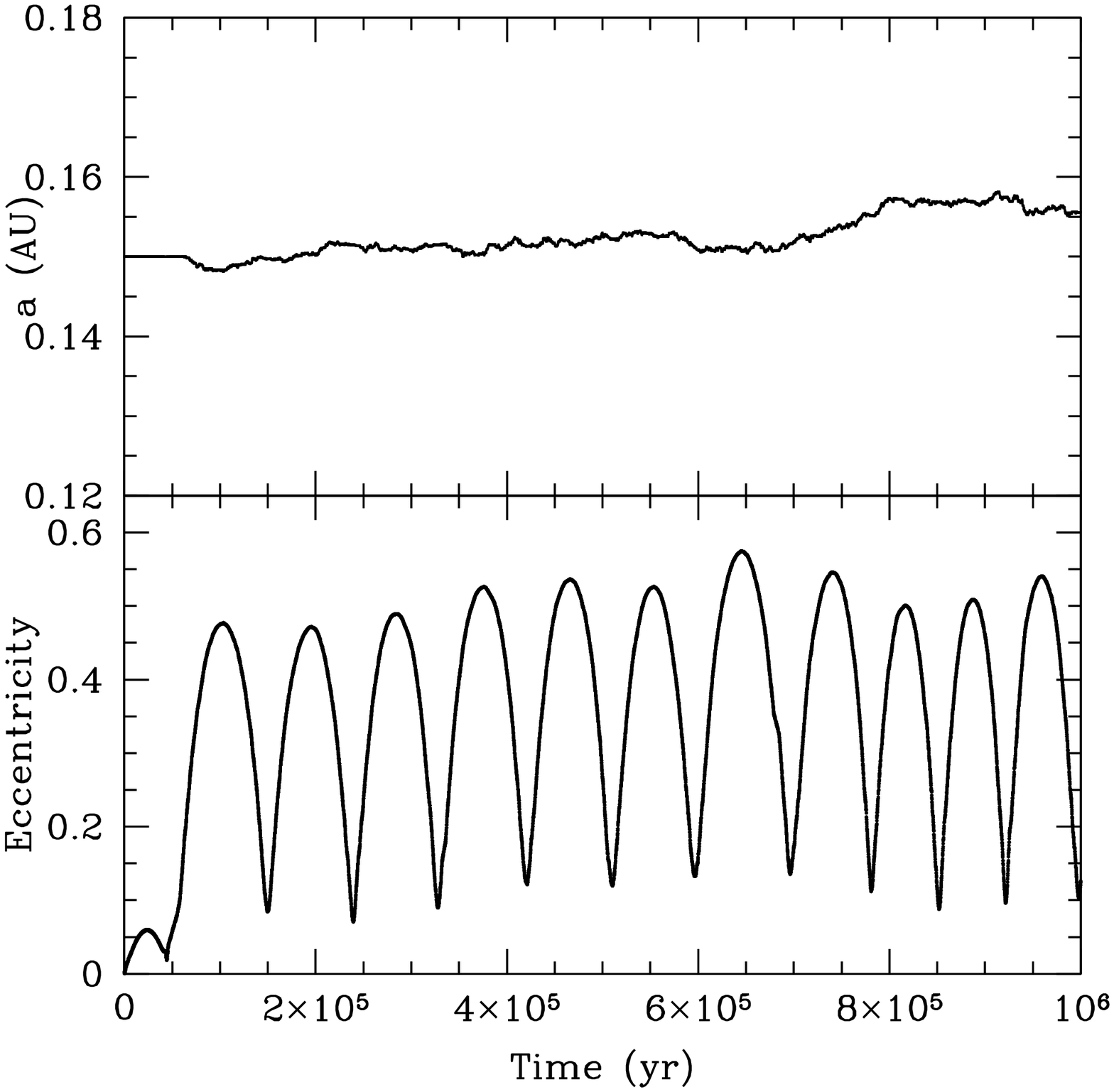}
\caption{Left panel: Behaviour of the semimajor axis (top) and 
eccentricity (bottom) of a Saturn-mass test planet starting at 
$a=0.10$~AU in the HD~80606 system over a $10^6$~yr period.  Right 
panel: Same, but for an object starting at $a=0.15$~AU, which was then 
ejected at $t=5.7\times 10^6$~yr. }
\label{80606}
\end{figure}

The results of the massive-body simulations are shown in 
Figure~\ref{massivebodies}.  The filled circles indicate test planets 
which remained throughout the $10^6$~yr integration.  For most of the 
systems, the regions stable for test particles are also stable for 
Saturn-mass planets.  For HD~3651 and HD~80606, some test planets which 
crossed orbits with the known planet survived.  The 2:1 resonance of 
HD~3651b ($a\sim 0.45$~AU) retained the Saturn-mass planet for 
$10^6$~yr, although its eccentricity varied chaotically, reaching 
$e\sim$0.22.  The HD~80606 system gave the most unexpected result: 
Saturn-mass planets remained in the region $a\le$0.15~AU, despite 
crossing orbits with HD~80606b.  The test planets at 0.05, 0.10, and 
0.15~AU reached maximum eccentricities of 0.13, 0.26, and 0.57, 
respectively.  For the test planets at 0.05 and 0.10~AU, the 
oscillations in eccentricity were regular in period and constant in 
amplitude, whereas for $a=0.15$~AU, the oscillations varied in period 
and increased in amplitude toward the end of the $10^6$~yr simulation 
(Figure~\ref{80606}).  For the two cases in which the test planets 
exhibited irregular variations in eccentricity, the simulations were 
continued for $10^7$~yr, anticipating the eventual destruction of the 
system.  The test planet at 0.45~AU in the HD~3651 system caused the 
ejection of HD~3651b after 1.8$\times 10^6$~yr.  Likewise, the test 
planet at 0.15~AU in the HD~80606 system was ejected after 5.7$\times 
10^6$~yr.

\section{Detection Limits}

\subsection{Methods}

In \citet{limitspaper}, we described a detection-limits algorithm 
implemented on the sample of constant stars from the long-term planet 
search at McDonald Observatory.  This approach was based on that used by 
\citet{endl02} to derive detection limits from their survey with the ESO 
Coude Echelle Spectrometer.  In brief, we add a Keplerian signal to the 
existing velocity data, then attempt to recover that signal using a 
Lomb-Scargle periodogram.  The mass of the simulated planet is increased 
until 99\% of the injected signals are recovered with FAP$<$0.1\%.  For 
the constant stars in \citet{limitspaper}, the null hypothesis is that 
no planets are present, and so the detection-limit algorithm can be 
applied directly to the velocity data.  In the case of the known planet 
hosts, this null hypothesis no longer applies, and it would not do to 
``pre-whiten'' those data by removing the known planet's orbit as if its 
parameters were known perfectly.  The presence of an additional planet 
will act to modify the fitted parameters of the known planet.  If two or 
more planets are present, and only one has been fitted, then part of the 
signal from the additional planets can be absorbed into the orbital 
elements of the 1-planet fit.  To approach this task with the maximum 
rigor, these effects must be accounted for.  Hence, the detection-limit 
algorithm was modified in the following way: the test Keplerian signal 
was added to each of the \textit{original data sets}, then these 
modified data sets were fitted for the known planet(s) using GaussFit.  
A residuals file was generated and then subjected to the periodogram 
search as described above.  This process of fitting and removing the 
known planet occurred for \textit{every} injected test signal.  This 
method has the advantage of being essentially identical to the 
planet-search method described in \S~3.1.

\subsection{Results}

\begin{figure}
\plottwo{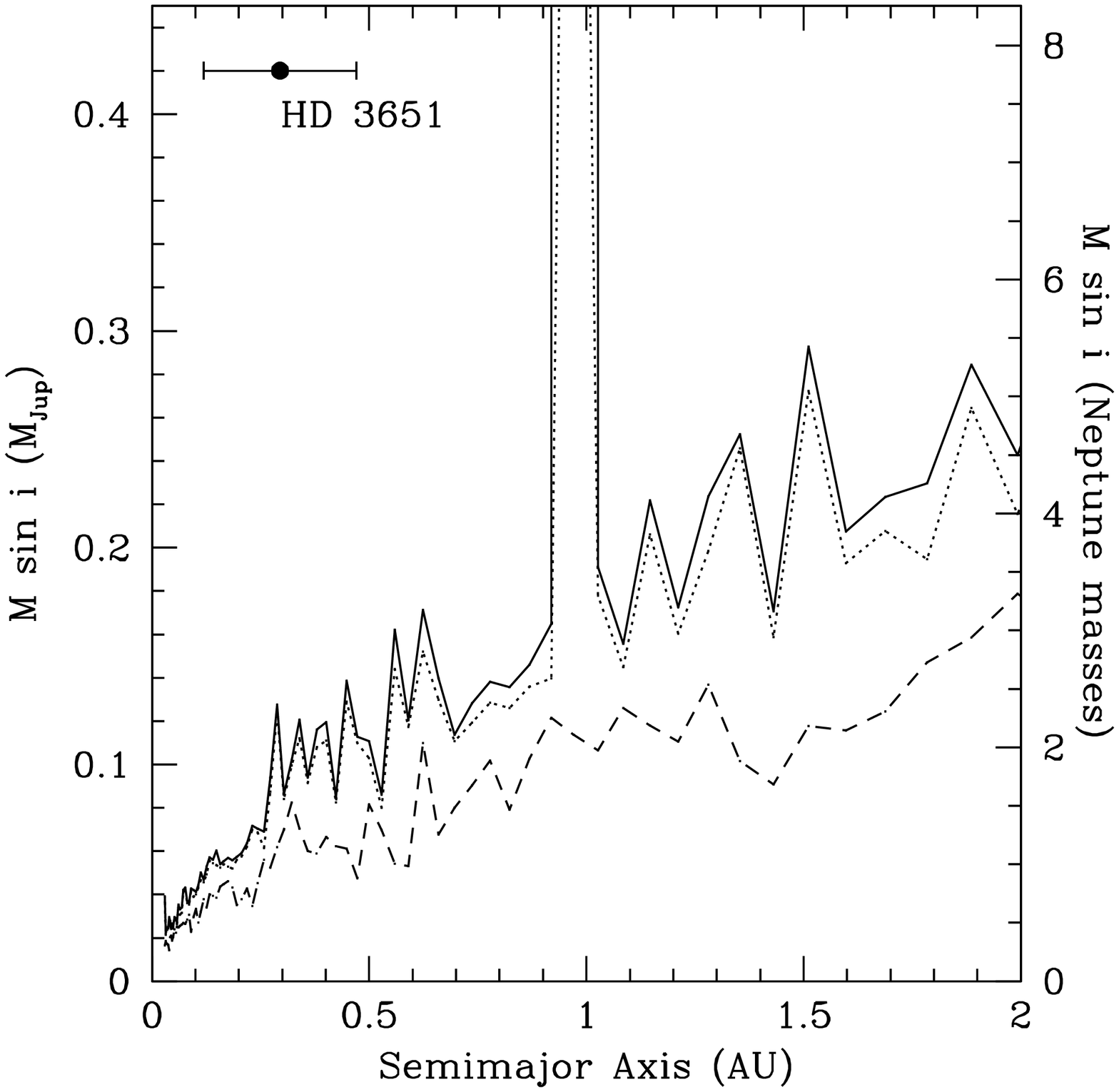}{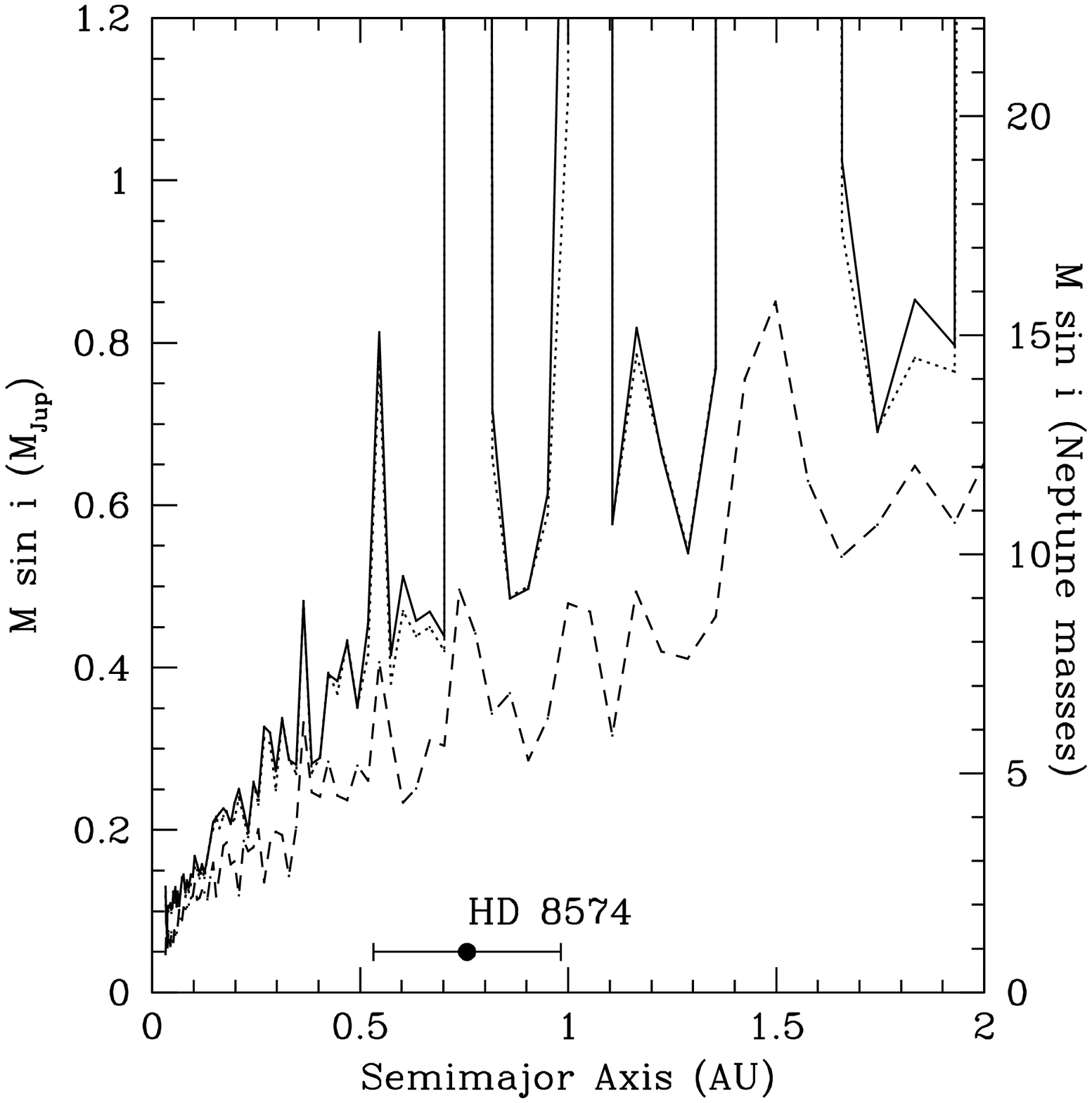}
\caption{Left panel: Detection limits for additional planets in orbits 
with $e=0.20$ in the HD~3651 system (solid line).  This value represents 
the mean eccentricity of surviving test particles from the dynamical 
simulations discussed in \S~4.  Planets in the parameter space above the 
solid line are excluded at the 99\% confidence level.  Limits for 
planets in circular orbits are shown as dotted (99\% recovery) and 
dashed (50\% recovery) lines.  Right panel: Same, but for HD~8574 (solid 
line: $e=0.10$).}
\label{limits1}
\end{figure}

\begin{figure}
\plottwo{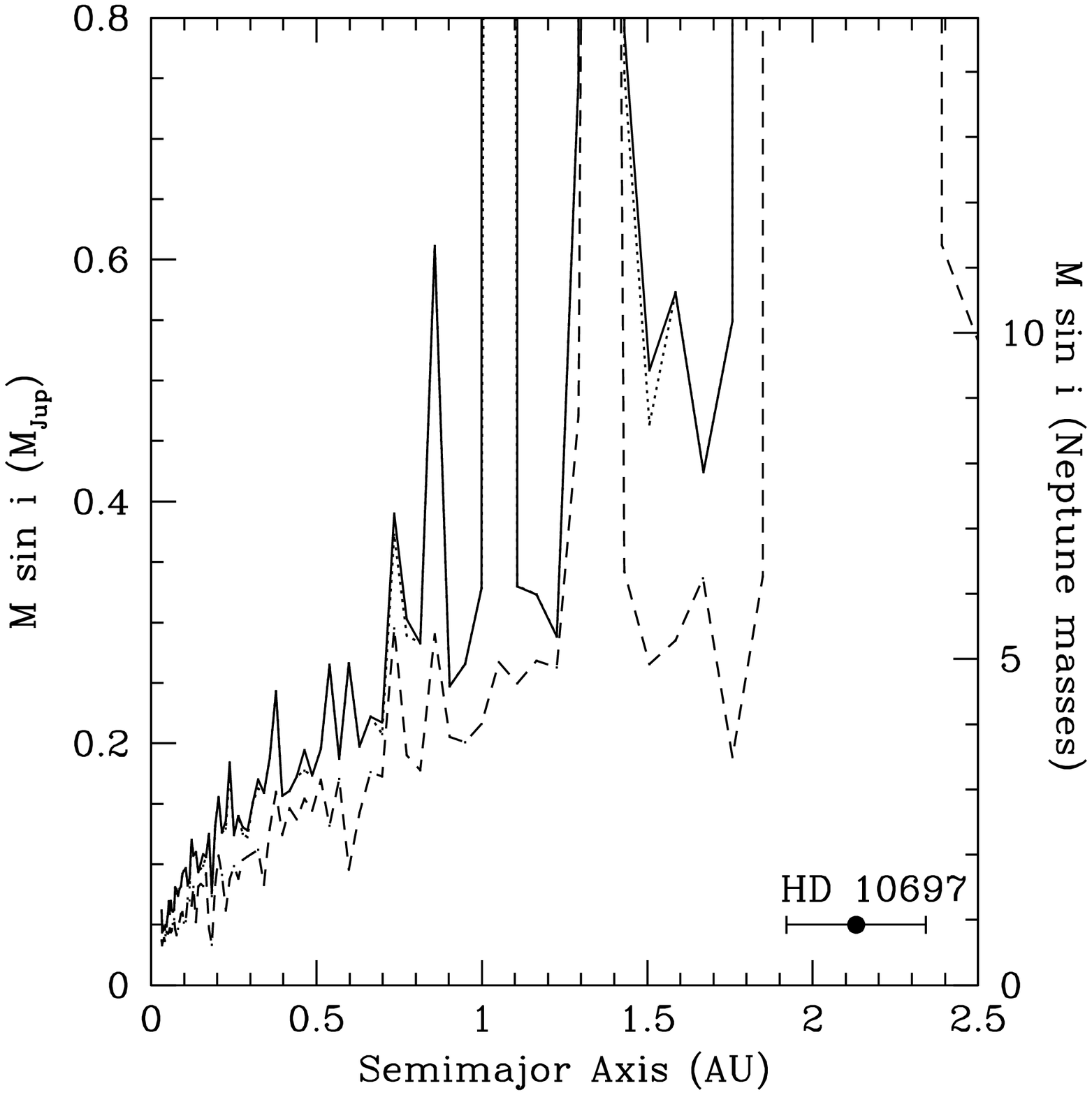}{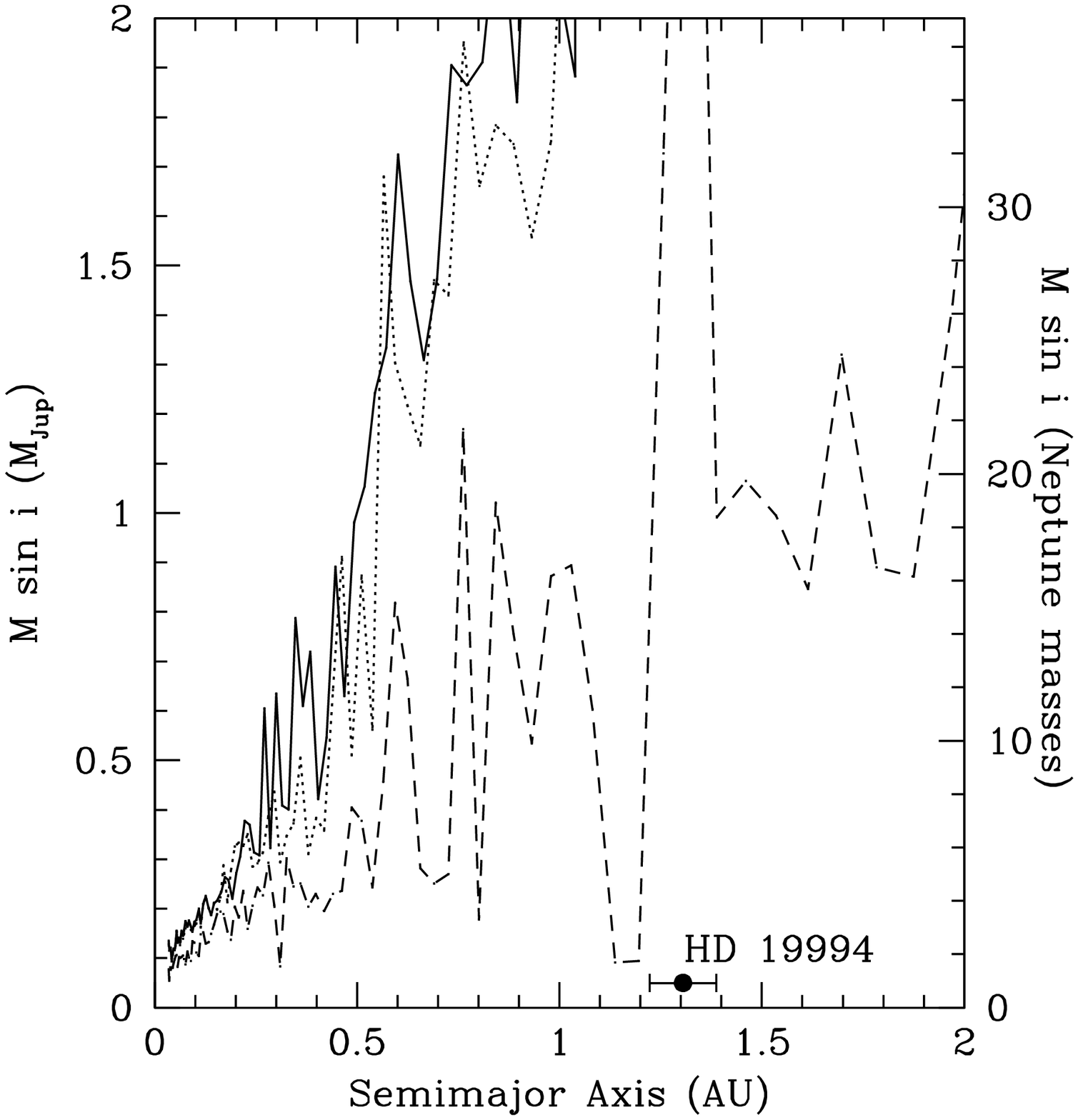}
\caption{Left panel: Same as Fig.~\ref{limits1}, but for HD~10697 (solid 
line: $e=0.04$).  Right panel: HD~19994 (solid line: $e=0.09$).} 
\label{limits2}
\end{figure}

\begin{figure}
\plottwo{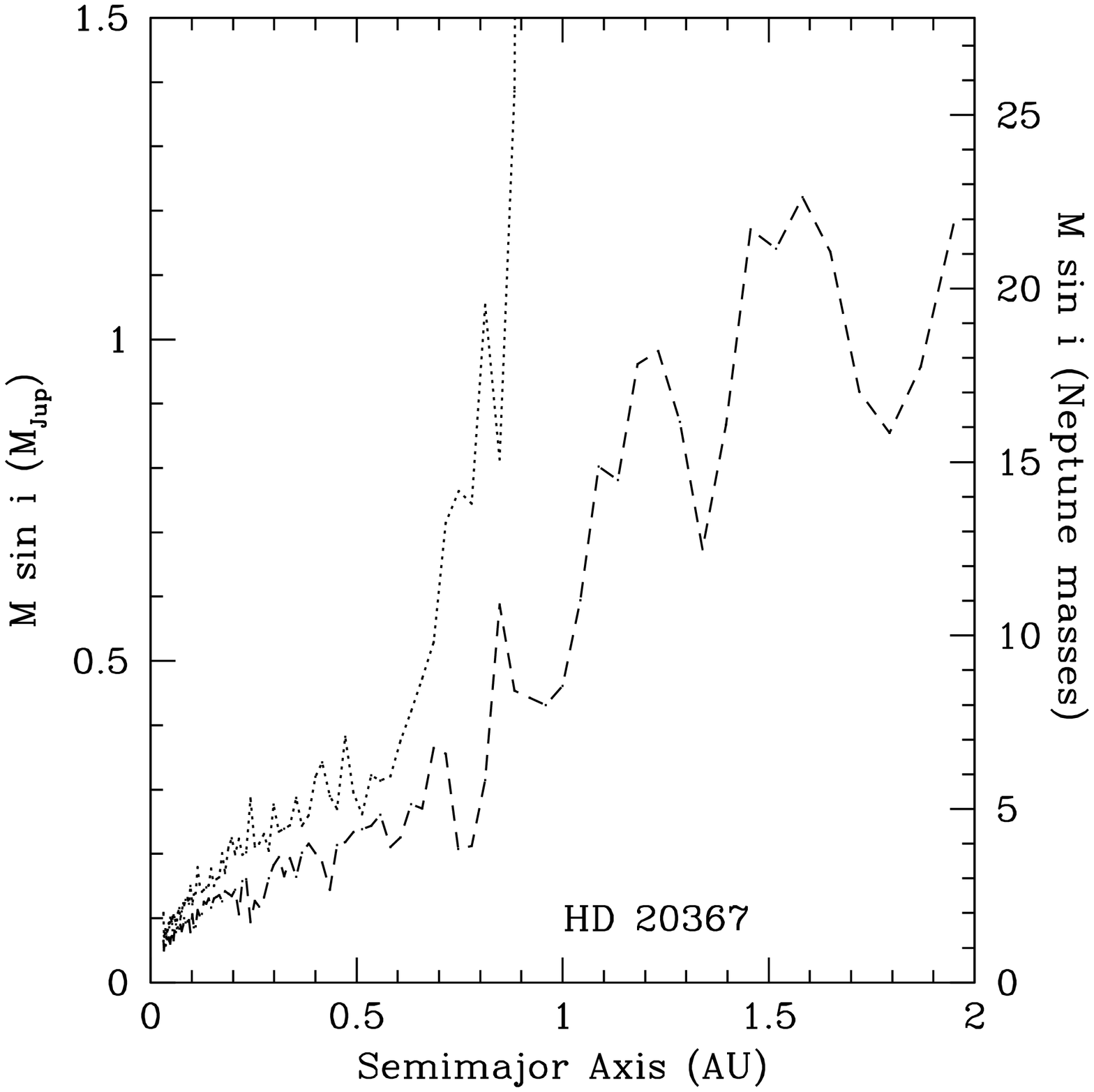}{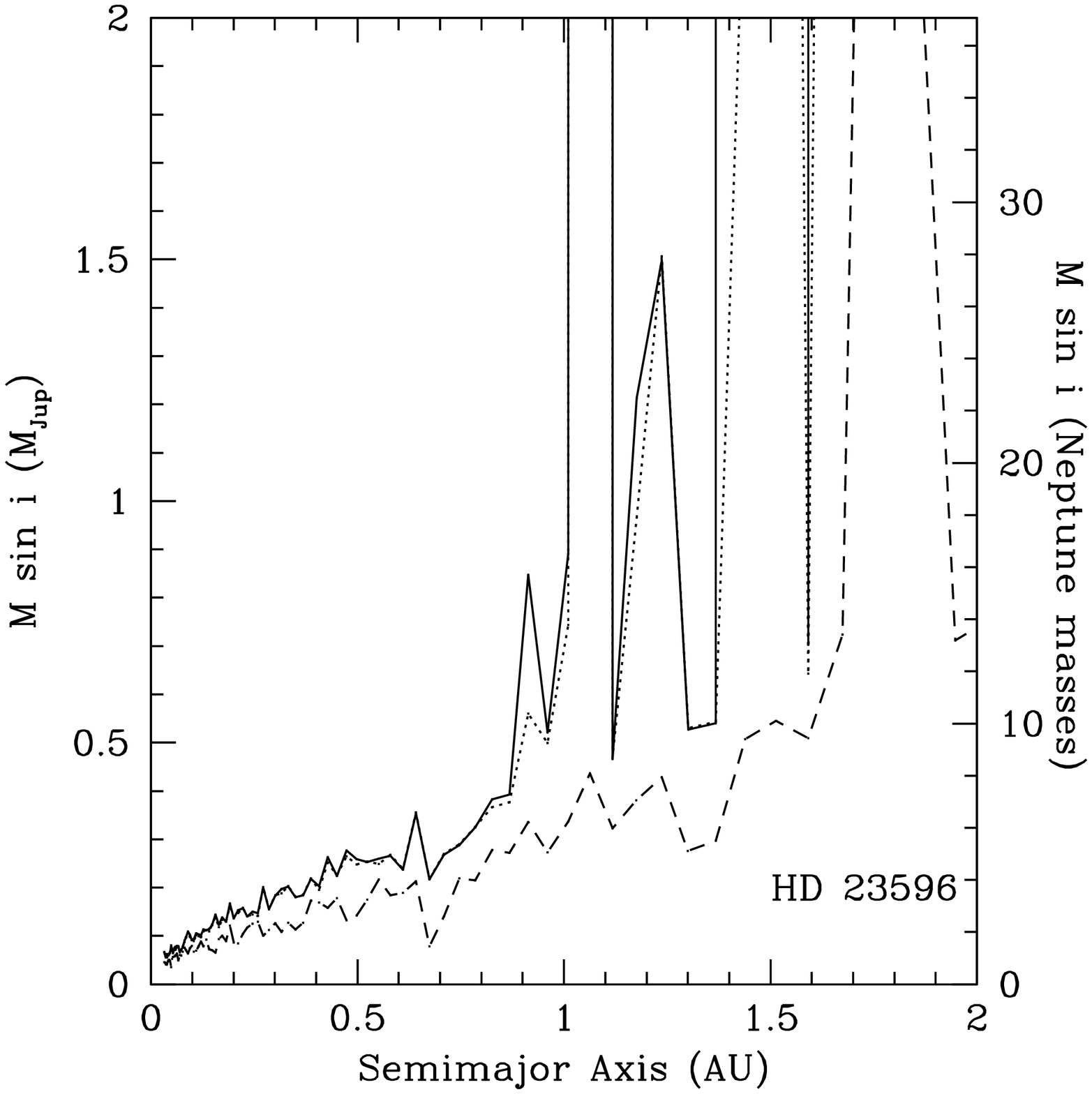}
\caption{Left panel: Same as Fig.~\ref{limits1}, but for HD~20367.  
These results were obtained without attempting to fit an existing 
planet, as no planet was confirmed in this system.  Right panel: 
HD~23596 (solid line: $e=0.10$).}
\label{limits3} 
\end{figure}

\begin{figure}
\plottwo{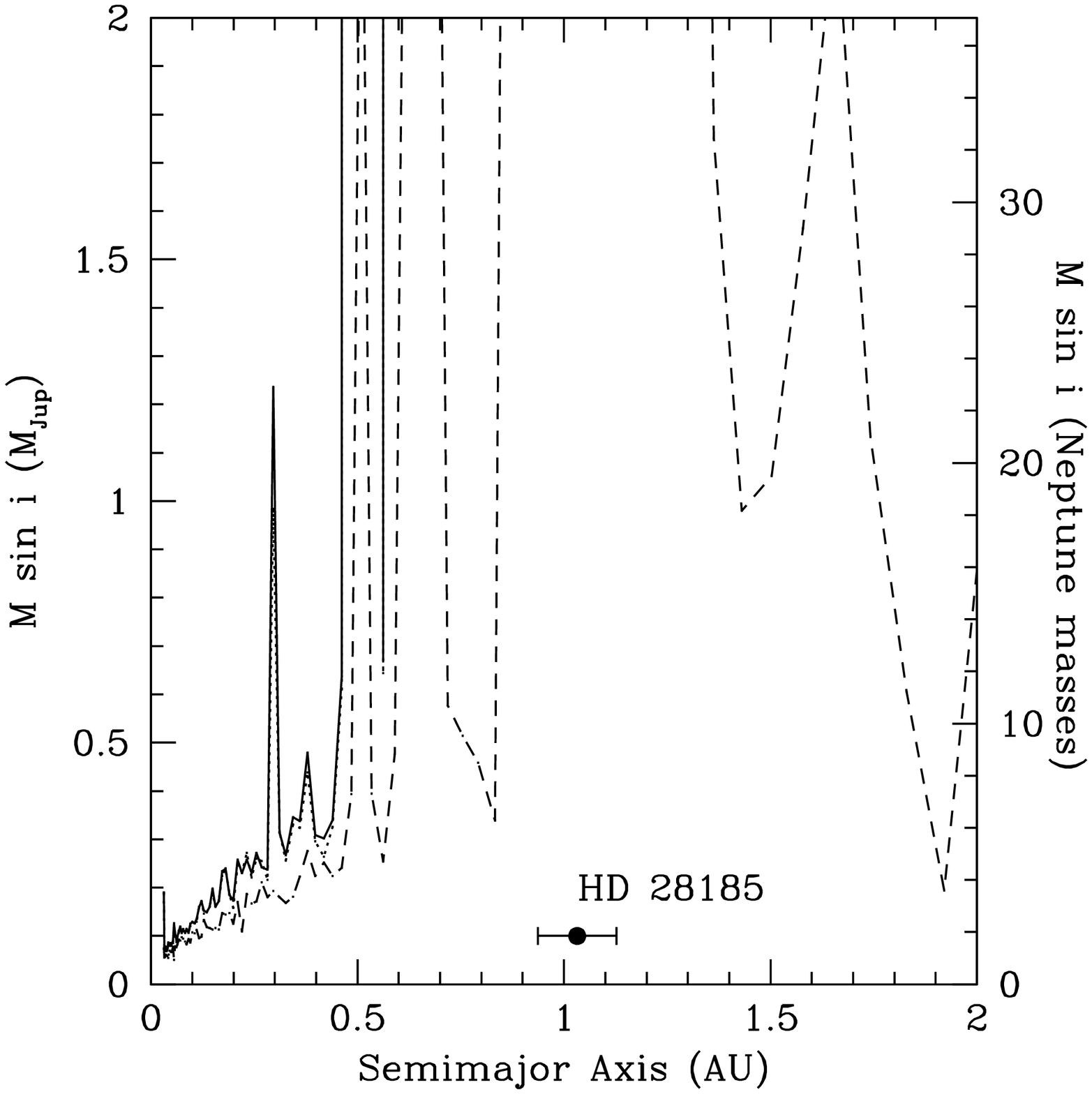}{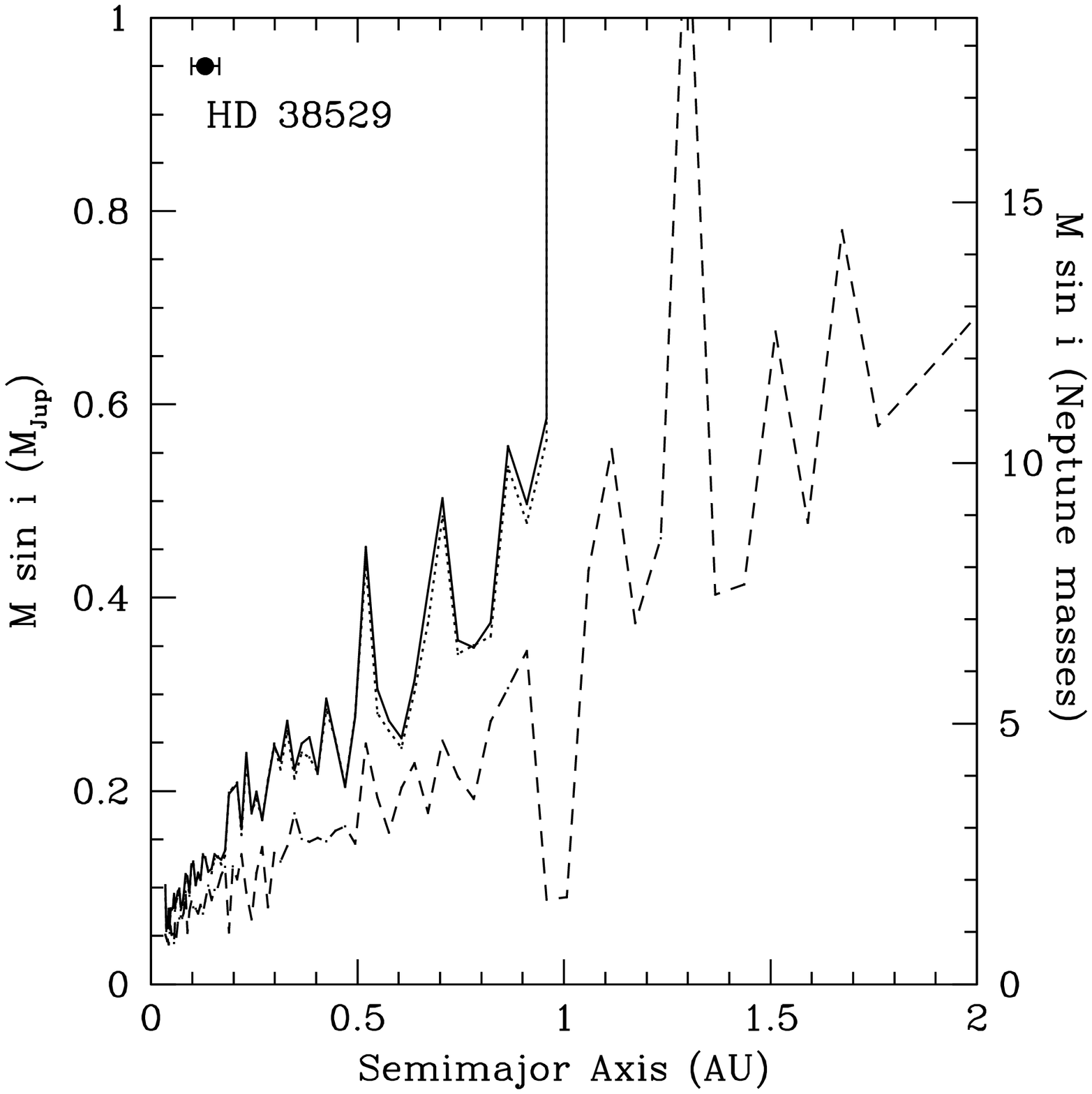}
\caption{Left panel: Same as Fig.~\ref{limits1}, but for HD~28185 (solid 
line: $e=0.09$).  Right panel: HD~38529 (solid line: $e=0.12$).}
\label{limits4}
\end{figure}

\begin{figure}
\plottwo{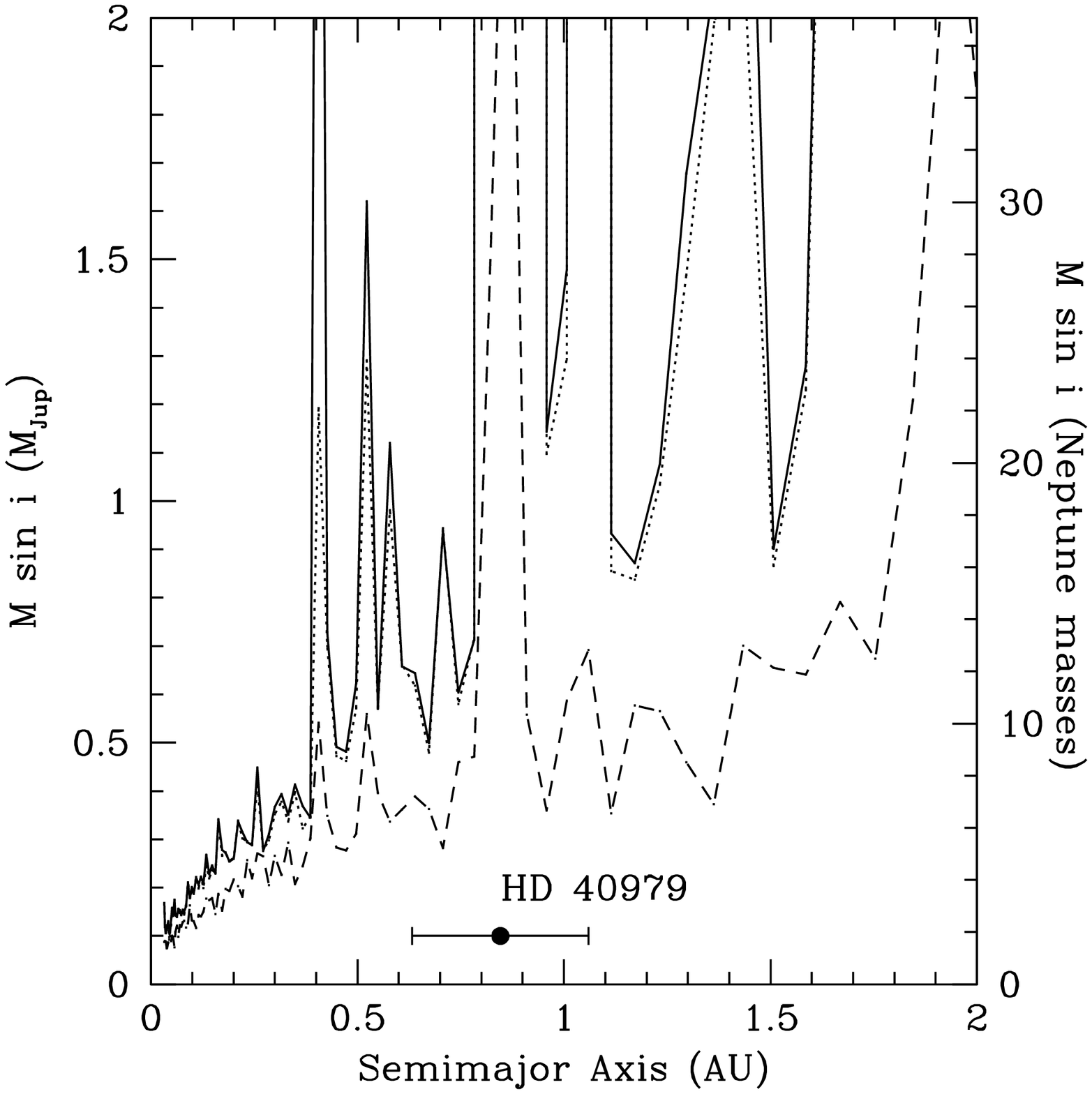}{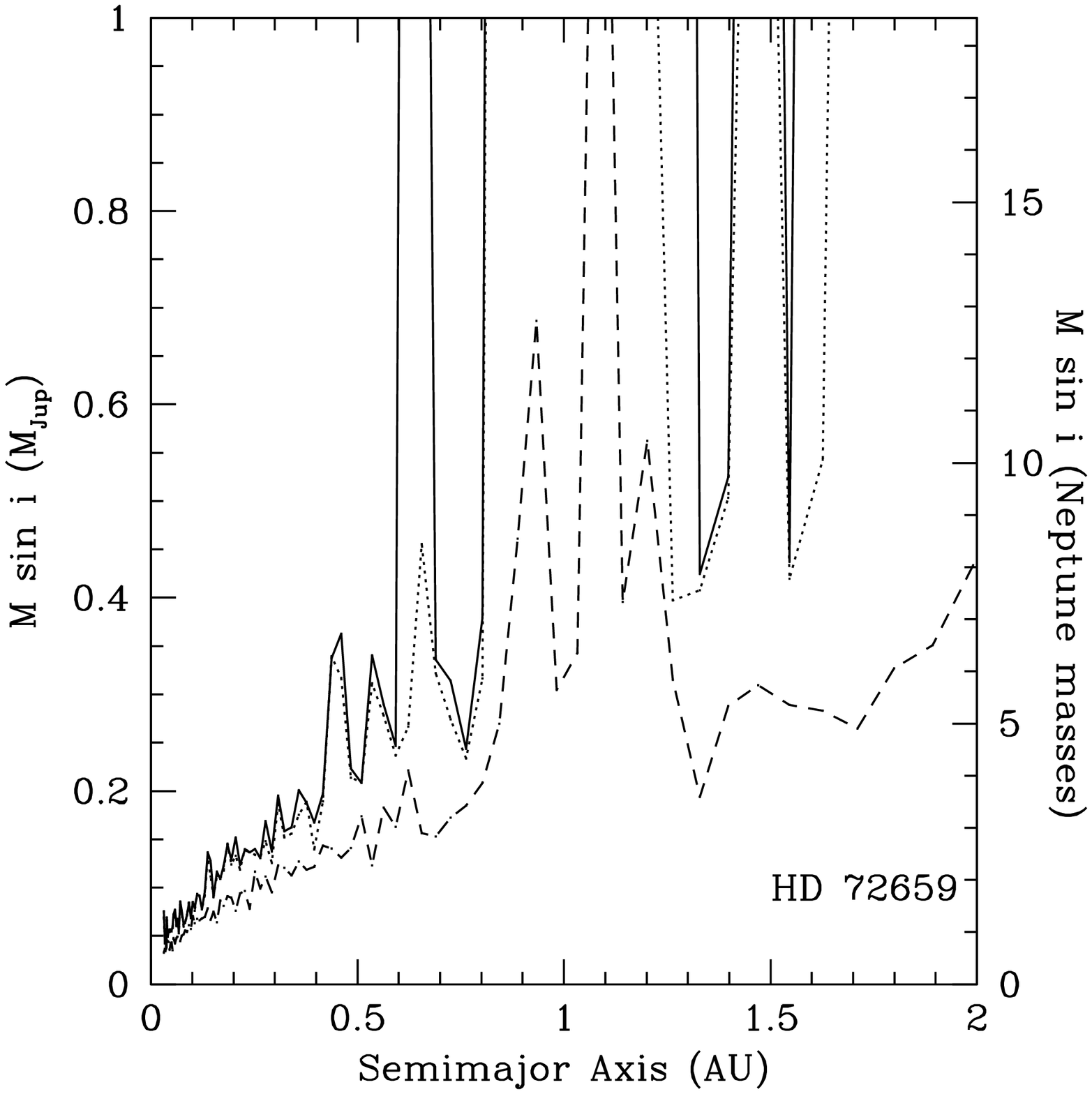}
\caption{Left panel: Same as Fig.~\ref{limits1}, but for HD~40979 (solid 
line: $e=0.11$).  Right panel: HD~72659 (solid line: $e=0.10$).}
\label{limits5}
\end{figure}

\begin{figure}
\plottwo{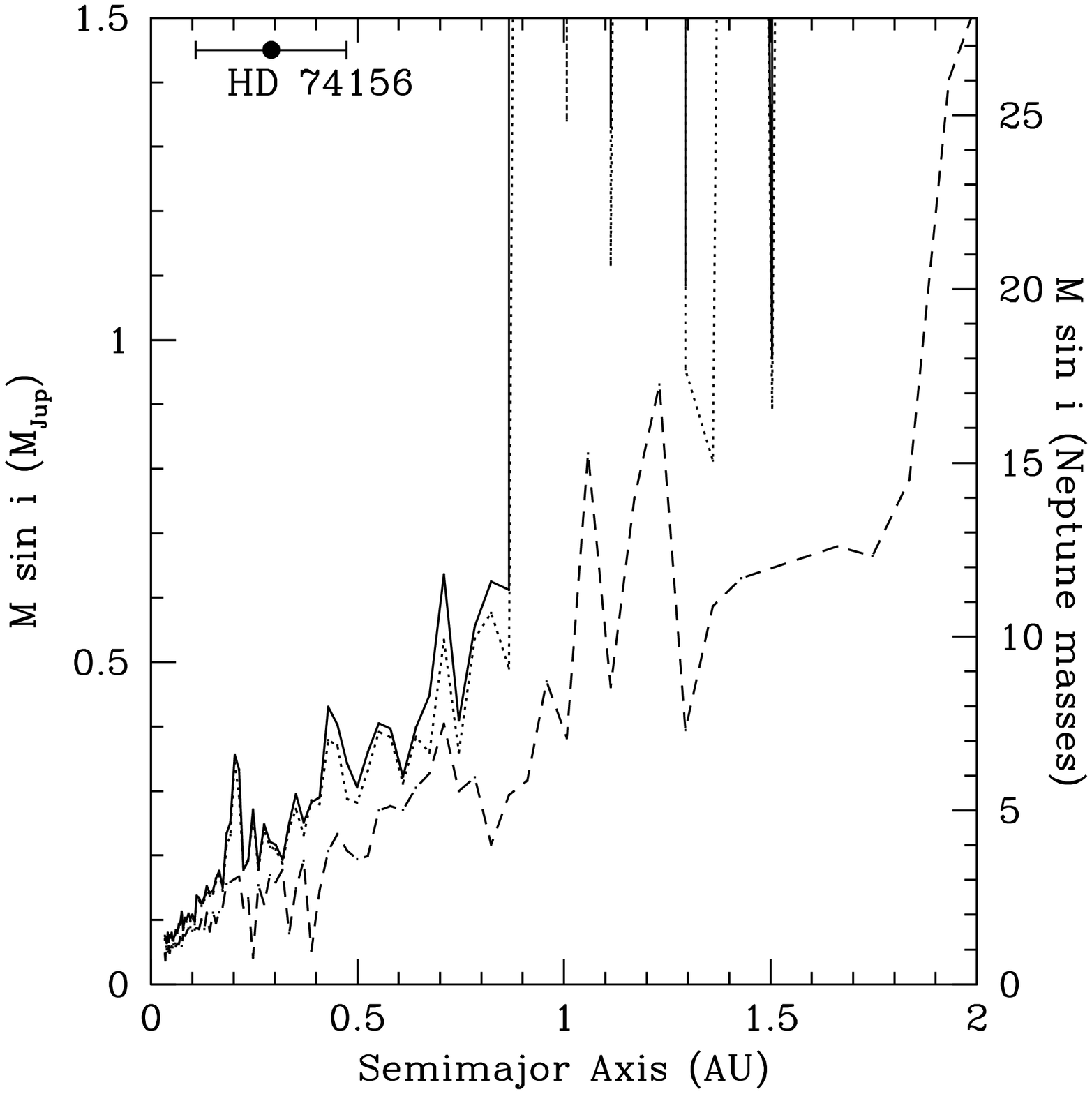}{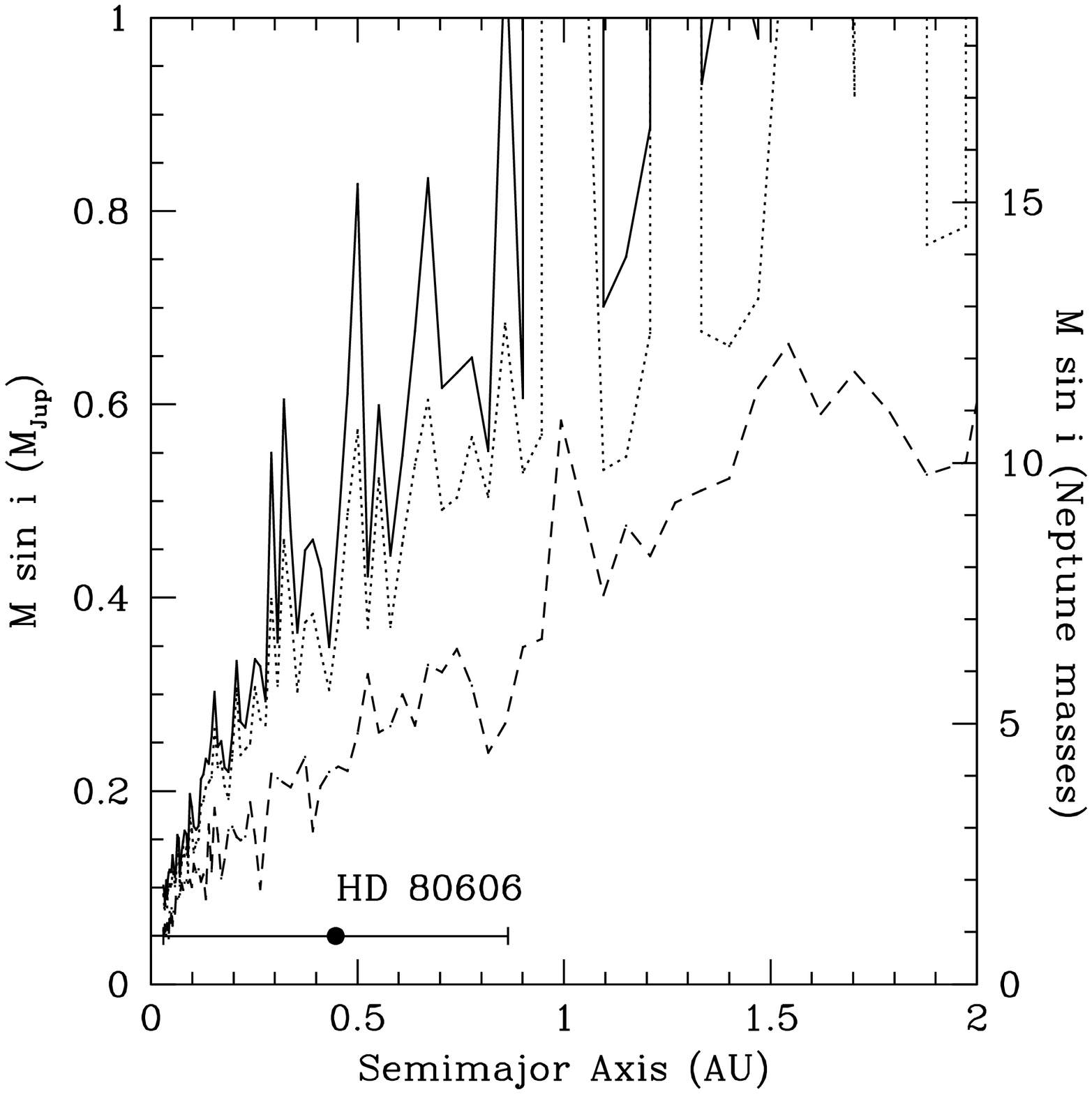}
\caption{Left panel: Same as Fig.~\ref{limits1}, but for HD~74156 (solid 
line: $e=0.15$).  Right panel: HD~80606 (solid line: $e=0.31$).}
\label{limits6}
\end{figure}

\begin{figure}
\plottwo{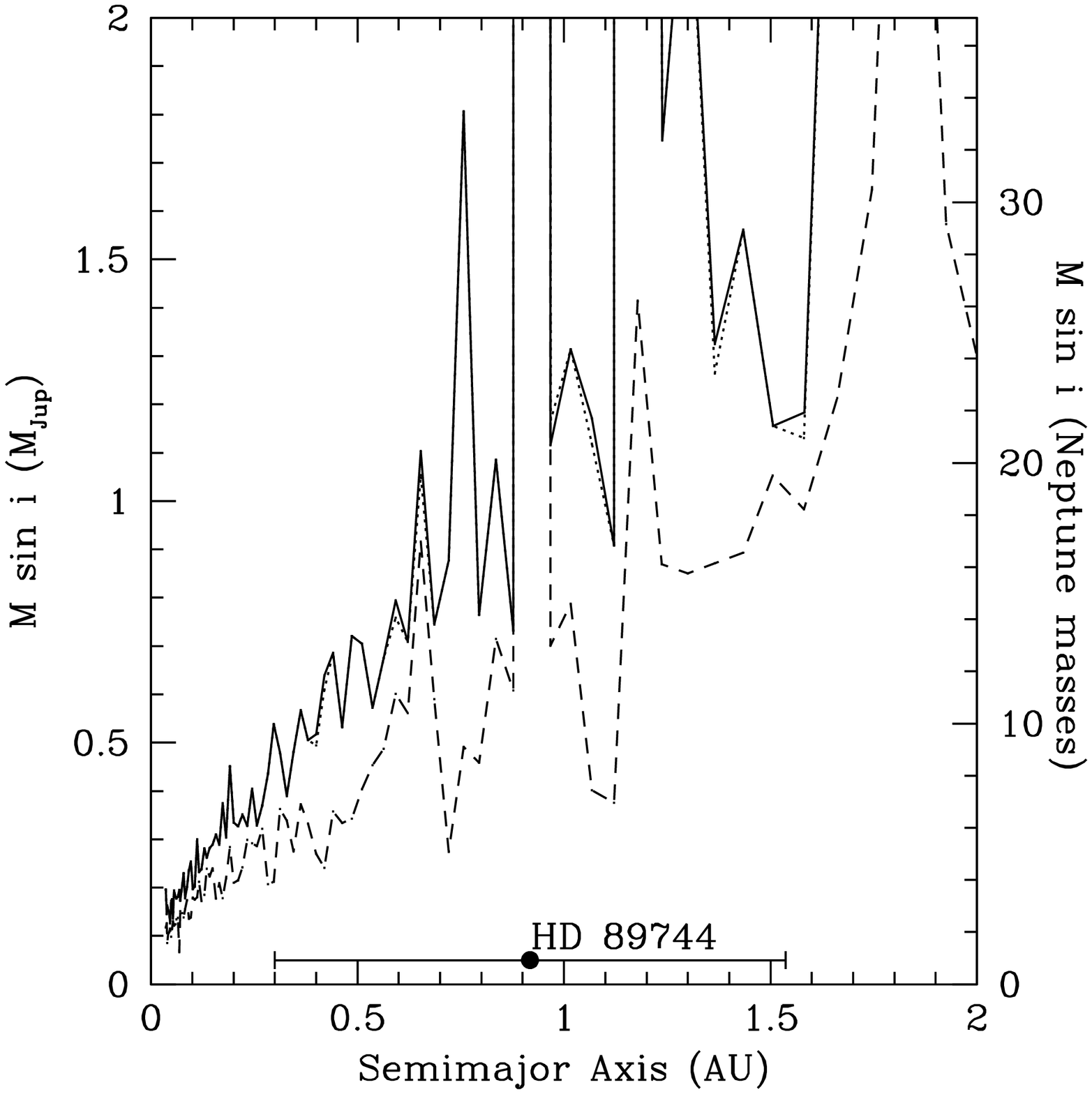}{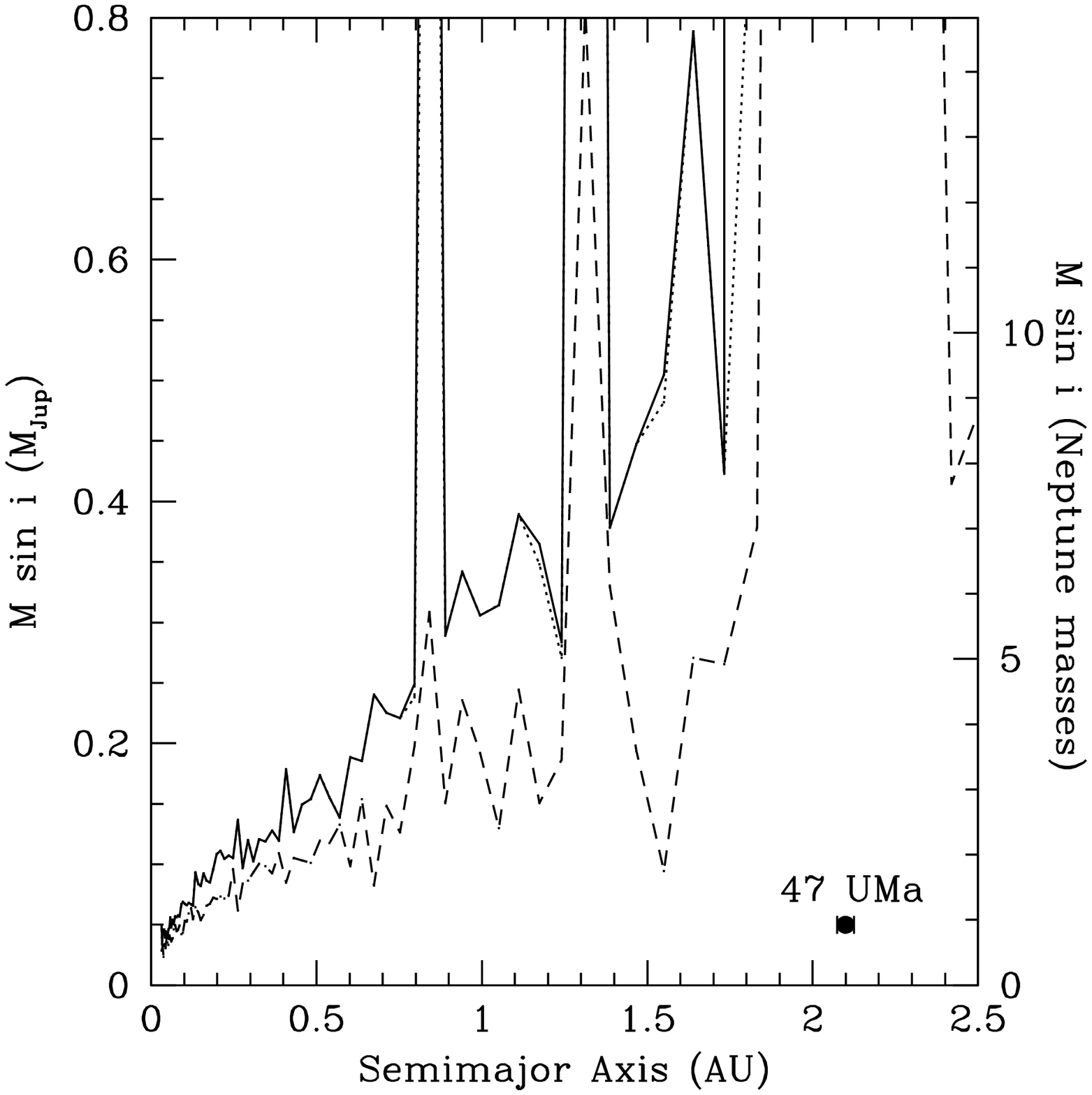}
\caption{Left panel: Same as Fig.~\ref{limits1}, but for HD~89744 (solid 
line: $e=0.01$).  Right panel: 47~UMa (solid line: $e=0.02$).  Only 
47~UMa~b was included in the limits computations. }
\label{limits7}
\end{figure}

\begin{figure}
\plottwo{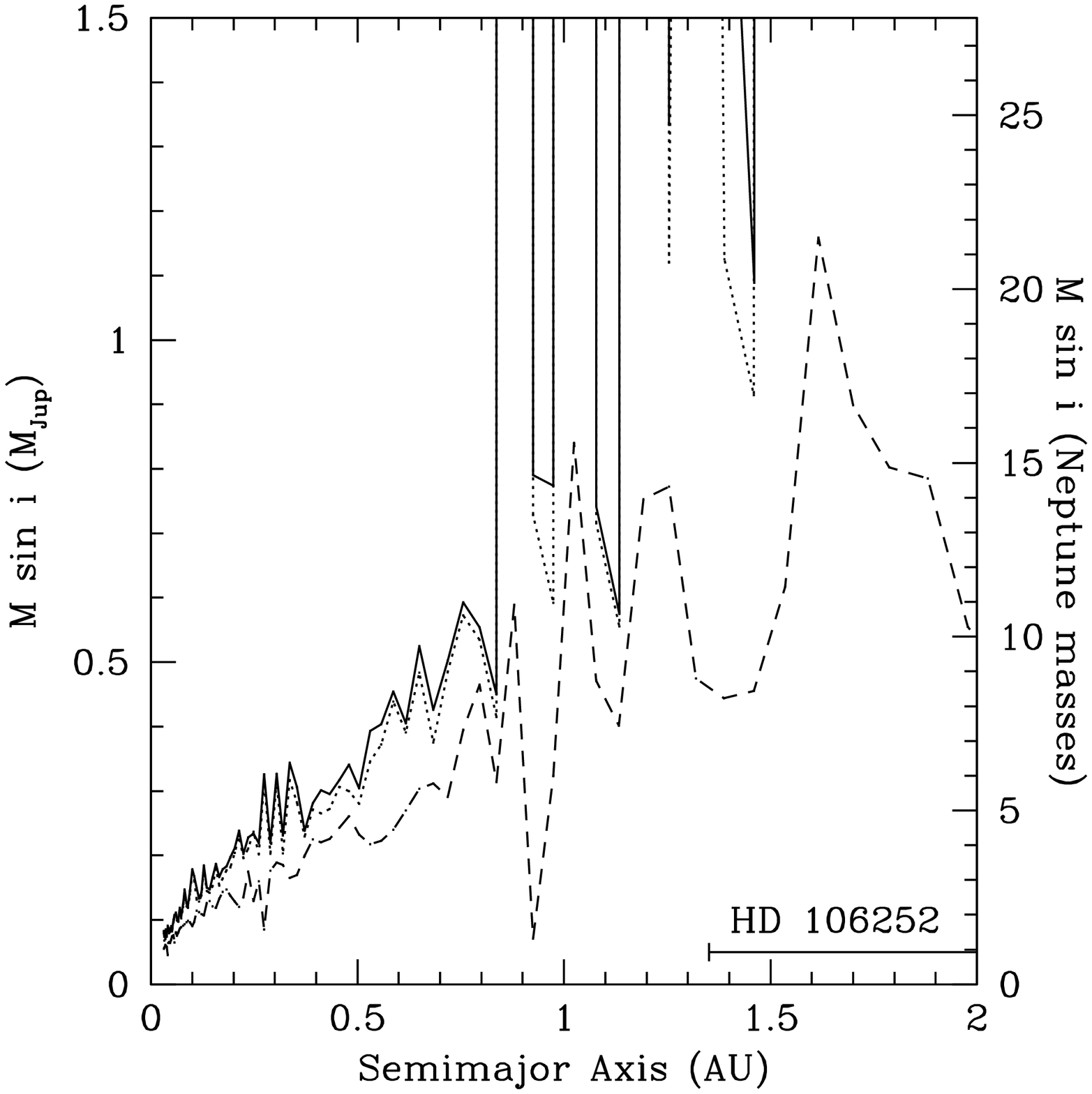}{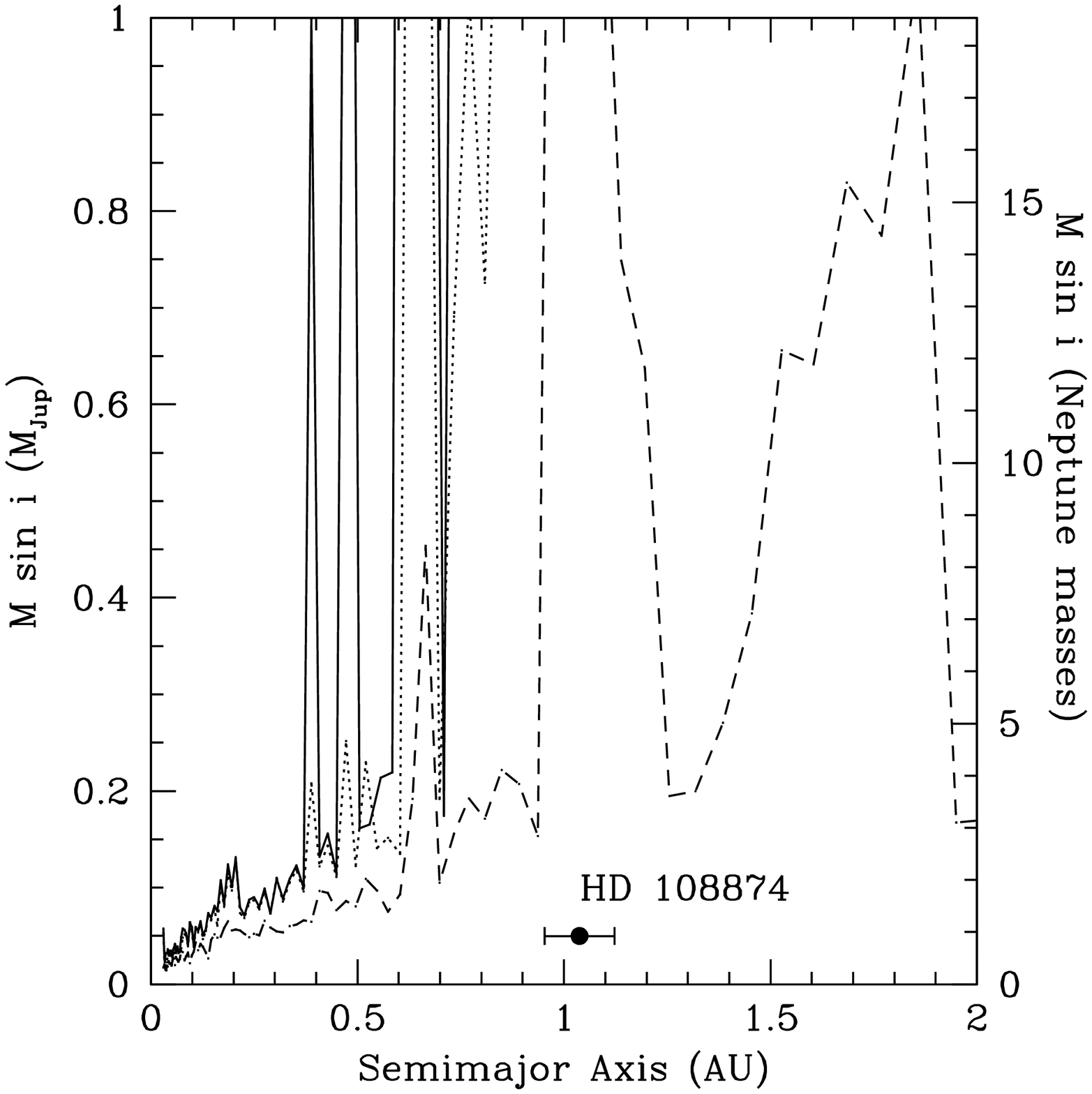}
\caption{Left panel: Same as Fig.~\ref{limits1}, but for HD~106252 
(solid line: $e=0.15$).  Right panel: HD~108874 (solid line: $e=0.15$).}
\label{limits8}
\end{figure}

\begin{figure}
\plottwo{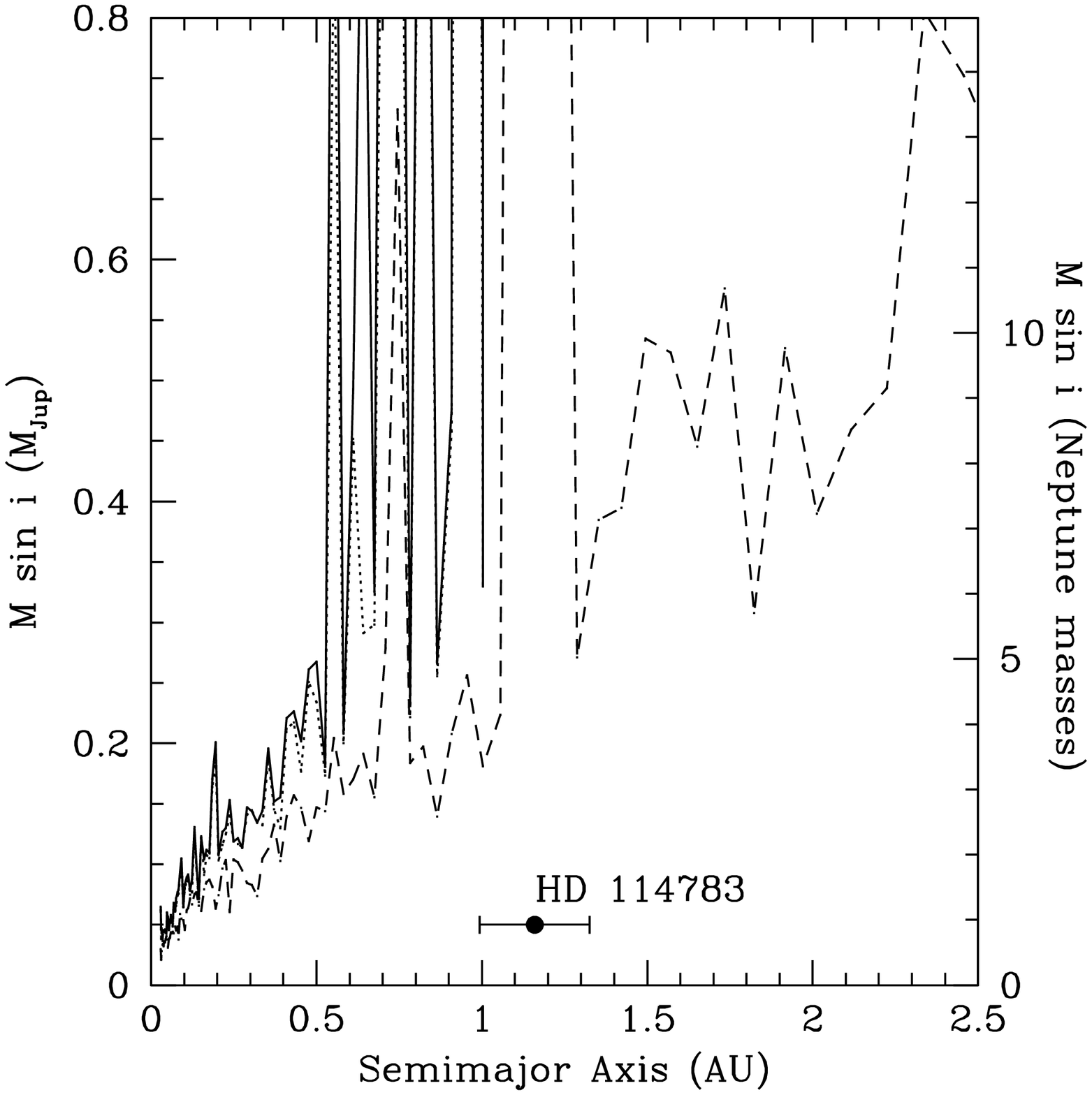}{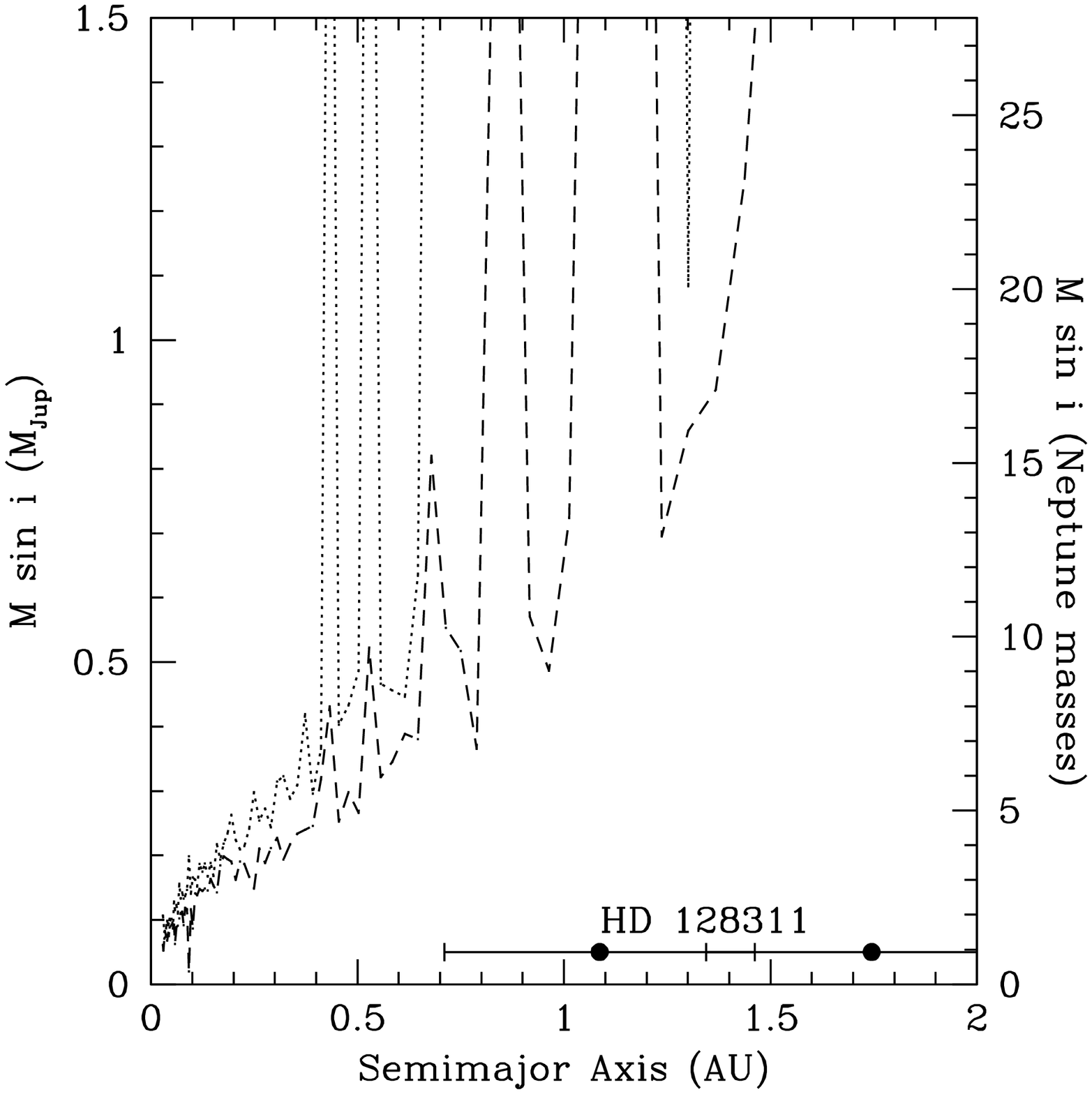}
\caption{Left panel: Same as Fig.~\ref{limits1}, but for HD~114783 
(solid line: $e=0.11$).  Right panel: HD~128311.  Only circular orbits 
are considered since no test-particle simulations were conducted for 
this system.}
\label{limits9}
\end{figure}

\begin{figure}
\plottwo{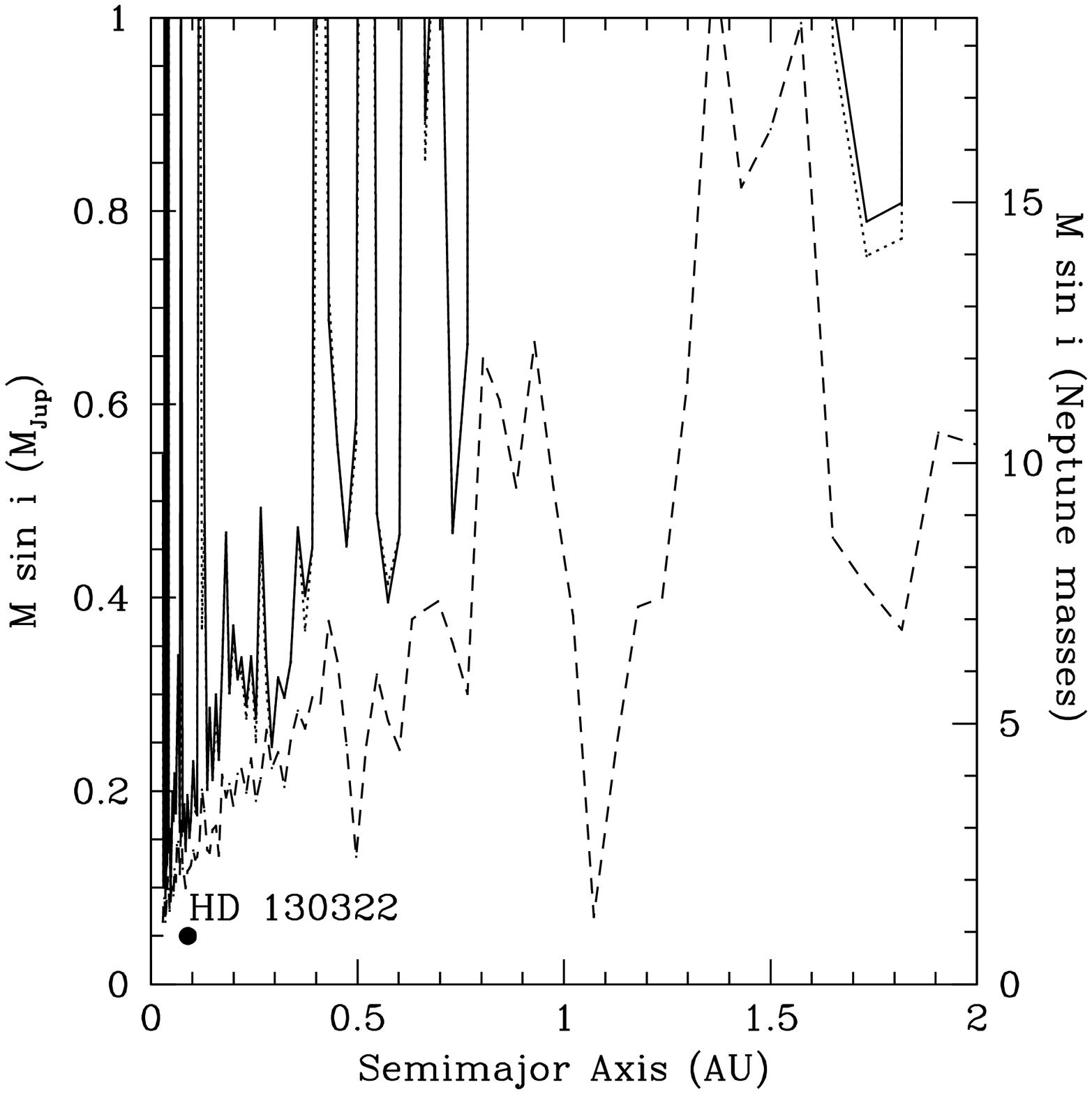}{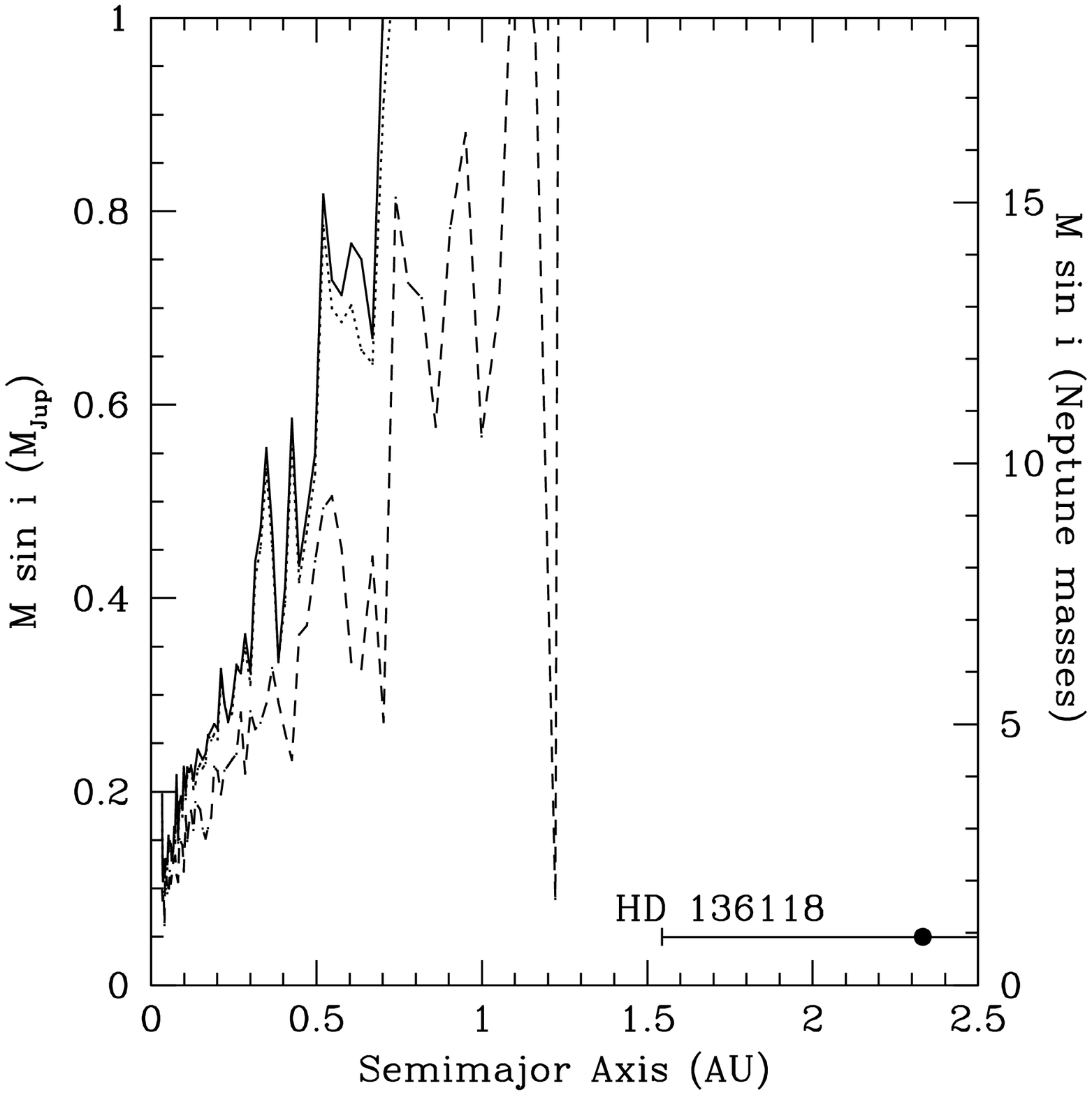}
\caption{Left panel: Same as Fig.~\ref{limits1}, but for HD~130322 
(solid line: $e=0.02$).  These results do not include data from 
\citet{udry00}.  Right panel: HD~136118 (solid line: $e=0.11$).}
\label{limits10}
\end{figure}

\begin{figure}
\plottwo{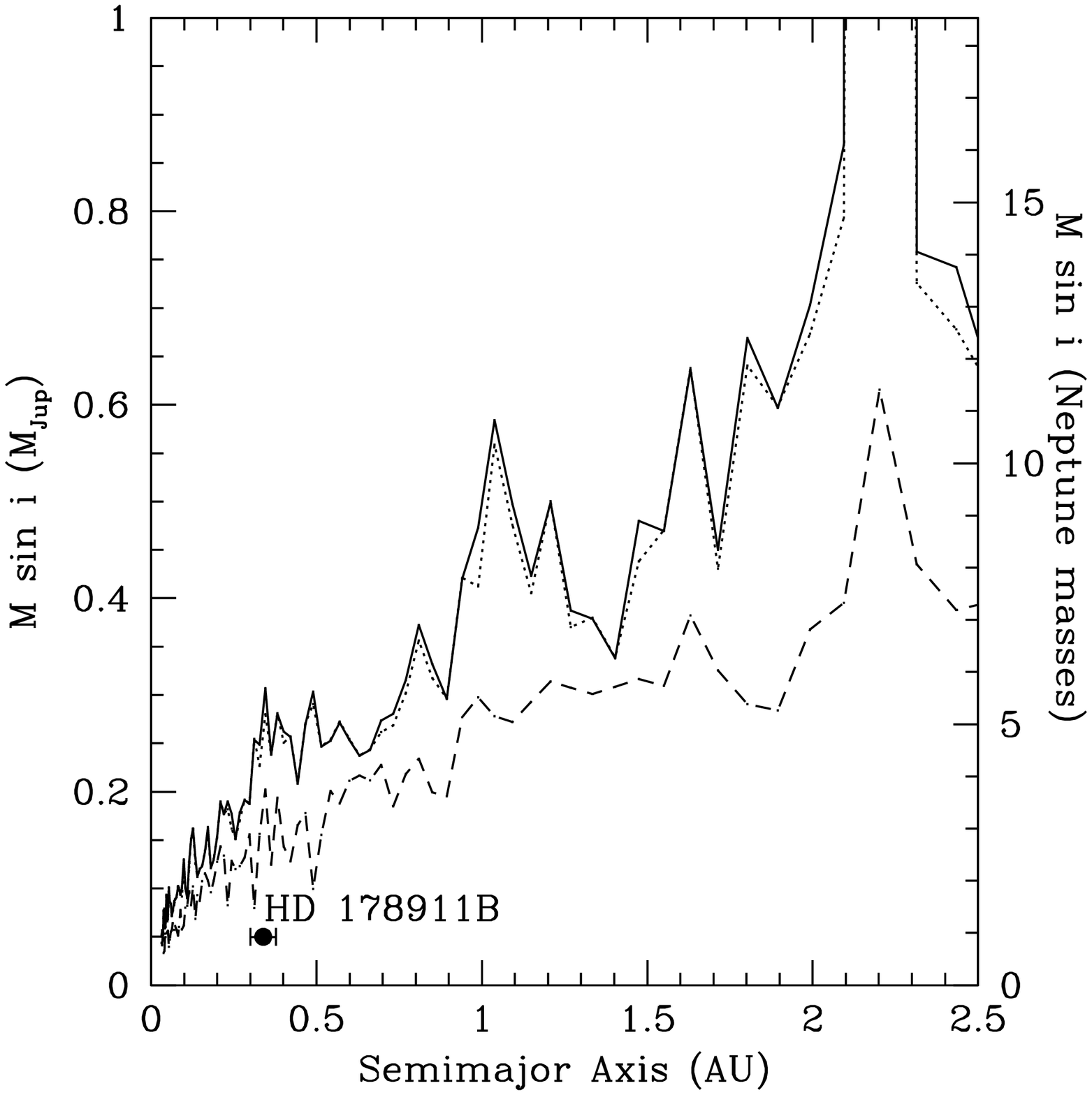}{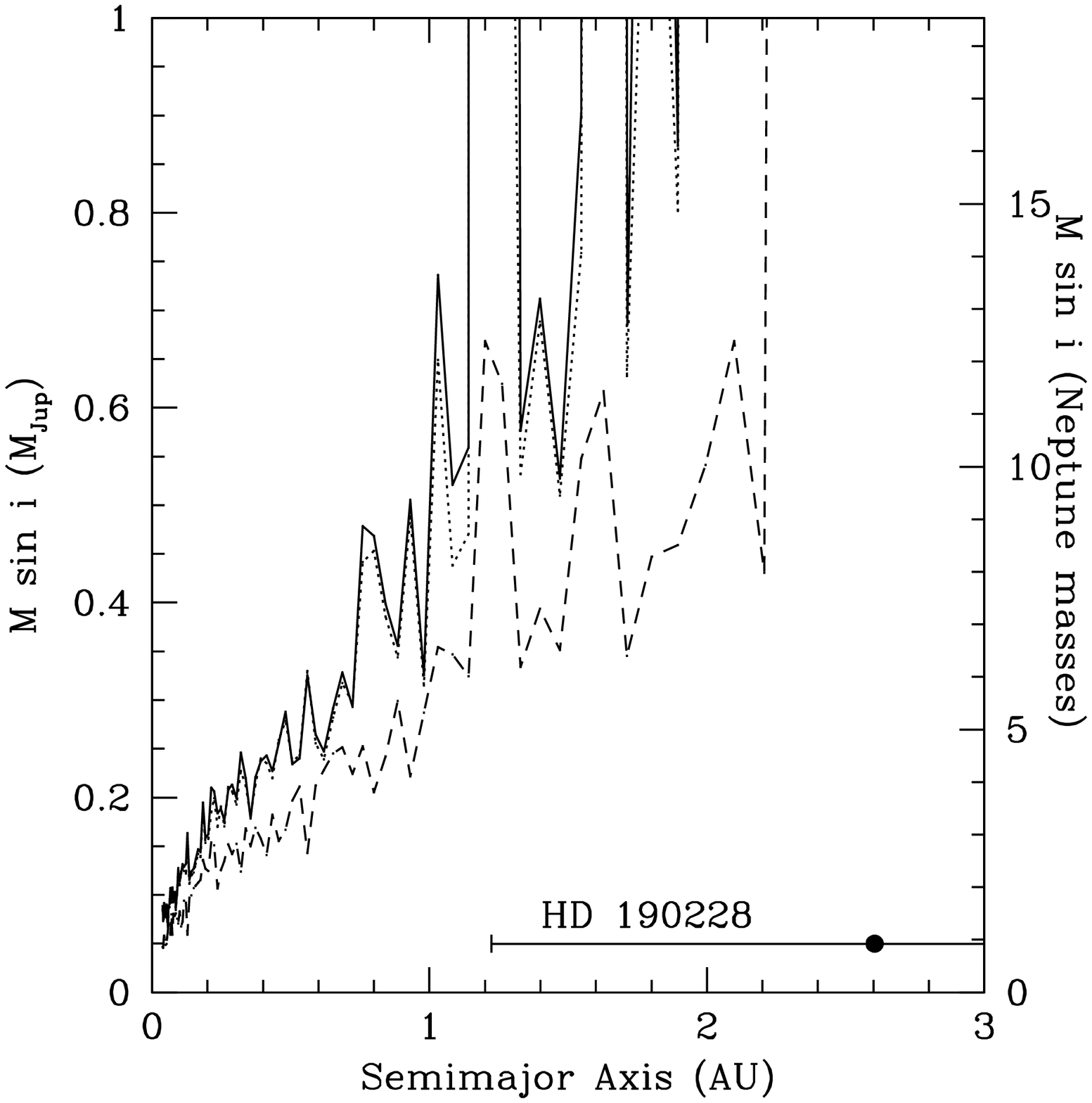}
\caption{Left panel: Same as Fig.~\ref{limits1}, but for HD~178911B 
(solid line: $e=0.07$).  Right panel: HD~190228 (solid line: $e=0.16$).}
\label{limits11}
\end{figure}

All data used in the fits for each planet host were subjected to the 
limits-determination routine, using 100 trial periods at even steps in 
the logarithm between 2 days and the total duration of observations.  
The results are plotted in Figures~\ref{limits1}-\ref{limits11}; planets 
with masses above the lines were recovered in 99\% of trials (solid and 
dotted lines), or 50\% of trials (dashed lines), and hence can be ruled 
out by the data at those confidence levels, respectively.  To match the 
parameter space specifically targeted in this study, and to match that 
of the test-particle simulations, the detection-limits plots show the 
inner 2~AU only.  For the eccentric trials (solid lines), the 
eccentricity of the injected test signals was chosen to be the mean 
eccentricity of the surviving test particles from the N-body simulations 
described in \S~4.2.  This approach was chosen because the dynamical 
simulations demonstrated that objects placed in circular orbits do not 
stay that way; the eccentricity of an undetected low-mass planet is 
expected to be influenced to nonzero values by the known giant planet.

It is important to note that the limits presented here represent the 
companions that can be ruled out by the data with 99\% confidence.  
Lower-mass planets could have been detected in this survey, but not 
necessarily at all (or 99\% of all) possible configurations.  It is 
likely that a particular combination of parameters for a simulated 
planet makes that signal fiendishly difficult to recover by this method, 
owing to the known planet's radial-velocity signal and the sampling of 
the data.  This is particularly important for simulated eccentric 
planets, where the velocity signal becomes markedly non-sinusoidal.  The 
50\% limits are also shown to illustrate the effect of relaxing the 
recovery criteria in order to reduce the impact of especially 
unfortunate configurations.


Table~\ref{masslimits} summarizes the results of the detection limits 
computations.  The mean detection limits shown in Table~\ref{masslimits} 
show that we could have detected 99\% of planets with M~sin~$i\sim$1.6 
Neptune masses at 0.05~AU, and M~sin~$i\sim$2.4 Neptune masses at 
0.1~AU.  The tightest limits were obtained for HD~3651, HD~108874, and 
47~UMa, in which we are able to rule out Neptune-mass planets within 
0.1~AU at the 99\% level.  For all of the systems, the limits shown in 
Figures~\ref{limits1}-\ref{limits11} exhibit some ``blind spots'' 
evident where the periodogram method failed to recover the injected 
signals with FAP$<$0.1\%.  Typically this occurs at certain trial 
periods for which the phase coverage of the observational data is poor, 
and often at the 1-month and 1-year windows.  Using this method of 
fitting the known planet for each injected trial signal, such regions of 
ignorance are also present at periods close to that of the existing 
planet.


\section{Discussion}

The aim of this project has been to intensively monitor known planetary 
systems in search of additional planets.  However, in the sample of 22 
planet hosts, the results have been quite the opposite.  These new data 
cast doubt on the existence of two of the previously known planets, 
HD~20367b and 47~UMa~c \citep{destructor}.  The announcement by 
\citet{bean08} of a third planet in the HD~74156 system, one of the 
targets of this study, prompted a detailed investigation; at present we 
cannot confirm this object.  These results suggest that systems with 
multiple giant planets are considerably more rare, or harder to detect, 
than anticipated at the outset of this project.

In this section, we will explore some reasons why no new multiple-planet 
systems were detected.  Four possibilities are: 1) Biases in the target 
selection conspire against detection of weak signals, 2) There exist 
fundamental physical differences between single- and multiple-planet 
systems, 3) We did not obtain a sufficient quantity of high-quality 
data, and 4) Apparent single-planet systems may contain terrestrial-mass 
planets below the detection threshold.

\subsection{Biases in the Sample}

As with any scientific experiment, it is important to determine whether 
the sample selection resulted in unforeseen biases which affected the 
results.  The target-selection process for this study, described in 
\S~2, included an intentional bias in favor of planet hosts with 
``large'' (10-20 \ms) radial-velocity scatter about the orbital 
solution.  The reasoning for this choice is straightforward: if a single 
planet can be fit with minimal scatter, there is little room for 
additional undetected planets to hide in the residuals.  An unintended 
consequence of this selection criterion is that the excess scatter may 
be intrinsic to the star rather than indicative of additional planets.  
The achievable velocity precision improves with the number and strength 
of photospheric lines \citep{butler96}.  Stars with higher temperatures 
or lower metallicities would have fewer and weaker lines, and result in 
lower velocity precision.  In rapidly rotating stars, the spectral lines 
are broadened, which also degrades the radial-velocity precision.  
\citet{fv05} showed that the probability of a given star hosting a 
planet is positively correlated with its metallicity.  In addition, 
those authors suggested that among planet host stars, metal-rich stars 
are more likely to host multiple planets.  To check for these sorts of 
biases, we can perform a Kolmogorov-Smirnov (K-S) test to determine the 
probability that two samples are drawn from the same distribution.  
Comparing our sample of 22 planet host stars with other planet hosts not 
targeted ($N=200$), the K-S test shows no significant differences in 
$T_{eff}$ ($P=0.698$), [Fe/H] ($P=0.841$), or V~sin~$i$ ($P=0.323$).  A 
comparison of the mean and median values of these quantities is shown in 
Table~\ref{stars}.  The uncertainties are too large to make 
statistically meaningful comparisons, but the K-S test results suggest 
that there are no significant differences between the 22 planet hosts 
targeted here and those planet hosts not chosen.

\subsection{Fundamental Differences}

In this section, we ask the question, ``Is there something special about 
the multi-planet systems''?  Physical differences between single and 
multiple planet systems could arise either from the host star or from 
the processes of formation and dynamical evolution.  Table~\ref{stats} 
gives statistics on the planetary and stellar parameters for single and 
multiple-planet systems.  Only those planets detected by radial-velocity 
with M~sin~$i<$13 \Mjup\ were considered in the compilation of these 
statistics.  Table~\ref{kstest} shows the results of K-S tests on the 
planetary and stellar characteristics listed in Table~\ref{stats}.  None 
of the parameters tested showed statistically significant differences 
between single and multiple planet systems.  There are hints from the 
data in Table~\ref{stats}, and the K-S test results in 
Table~\ref{kstest} that planets in multiple systems have larger $a$ and 
smaller M~sin~$i$ than those in single-planet systems.  Both of these 
trends would work against the radial-velocity detection of planets in 
multiple systems.  As the semimajor axis $a$ increases by a factor of 
$N$, the velocity semiamplitude $K$ decreases by $\sqrt{N}$, and as the 
planet mass decreases by a factor of $N$, $K$ also drops by a factor of 
$N$.  It is also possible that a tendency toward lower mass and larger 
semimajor axis in multi-planet systems is the result of a selection 
effect.  Once a single planet is found, follow-up observations may 
reveal longer-period (larger $a$) planets, and intensive monitoring 
programs such as this work may then find lower-mass planets.  We can 
test whether a selection effect is at work by computing the statistics 
in Table~\ref{stats} for the \textit{first} planet discovered in the 
known multi-planet systems.  These results are also given in 
Table~\ref{kstest}; by comparing only the first planet found in the 
multiple systems with single planets, any significant difference between 
the distributions vanishes.  Recently, \citet{wright08} have presented a 
detailed investigation of multiple-planet systems, and they find that 
planets in multiple systems tend to have lower eccentricities than 
single planets.  We discuss this possibility in \S~6.5.  
\citet{wright08} also note that the orbital distances of planets in 
multiple systems are more evenly distributed in log-period, whereas 
single planets are more frequent at $a\sim$0.05~AU and near 1~AU.

\subsection{Observing Strategy}


In considering whether there are important differences between the 
objects targeted in this work and known multi-planet systems, we can 
focus the comparison on the type of planetary system this survey was 
aimed at finding.  The original motivation for this work was to 
investigate the possibility that systems containing a Jovian planet also 
contain Neptune-mass planets (1 Neptune mass=0.054 \Mjup).  At this 
writing, there are four such systems: 55~Cnc, GJ~876, $\mu$~Ara 
(=HD~160691), and GJ~777A (=HD~190360).  With a sample size of only 
four, a meaningful statistical comparison of the host stars is not 
possible, but one can look at the characteristics of the body of 
radial-velocity data for these systems.  In so doing, we ask whether 
those data are of exceptional quality or quantity which facilitated the 
detection of the additional low-mass planets in those systems.  The 
recent detection of a fifth planet in the 55~Cnc system by 
\citet{fischer08} used 636 measurements, binned into 250 Lick visits and 
70 Keck visits.  The detection of the fourth planet by 
\citet{mcarthur04} used 138 HET observations combined with 143 Lick data 
points \citep{marcy02} and 48 data points from \citet{naef04}.  For 
$\mu$~Ara, the Neptune-mass planet was discovered using the HARPS 
spectrograph, which consistently delivers velocity precision of $\sim$1 
\ms\ \citep{santos04b, pepe07}.  The fourth planet in the $\mu$~Ara 
system \citep{pepe07} was discovered using a total of 86 HARPS 
measurements combined with data from CORALIE and the AAT.  The 
18M$_{\oplus}$ planet GJ~777Ac was discovered by \citet{vogt05} using 87 
Keck velocities, and \citet{rivera05} found the 7.5M$_{\oplus}$ GJ~876d 
after 155 Keck observations.  All four of these systems appear to have 
required an unusually large amount of the highest-quality data from Keck 
and HARPS, with a mean of 107 data points.  By contrast, the targets in 
this work each received an average of 53 HET visits.  It is possible 
that the number of visits required to detect a hot Neptune was 
underestimated.

\subsection{Swarms of Earths}

Another possibility is that multiple-planet systems are indeed common, 
but, like our own Solar system, contain many terrestrial-mass objects 
which are undetectable by current radial-velocity surveys. 
Core-accretion simulations by \citet{idalin04a} predict a preponderance 
of 1-10M$_{\oplus}$ planets inside of 1~AU, and a ``planet desert'' in 
the range of 10-100 M$_{\oplus}$, arising due to rapid gas accretion by 
cores once they reach about 10 M$_{\oplus}$.  The current survey is not 
sensitive to the terrestrial-mass objects, but planets within the 
``desert'' could have been detected.  Interestingly, 
\citet{schlaufman08} show that the presence of the planet desert could 
be confirmed by a radial-velocity survey with 1 \ms\ precision and 
$\sim$700 observations, which is similar in scope to the present work.  
Of course, many more than 22 systems need to be studied before 
conclusions can be made, but the characterization of hundreds of new 
systems by the \textit{Kepler} spacecraft \citep{borucki03} will help to 
define the upper and lower mass boundaries of the planet desert.  
\citet{idalin04a} note that the lower mass boundary would indicate the 
core mass required for rapid gas accretion, while the upper mass 
boundary would give insight into the mechanism by which gas accretion 
stops.  \textit{Kepler} discoveries of short-period super-Earths with 
masses 1-10M$_{\oplus}$ would lend further support to the core-accretion 
mechanism.

Simulations of planetesimal formation and migration also provide support 
for the existence of terrestrial-mass planets in systems with a gas 
giant planet.  The GJ~876 system \citep{rivera05}, which contains two 
giant planets and an interior ``super-Earth'' (M~sin~$i$=7.5 
M$_{\oplus}$), is thought to have originated by the shepherding of 
material as the giant planets migrated inward \citep{zhou05}.  200~Myr 
simulations by \citet{raymond06} and \citet{mandell07} resulted in the 
formation of planets with 1-5 Earth masses interior and exterior to the 
migrating hot Jupiter.  Those models included only Type~II migration, in 
which the migrating giant planet opens a gap in the protoplanetary disk.  
The models of \citet{fogg07} consider the effects of Type~I migration, 
in which the giant planet does not open a gap in the disk and inward 
drift is driven by differential torques on the planet.  Inclusion of 
Type~I migration did not alter the general outcome, that planets of 
several Earth masses are shepherded inward by the hot Jupiter, and some 
remain exterior to it.  These models indicate that the inner regions of 
planetary systems may be populated with terrestrial-mass planets which 
would remain wholly undetectable by current radial-velocity surveys.  
Although this work achieved detection limits of 15-30 Earth masses, 
rocky planets in the range of 1-5 Earth masses could easily have been 
missed.

\subsection{Broader Implications for Planetary Systems }

We now take a step back and look at the bigger picture of planetary 
system formation and evolution.  Based on the target selection and the 
resulting detection limits, this survey was most sensitive to systems 
with two giant planets (larger than Saturn mass).  More specifically, 
our ``key demographic'' is a system with a ``cold'' Jupiter 
($a\sim$1~AU) and a close-in planet with M~sin~$i~\gtsimeq$1-2 Neptune 
masses (0.05-0.1 \Mjup).  The detection limits given in \S~5 exclude 
such configurations at the 99\% level for all of the planetary systems 
considered here.  Systems containing a long-period, massive planet could 
also have been detected by trends or curvature in the velocity 
residuals; no such trends were present for any of the targets.  This 
survey was much less sensitive to planetary systems like our own, with 
multiple terrestrial-mass planets and long-period giants, for the 
reasons discussed in \S~6.4.  Planetary systems with architectures like 
our own Solar system may yet be common, but we will need to wait for the 
results from \textit{Kepler} to begin making quantitative statements.

The results of this work are most useful in assessing the frequency of 
planetary systems in which extensive migration has occurred, to bring 
two gas giant planets interior to the ``snow line.'' In the 
core-accretion theory of giant planet formation \citep{pollack96, 
lissauer95}, surface-density enhancement by ices facilitates the 
formation of $\sim$10-15 M$_{\oplus}$ cores.  The snow line, beyond 
which ices are present in the protoplanetary disk, has been estimated to 
lie at 1.6-1.8~AU in a minimum-mass solar nebula \citep{lecar06}.  
Perhaps the extensive migration required to construct systems with 
multiple giant planets with $a\ltsimeq$2~AU is uncommon; the typical 
timescale in which a system is undergoing migration may be short.  In 
other words, migration may be fast, a hypothesis which has led to 
theoretical scenarios in which the observed planets are the last of many 
``batches'' of planets which migrated onto the host star 
\citep{trilling02, idalin04a, narayan05}.  Type~I migration, in which a 
net viscous torque on the protoplanet changes its orbit \citep{ward97}, 
results in very fast migration with a timescale proportional to 
$M_{planet}^{-1}$.  When a planet is massive enough (0.3-1.0 \Mjup: 
Armitage~2007) to clear a gap in the disk, the slower Type~II migration 
begins.  The results of this work, showing a deficit of systems with 
multiple giant planets inside of 2-3~AU, suggest that they are dominated 
by Type~I migration and rapidly accrete onto the star.  \citet{tanaka02} 
showed that the Type~I migration timescale is inversely proportional to 
the disk mass: planets in more massive disks migrate faster.  If we make 
the reasonable assumption that multiple giant planets form from 
unusually massive disks, then Type~I migration works against these 
planets surviving the migration if they remain below the gap-opening 
mass.  To generate systems with multiple giant planets inside of 2-3~AU, 
migration should then be rapid enough to bring them there, but not so 
fast as to send the planets into the star.  The results presented here 
suggest that such a scenario is uncommon.

In addition to migration, the dynamical history of planetary systems is 
an important factor in producing the observed architectures.  The 
eccentricity distribution of extrasolar planets suggests that 
dynamically active histories are common.  Interactions between giant 
planets can result in the ejection of one while imparting a significant 
eccentricity on the remaining planet \citep{rasio96, ford05, 
malmberg08}.  Systems containing a single giant planet on a moderately 
eccentric orbit may be the result of such encounters, and thus less 
likely to host the sort of planets this survey was seeking.  The median 
eccentricity of the planets targeted in this work is 0.29, compared to a 
median $e$ of 0.15 for all other planets.  Comparing the distributions 
by the K-S test gives a probability of 0.048, indicating a marginally 
significant difference between the two.  \citet{fischer08} use the 
relatively low eccentricities ($e<0.2$) of the five 55~Cnc planets to 
suggest that a benign dynamical history allowed so many planets to 
remain.  The GJ~876, HD~37124, HD~73526, and GJ~581 systems also have 
multiple planets with $e<0.2$, but counterexamples are found in 
HD~160691, HD~74156, and HD~202206 ($e_{max}$=0.57, 0.64, and 0.44, 
respectively).  An uneventful dynamical history contributes to a 
planetary system's observed end state, but comprises only a part of the 
picture in combination with its formation history.

A primary goal of the search for extrasolar planets is to estimate how 
common the architecture of our own Solar system might be.  If the 
processes of planet formation and migration form many systems similar to 
our own, it becomes more likely that Earth-like planets may be present.  
The results of this work indicate that planetary systems like our own 
may be common if 1) terrestrial-mass planets are present but undetected, 
or 2) Type~I migration timescales are so short that multiple giant 
planets rarely end up within 2-3~AU.  Conversely, our Solar system may 
be rare if the dynamical history of most planetary systems results in 
many ejections and high eccentricities.


\section{Summary}

We have carried out an intensive radial-velocity campaign to monitor 22 
known planetary systems for additional planets.  No new planets were 
found, and these new data do not support the proposed planets HD~20367b, 
HD~74156d, and 47~UMa~c.  We have used test particles and Saturn-mass 
bodies to probe 20 planetary systems for regions in which additional 
planets could exist.  The massive-body results are consistent with the 
test-particle results: each of these systems has regions, sometimes 
quite large, where additional planets may remain in stable orbits.  
Finally, we show that this campaign could have detected 99\% of planets 
with M~sin~$i$~\ltsimeq 2.6 Neptune masses within 0.10~AU.

\acknowledgements

This material is based on work supported by the National Aeronautics and 
Space Administration under Grants NNG04G141G, NNG05G107G issued through 
the Terrestrial Planet Finder Foundation Science program and Grant 
NNX07AL70G issued through the Origins of Solar Systems Program. We are 
grateful to the HET TAC for their generous allocation of telescope time 
for this project.  Much of the computing for the dynamical simulations 
used the Lonestar cluster at the Texas Advanced Computing Center.  This 
research has made use of NASA's Astrophysics Data System (ADS), and the 
SIMBAD database, operated at CDS, Strasbourg, France.  The Hobby-Eberly 
Telescope (HET) is a joint project of the University of Texas at Austin, 
the Pennsylvania State University, Stanford University, 
Ludwig-Maximilians-Universit{\" a}t M{\"u}nchen, and 
Georg-August-Universit{\" a}t G{\" o}ttingen The HET is named in honor 
of its principal benefactors, William P.~Hobby and Robert E.~Eberly.



\begin{deluxetable}{llllllll}
\tabletypesize{\scriptsize}
\tablecolumns{8}
\tablewidth{0pt}
\tablecaption{Stellar Parameters }
\tablehead{
\colhead{Star} & \colhead{Spec.~Type} & \colhead{Distance} & 
\colhead{Mass} &
\colhead{[Fe/H]} & \colhead{$T_{eff}$} & \colhead{V sin $i$} & 
\colhead{log$R'_{HK}$}\\
\colhead{} & \colhead{} & \colhead{(pc)} & \colhead{(\Msun)} &
\colhead{} & \colhead{(K)} & \colhead{(\kms)} &
\colhead{}
}
\startdata
\label{stellarparms}
HD 3651 & K0V & 11.1$\pm$0.1 & 0.882$\pm$0.026 & 0.24$\pm$0.03 & 
5221$\pm$44 & 1.1 
& -4.99$\pm$0.05 \\
HD 8574 & F8 & 44.2$\pm$1.6 & 1.122$\pm$0.022 & -0.03$\pm$0.03 & 
6050$\pm$44 & 4.5 
& -4.88$\pm$0.04 \\
HD 10697 & G5IV & 32.6$\pm$0.9 & 1.112$\pm$0.026 & 0.17$\pm$0.03 & 
5680$\pm$44 & 
2.5 & -5.07$\pm$0.15 \\
HD 19994 & F8V & 22.4$\pm$0.4 & 1.365$\pm$0.042 & 0.27$\pm$0.03 & 
6188$\pm$44 & 
8.6 & -4.93$\pm$0.04 \\
HD 20367 & G0 & 27.1$\pm$0.8 & 1.04$\pm$0.06\tablenotemark{a} & 
-0.09$\pm$0.10\tablenotemark{b} & 
5998$\pm$75 & 3.0 & -4.50$\pm$0.05 \\
HD 23596 & F8V & 52.0$\pm$2.3 & 1.159$\pm$0.062 & 0.33$\pm$0.03 & 
5904$\pm$44 & 
4.2 & -4.96$\pm$0.05 \\
HD 28185 & G5 & 39.6$\pm$1.7 & 0.98$\pm$0.05\tablenotemark{c} & 
0.12$\pm$0.10\tablenotemark{b} & 
5546$\pm$75 & 3.0 & -5.37$\pm$0.40 \\
HD 38529 & G4IV & 42.4$\pm$1.7 & 1.477$\pm$0.052 & 0.51$\pm$0.03 & 
5697$\pm$44 & 
3.9 & -5.01$\pm$0.03 \\
\enddata
\tablenotetext{a}{Mass obtained from \citet{holmberg07}.}
\tablenotetext{b}{[Fe/H], $T_{eff}$, and V~sin~$i$ obtained from 
\citet{holmberg07}.}
\tablenotetext{c}{Mass obtained from \citet{santos04a}. }
\end{deluxetable}

\begin{deluxetable}{llrrr}
\tabletypesize{\scriptsize}
\tablecolumns{5}
\tablewidth{0pt}
\tablecaption{Summary of Published Radial-Velocity Data }
\tablehead{
\colhead{Star} & \colhead{Reference} & \colhead{$N$} &
\colhead{$<\sigma>$} & \colhead{RMS about fit}\\
\colhead{} & \colhead{} & \colhead{} & \colhead{(\ms)} & \colhead{(\ms)}
 }
\startdata
\label{rvdata1}
HD 3651 & \citet{butler06} & 163 & 3.4 & 6.6 \\
HD 8574 & \citet{perrier03} & 41 & 10.3 & 13.1 \\
HD 8574 & \citet{butler06} & 26 & 10.4 & 23.0 \\
HD 10697 & \citet{butler06} & 59 & 2.7 & 6.8 \\
HD 19994 & \citet{mayor04} & 48 & 6.7 & 8.1 \\
HD 23596 & \citet{perrier03} & 39 & 9.1 & 9.2 \\
HD 28185 & \citet{santos01} & 40 & 6.5 & 10.0 \\
HD 38529 & \citet{butler06} & 162 & 5.3 & 13 \\
HD 40979 & \citet{butler06} & 65 & 9.1 & 23 \\
HD 72659 & \citet{butler06} & 32 & 3.2 & 4.2 \\
HD 74156 & \citet{naef04} & 95 & 10.8 & 10.6 \\
HD 80606 & \citet{naef01b} & 61 & 13.7 & 17.7 \\
HD 89744 & \citet{butler06} & 50 & 11.2 & 16.0 \\
47 UMa & \citet{fischer02} & 91 & 5.7 & 7.4 \\
47 UMa & \citet{naef04} & 44 & 7.3 & 7.4 \\
HD 106252 & \citet{perrier03} & 40 & 10.7 & 10.5 \\
HD 106252 & \citet{butler06} & 15 & 11.4 & 9.1 \\
HD 108874 & \citet{vogt05} & 49 & 3.4 & 3.7 \\
HD 114783 & \citet{butler06} & 54 & 2.7 & 4.7 \\
HD 128311 & \citet{vogt05} & 76 & 3.3 & 18.0 \\
HD 130322 & \citet{udry00} & 118 & 12.4 & 16.1 \\
HD 130322 & \citet{butler06} & 12 & 2.7 & 11.0 \\
HD 136118 & \citet{butler06} & 37 & 16.1 & 22.0 \\
HD 178911B & \citet{zucker02} & 51 & 10.4 & 11.0 \\
HD 178911B & \citet{butler06} & 14 & 2.7 & 7.7 \\
HD 190228 & \citet{perrier03} & 51 & 8.7 & 8.0 \\
\enddata
\end{deluxetable}

\begin{deluxetable}{lr@{$\pm$}lr@{$\pm$}lr@{$\pm$}lr@{$\pm$}lr@{$\pm$}lr@{$\pm$}
lr@{$\pm$}lll}
\rotate
\tabletypesize{\scriptsize}
\tablecolumns{10}
\tablewidth{0pt}
\tablecaption{Keplerian Orbital Solutions }
\tablehead{
\colhead{Planet} & \multicolumn{2}{c}{Period} & \multicolumn{2}{c}{$T_0$}
&
\multicolumn{2}{c}{$e$} & \multicolumn{2}{c}{$\omega$} &
\multicolumn{2}{c}{K } & \multicolumn{2}{c}{M sin $i$ } &
\multicolumn{2}{c}{$a$ } & \colhead{$\chi^2_{\nu}$} & \colhead{rms}\\
\colhead{} & \multicolumn{2}{c}{(days)} & \multicolumn{2}{c}{(JD-2400000)}
&
\multicolumn{2}{c}{} &
\multicolumn{2}{c}{(degrees)} & \multicolumn{2}{c}{(\ms)} &
\multicolumn{2}{c}{(\Mjup)} & \multicolumn{2}{c}{(AU)} & \colhead{} & 
\colhead{\ms}
 }
\startdata
\label{planetparms}   
HD 3651 b & 62.218 & 0.015 & 53932.6 & 0.6 & 0.596 & 0.036 & 242.5 & 4.5 & 
15.9 & 0.7 & 0.229 & 0.008 & 0.295 & 0.003 & 3.82 & 6.3 \\
HD 8574 b & 227.0 & 0.2 & 53981.0 & 3.2 & 0.297 & 0.026 &
26.6 & 5.4 & 58.3 & 1.8 & 1.80 & 0.06 & 0.757 & 0.005 & 2.21 & 14.2 \\
HD 10697 b & 1075.2 & 1.5 & 51480 & 18 & 0.099 & 0.007 &
111.2 & 6.3 & 115.4 & 1.1 & 6.21 & 0.15 & 2.131 & 0.018 & 3.39 & 8.1 \\
HD 19994 b & 466.2 & 1.7 & 53757 & 72 & 0.063 & 0.062 &
346 & 55 & 29.3 & 2.1 & 1.37 & 0.12 & 1.305 & 0.016 & 5.27 & 14.0 \\
HD 23596 b & 1561 & 12 & 53163 & 22 & 0.266 & 0.014 & 272.6 & 3.3 & 
127.0 & 2.0 & 7.71 & 0.39 & 2.772 & 0.062 & 0.88 & 8.7 \\
HD 28185 b & 385.9 & 0.6 & 53793.6 & 8.8 & 0.092 & 0.019 &
351.9 & 8.2 & 158.8 & 4.2 & 5.59 & 0.33 & 1.032 & 0.019 & 2.28 & 9.5 \\
HD 38529 b & 14.3098 & 0.0005 & 54012.64 & 0.16 & 0.257 & 0.015 &
92.5 & 3.9 & 56.1 & 0.9 & 0.839 & 0.030 & 0.131 & 0.002 & 6.32 & 11.8 \\
HD 38529 c & 2140.2 & 5.7 & 52256.4 & 6.4 & 0.341 & 0.005 &
17.8 & 1.2 & 173.2 & 1.2 & 13.38 & 0.39 & 3.712 & 0.048 & 6.32 & 11.8 \\
\enddata
\tablenotetext{a}{Results from two-planet fit.}
\tablenotetext{b}{Results from one-planet fit.}
\tablenotetext{c}{Results for HD~130322 exclude data from Udry et 
al.~(2000).}
\end{deluxetable}

\begin{deluxetable}{lllll}
\tabletypesize{\scriptsize}
\tablecolumns{5}
\tablewidth{0pt}
\tablecaption{Summary of Radial-Velocity Data }
\tablehead{
\colhead{Star} & \colhead{$N$} & \colhead{RMS about fit (\ms)} &
\colhead{$\Delta T$ (days)} & \colhead{Source}
}
\startdata
\label{datasummary}
HD 3651 & 163 & 6.5 &  & \citet{butler06} \\
HD 3651 & 35 & 5.1 &  & HET\tablenotemark{a} \\
HD 3651 & 4 & 9.3 &  & 2.7m\tablenotemark{b} \\
HD 3651 (total) & 202 & 6.3 & 7376 & \\
\hline
HD 8574 & 41 & 14.8 &  & \citet{perrier03} \\
HD 8574 & 44 & 8.7 &  & HET \\
HD 8574 & 16 & 13.4 &  & 2.7m \\
HD 8574 & 26 & 20.7 &  & \citet{butler06} \\
\enddata
\tablenotetext{a}{9.2~m Hobby-Eberly Telescope.}
\tablenotetext{b}{McDonald Observatory 2.7~m Harlan J.~Smith Telescope.}
\tablenotetext{c}{No planet was fit.}
\end{deluxetable}

\begin{deluxetable}{lrr}
\tabletypesize{\scriptsize}
\tablecolumns{3}
\tablewidth{0pt}
\tablecaption{Results of Periodogram Analysis }
\tablehead{
\colhead{Star} & \colhead{Period (days)} & \colhead{FAP}}   
\startdata
\label{faps}
HD 3651 & 44.17 & 0.707 \\ 
HD 8574 & 2272.73 & 0.687 \\
HD 10697 & 26.68 & 0.028 \\
HD 19994 & 54.88 & 0.399 \\
HD 20367 & 5.58 & 0.085 \\
HD 23596 & 25.13 & 0.141 \\
HD 28185 & 4.76 & 0.224 \\
HD 38529 & 294.12 & 0.023 \\
HD 40979 & 2.26 & 0.795 \\
HD 72659 & 6.99 & 0.758 \\
HD 74156 & 80.39 & 0.035 \\
HD 80606 & 357.14 & 0.616 \\
HD 89744 & 23.27 & 0.075 \\
47 UMa\tablenotemark{a} & 2380.95 & 0.045 \\
47 UMa\tablenotemark{b} & 2.91 & 0.341 \\
HD 106252 & 322.58 & 0.126 \\
HD 108874 & 12.39 & 0.857 \\
HD 114783 & 8.44 & 0.925 \\
HD 128311 & 11.21 & $<$0.0001 \\
HD 130322 & 438.60 & 0.002 \\
HD 136118 & 442.48 & 0.014 \\
HD 178911B & 7.88 & 0.925 \\
HD 190228 & 2.57 & 0.777 \\
\enddata
\tablenotetext{a}{Residuals from one-planet fit.}
\tablenotetext{b}{Residuals from two-planet fit.}
\end{deluxetable}

\begin{deluxetable}{llllll}
\tabletypesize{\scriptsize}
\tablecolumns{4}
\tablewidth{0pt}
\tablecaption{Companion Limit Summary}
\tablehead{
\colhead{Star} & \colhead{Eccentricity} & \colhead{M sin $i$} & 
\colhead{M sin $i$} & \colhead{Median $K$} & \colhead{Median $K$}\\
\colhead{} & \colhead{} & \colhead{0.05 AU} & \colhead{0.1 AU} & 
\colhead{99\% recovery} & \colhead{50\% recovery} \\
\colhead{} & \colhead{} & \colhead{(\Mjup)} & \colhead{(\Mjup)} & 
\colhead{\ms} & \colhead{\ms}
}
\startdata
\label{masslimits}
HD 3651 & 0.20 & 0.025 & 0.041 & 4.8 & \nodata \\
HD 3651 & 0    & 0.024 & 0.040 & 4.4 & 3.2 \\
HD 8574 & 0.10 & 0.124 & 0.142 & 14.1 & \nodata \\
HD 8574 & 0    & 0.124 & 0.143 & 14.1 & 10.2 \\ 
HD 10697 & 0.04 & 0.059 & 0.094 & 7.4 & \nodata \\
HD 10697 & 0    & 0.059 & 0.094 & 7.4 & 5.6 \\
HD 19994 & 0.09 & 0.116 & 0.173 & 16.2 & \nodata \\
HD 19994 & 0    & 0.117 & 0.166 & 16.2 & 10.7 \\
\enddata
\end{deluxetable}

\begin{deluxetable}{lrrr}
\tabletypesize{\scriptsize}
\tablecolumns{4}
\tablewidth{0pt}
\tablecaption{Comparison of Stellar Characteristics}
\label{stats0}
\tablehead{
\colhead{Quantity} & \colhead{Targets} & \colhead{Non-Targets} &
\colhead{Units} \\
}
\startdata
\label{stars}
$[Fe/H]$ (mean) & 0.12$\pm$0.18 & 0.07$\pm$0.23 & dex \\
$[Fe/H]$ (median) & 0.13   &  0.14  & dex \\
$T_{eff}$ (mean) & 5741$\pm$361  & 5608$\pm$496 & K \\
$T_{eff}$ (median) & 5697  & 5704  & K \\
$(B-V)$ (mean) & 0.67$\pm$0.11  & 0.74$\pm$0.20  & mag \\
$(B-V)$ (median) & 0.63  & 0.69  & mag \\
V sin $i$ (mean) & 3.72$\pm$2.50  & 2.75$\pm$1.72  & \kms \\
V sin $i$ (median) & 2.48  & 2.40  & \kms \\
\enddata
\end{deluxetable}

\begin{deluxetable}{lrrr}
\tabletypesize{\scriptsize}
\tablecolumns{4}
\tablewidth{0pt}
\tablecaption{Characteristics of Single and Multiple Planet Systems }
\tablehead{
\colhead{Quantity} & \colhead{Single} & \colhead{Multiple} & 
\colhead{Units} \\
}
\startdata
\label{stats}
$a$ (mean)  & 0.95$\pm$1.05 & 1.19$\pm$1.38 & AU \\
$a$ (median)  & 0.49  & 0.63  & AU \\
$e$ (mean)  & 0.24$\pm$0.23 & 0.19$\pm$0.17 & \\
$e$ (median)  & 0.18  &  0.16  & \\
M~sin~$i$ (mean)  & 2.72$\pm$3.16 & 1.93$\pm$2.38 & \Mjup \\
M~sin~$i$ (median) &  1.60  &  1.03  & \Mjup \\
\hline
Star mass (mean) & 1.14$\pm$0.41 & 1.06$\pm$0.32  & \Msun \\
Star mass (median) & 1.07  &  1.04  & \Msun \\
$[Fe/H]$ (mean) & 0.09$\pm$0.21 & 0.05$\pm$0.30 & dex \\
$[Fe/H]$ (median) & 0.14  &  0.14  & dex \\
$T_{eff}$ (mean) & 5640$\pm$473  & 5532$\pm$529 & K \\
$T_{eff}$ (median) & 5724  & 5584  & K \\
$(B-V)$ (mean) & 0.73$\pm$0.18  & 0.77$\pm$0.22  & mag \\
$(B-V)$ (median) & 0.68  & 0.72  & mag \\
\enddata
\end{deluxetable}

\begin{deluxetable}{lr}
\tabletypesize{\scriptsize}
\tablecolumns{2}
\tablewidth{0pt}
\tablecaption{K-S Tests on Single and Multiple Planet Systems }
\tablehead{
\colhead{Quantity} & \colhead{K-S Probability\tablenotemark{a}} \\
}
\startdata
\label{kstest}
$a$  &  0.004 \\
$a$ (first planet) & 0.249 \\
M~sin~$i$  &  0.015 \\
M~sin~$i$ (first planet)  &  0.349 \\
$e$  &  0.125 \\
\hline
Star mass &  0.644  \\
$[Fe/H]$  &  0.841  \\
$T_{eff}$  & 0.135 \\
$(B-V)$  &  0.383  \\
\enddata
\tablenotetext{a}{Probability that the two samples are drawn from the same 
distribution.}
\end{deluxetable}

\clearpage

\clearpage

\clearpage

\begin{deluxetable}{lrr}
\tabletypesize{\scriptsize}
\tablecolumns{3}
\tablewidth{0pt}
\tablecaption{HET Radial Velocities for HD 3651}
\tablehead{
\colhead{JD-2400000} & \colhead{Velocity (\ms)} & \colhead{Uncertainty
(\ms)}}
\startdata
\label{3651vels}
53581.21162  &     14.4  &    3.7  \\
53600.79860  &      0.4  &    4.7  \\
53604.79357  &     -6.7  &    3.5  \\
53606.78360  &     -9.3  &    3.9  \\
53608.77426  &    -10.9  &    3.6  \\
53615.96471  &    -18.5  &    3.6  \\
53628.74240  &      3.8  &    3.3  \\
53669.61203  &    -12.2  &    3.6  \\
53678.79142  &    -10.0  &    3.6  \\
53682.78611  &    -18.1  &    3.3  \\
53687.77875  &     17.0  &    3.7  \\
53691.76158  &     12.3  &    3.9  \\
53694.75466  &     16.7  &    3.7  \\
53696.76029  &     15.5  &    3.4  \\
53955.83593  &     12.3  &    3.8  \\
53956.83044  &      7.3  &    3.9  \\
53957.82392  &     10.6  &    3.5  \\
53973.80980  &     -4.3  &    4.7  \\
53976.78586  &    -13.2  &    3.4  \\
53978.97197  &     -2.2  &    5.9  \\
53985.95982  &    -15.2  &    4.4  \\
53987.95527  &    -13.6  &    3.0  \\
53989.74009  &    -20.6  &    2.8  \\
54003.70817  &     13.2  &    4.4  \\
54005.68492  &     17.2  &    3.6  \\
54056.78111  &     -9.9  &    3.5  \\
54062.55312  &     19.0  &    3.4  \\
54064.54902  &     13.0  &    3.3  \\
54130.55508  &     15.4  &    3.4  \\
54282.92879  &     -1.7  &    4.4  \\
54352.96182  &     -9.8  &    3.1  \\
54394.64607  &     -1.9  &    3.8  \\
54399.61380  &     -9.2  &    3.5  \\
54414.77832  &     -1.0  &    3.6  \\
54423.75714  &    -14.9  &    4.4  \\
\enddata
\end{deluxetable}

\begin{deluxetable}{lrr}
\tabletypesize{\scriptsize}
\tablecolumns{3}
\tablewidth{0pt}
\tablecaption{2.7m Radial Velocities for HD 3651}
\tablehead{
\colhead{JD-2400000} & \colhead{Velocity (\ms)} & \colhead{Uncertainty
(\ms)}}
\startdata
\label{hjs3651}
53633.86853  &     -3.3  &    5.0  \\
53654.79777  &      6.6  &    6.6  \\
53690.69920  &     -3.4  &    6.5  \\
54020.84477  &      0.1  &    5.8  \\
\enddata
\end{deluxetable}

\begin{deluxetable}{lrr}
\tabletypesize{\scriptsize}
\tablecolumns{3}
\tablewidth{0pt}
\tablecaption{HET Radial Velocities for HD 8574}
\tablehead{
\colhead{JD-2400000} & \colhead{Velocity (\ms)} & \colhead{Uncertainty
(\ms)}}
\startdata
\label{8574vels}
53601.81736  &    -40.9  &    6.3  \\
53604.80314  &    -51.4  &    6.8  \\
53605.82189  &    -42.2  &    6.9  \\
53607.81271  &    -45.9  &    6.7  \\
53609.79513  &    -48.9  &    7.3  \\
53612.79858  &    -46.1  &    7.3  \\
53633.96072  &    -30.7  &    7.8  \\
53653.69022  &    -22.6  &    9.2  \\
53663.88291  &    -27.5  &    9.0  \\
53665.63807  &    -16.9  &    8.9  \\
53668.64072  &    -25.3  &    8.7  \\
53687.81363  &     -6.7  &    9.5  \\
53695.79449  &     -2.5  &   11.0  \\
53696.78211  &      6.2  &   10.6  \\
53703.77271  &     34.9  &    9.6  \\
53705.75396  &     37.0  &    9.8  \\
53936.90653  &     38.7  &    7.5  \\
53936.90653  &     42.1  &   11.1  \\
53969.80550  &     61.9  &    7.4  \\
53975.81126  &     64.7  &   10.6  \\
53987.99307  &     55.7  &    9.0  \\
53989.98424  &     66.3  &    9.6  \\
53997.96536  &     29.8  &    8.1  \\
54000.73798  &     18.7  &   10.0  \\
54013.69475  &      4.2  &    8.7  \\
54018.90830  &     -2.5  &    7.8  \\
54043.84966  &    -31.4  &   10.2  \\
54049.61516  &    -43.5  &    9.9  \\
54057.78673  &    -38.9  &    9.9  \\
54067.55166  &    -37.1  &    8.9  \\
54071.76076  &    -50.1  &    9.7  \\
54106.65631  &    -14.8  &    9.7  \\
54110.66352  &    -12.7  &   10.5  \\
54121.63331  &     -3.8  &    9.3  \\
54306.89112  &    -30.4  &    8.2  \\
54327.84854  &    -31.8  &    7.3  \\
54344.80292  &    -26.1  &    7.9  \\
54352.76329  &    -24.5  &    9.3  \\
54367.73942  &      0.4  &    8.2  \\
54402.85369  &     49.0  &    8.9  \\
54402.86084  &     42.4  &    8.1  \\
54404.84838  &     52.9  &    8.8  \\
54419.81583  &     65.1  &   10.8  \\
54434.54809  &     85.3  &    9.3  \\
\enddata
\end{deluxetable}

\begin{deluxetable}{lrr}
\tabletypesize{\scriptsize}
\tablecolumns{3}
\tablewidth{0pt}
\tablecaption{2.7m Radial Velocities for HD 8574}
\tablehead{
\colhead{JD-2400000} & \colhead{Velocity (\ms)} & \colhead{Uncertainty
(\ms)}}
\startdata
\label{hjs8574}
52116.95398  &     24.0  &    9.6  \\
52141.96262  &     70.0  &    8.0  \\
52219.89758  &    -10.8  &   10.0  \\
52249.70181  &    -30.2  &    9.6  \\
52331.61330  &      1.7  &    7.7  \\
52493.90858  &    -17.4  &    9.7  \\
52540.91557  &    -13.9  &    8.9  \\
52658.62787  &    -18.5  &    9.3  \\
52932.83883  &    -45.8  &    8.9  \\
53015.71309  &     29.6  &    9.2  \\
53564.94976  &      2.1  &   10.8  \\
53632.92472  &    -62.8  &    8.6  \\
53635.90969  &    -22.1  &    8.4  \\
53691.75684  &     -4.6  &   10.4  \\
53970.92894  &     95.0  &   13.5  \\
54018.87142  &      3.7  &    9.9  \\
\enddata
\end{deluxetable}

\begin{deluxetable}{lrr}
\tabletypesize{\scriptsize}
\tablecolumns{3}
\tablewidth{0pt}
\tablecaption{HET Radial Velocities for HD 10697}
\tablehead{
\colhead{JD-2400000} & \colhead{Velocity (\ms)} & \colhead{Uncertainty
(\ms)}}
\startdata
\label{10697vels}
53581.90281  &     22.3  &    9.3  \\
53606.84709  &    -21.0  &    8.7  \\
53653.91334  &    -49.4  &    7.9  \\
53663.69183  &    -43.8  &    8.9  \\
53665.67577  &    -51.5  &    8.4  \\
53681.83488  &    -67.6  &    8.4  \\
53681.83752  &    -67.4  &    8.6  \\
53690.81667  &    -60.4  &   10.3  \\
53694.60078  &    -77.7  &   10.0  \\
53696.79921  &    -73.3  &    9.2  \\
53701.77012  &    -82.7  &   10.4  \\
53703.79160  &    -70.7  &   10.0  \\
53923.95145  &    -65.8  &    8.3  \\
53954.87796  &    -47.2  &    8.5  \\
53956.86965  &    -47.3  &    8.8  \\
53958.88031  &    -46.6  &    9.0  \\
53969.83080  &    -69.9  &    9.4  \\
53971.83685  &    -61.7  &    8.8  \\
53984.79870  &    -34.4  &    8.3  \\
53988.80905  &    -26.3  &    8.5  \\
53990.98655  &    -37.3  &    9.1  \\
53999.74552  &    -28.7  &    7.8  \\
54041.64127  &    -10.3  &    8.3  \\
54042.65929  &    -10.5  &    8.4  \\
54049.62382  &    -19.4  &    8.7  \\
54056.63399  &     -3.7  &   10.6  \\
54069.56942  &     14.7  &    9.7  \\
54071.57888  &      6.6  &    9.8  \\
54105.67044  &     33.5  &    8.9  \\
54108.67092  &     23.4  &    9.2  \\
54130.60755  &     41.9  &   10.4  \\
54135.60424  &     60.0  &   10.3  \\
54330.86457  &    130.5  &    8.7  \\
54344.80964  &    108.3  &    8.8  \\
54346.79984  &    124.2  &    8.4  \\
54352.78595  &    136.5  &    9.3  \\
54357.77232  &    114.1  &    8.2  \\
54366.75077  &    119.1  &    9.1  \\
54419.83538  &    117.8  &   11.5  \\
54424.59411  &    121.5  &    9.1  \\
\enddata
\end{deluxetable}

\begin{deluxetable}{lrr}
\tabletypesize{\scriptsize}
\tablecolumns{3}
\tablewidth{0pt}
\tablecaption{2.7m Radial Velocities for HD 10697}
\tablehead{
\colhead{JD-2400000} & \colhead{Velocity (\ms)} & \colhead{Uncertainty
(\ms)}}
\startdata
\label{hjs10697}
51066.97570  &     76.4  &    6.0  \\
51152.79209  &    108.3  &    6.3  \\
51211.61496  &     98.3  &    6.5  \\
51239.60083  &    105.0  &   12.8  \\
51449.91000  &    -20.3  &    5.3  \\
51503.72131  &    -69.6  &   10.7  \\
51529.67754  &    -72.8  &    7.6  \\
51558.57566  &    -97.3  &    6.7  \\
51775.92530  &    -83.4  &    5.6  \\
51811.88858  &    -86.5  &    6.8  \\
51859.67414  &    -64.9  &    5.8  \\
51917.68431  &    -19.2  &    7.0  \\
51946.64764  &      3.0  &    7.5  \\
51987.56487  &     10.3  &    8.8  \\
52116.96710  &     84.9  &    6.7  \\
52247.79070  &    109.6  &    5.9  \\
52306.67720  &    101.0  &    5.7  \\
52493.92186  &      3.3  &    6.2  \\
52539.87531  &    -31.9  &    5.7  \\
52577.87821  &    -63.3  &    6.2  \\
52897.88453  &    -68.5  &    8.2  \\
52932.87904  &    -69.5  &    6.8  \\
53017.69801  &      4.1  &    6.5  \\
53215.85694  &     99.7  &   15.3  \\
53215.87160  &     89.8  &    6.7  \\
53320.75657  &    129.6  &   10.1  \\
53564.96191  &      3.4  &    8.1  \\
53566.91539  &     16.1  &    6.8  \\
53635.89769  &    -51.9  &    8.8  \\
53690.71397  &    -94.3  &    7.5  \\
53968.92927  &    -88.3  &   10.9  \\
54018.85867  &    -61.1  &    6.9  \\
\enddata
\end{deluxetable}

\begin{deluxetable}{lrr}
\tabletypesize{\scriptsize}
\tablecolumns{3}
\tablewidth{0pt}
\tablecaption{HET Radial Velocities for HD 19994}
\tablehead{
\colhead{JD-2400000} & \colhead{Velocity (\ms)} & \colhead{Uncertainty
(\ms)}}
\startdata
\label{19994vels}
53605.28626  &    -22.1  &    5.1  \\
53608.94050  &    -39.5  &    6.4  \\
53612.92809  &    -26.6  &    6.2  \\
53627.89670  &    -21.8  &    6.0  \\
53633.88721  &     -3.1  &    6.7  \\
53655.82407  &     11.6  &    6.3  \\
53663.80786  &      5.6  &    7.5  \\
53665.80005  &     10.1  &    7.1  \\
\enddata
\end{deluxetable}

\begin{deluxetable}{lrr}
\tabletypesize{\scriptsize}
\tablecolumns{3}
\tablewidth{0pt}
\tablecaption{2.7m Radial Velocities for HD 19994}
\tablehead{
\colhead{JD-2400000} & \colhead{Velocity (\ms)} & \colhead{Uncertainty
(\ms)}}
\startdata
\label{hjs19994}
53635.94301  &     -2.0  &    9.7  \\
53655.86581  &    -33.6  &   18.0  \\
53690.87053  &     34.6  &    8.0  \\
53747.67357  &     52.8  &   10.1  \\
54020.88376  &    -33.9  &    8.5  \\
54310.93898  &     22.2  &   10.4  \\
54346.88504  &      2.7  &   11.4  \\
54377.85942  &    -19.8  &    9.6  \\
54404.79890  &     -4.2  &    8.9  \\
54404.80349  &     -9.8  &   10.1  \\
54460.79143  &     14.2  &   12.5  \\
54496.57966  &    -23.3  &    8.5  \\
\enddata
\end{deluxetable}

\begin{deluxetable}{lrr}
\tabletypesize{\scriptsize}
\tablecolumns{3}
\tablewidth{0pt}
\tablecaption{HET Radial Velocities for HD 20367}
\tablehead{
\colhead{JD-2400000} & \colhead{Velocity (\ms)} & \colhead{Uncertainty
(\ms)}}
\startdata
\label{20367vels}
53581.93777  &      9.2  &    8.3  \\
53592.91499  &     13.0  &   11.9  \\
53594.92230  &     -4.3  &   11.7  \\
53605.87357  &     -1.8  &   10.8  \\
53607.87923  &     -3.5  &   10.3  \\
53608.87356  &     15.5  &   10.5  \\
53610.85696  &      7.5  &   10.8  \\
53612.85169  &     -8.8  &   12.3  \\
\enddata
\end{deluxetable}

\begin{deluxetable}{lrr}
\tabletypesize{\scriptsize}
\tablecolumns{3}
\tablewidth{0pt}
\tablecaption{2.7m Radial Velocities for HD 20367}
\tablehead{
\colhead{JD-2400000} & \colhead{Velocity (\ms)} & \colhead{Uncertainty
(\ms)}}
\startdata
\label{hjs20367}
53632.00290  &      2.7  &    7.4  \\
53635.96984  &     -3.7  &    8.2  \\
53691.80381  &     -6.8  &    7.9  \\
53808.62780  &     -7.9  &    6.2  \\
53967.90828  &     11.5  &    8.2  \\
53968.90725  &     11.4  &   10.8  \\
54018.98018  &     -2.3  &    6.3  \\
54158.59352  &     -5.9  &    6.7  \\
54189.59664  &     -2.8  &    7.4  \\
54189.61001  &     -6.9  &    7.7  \\
54190.61247  &    -13.3  &    7.9  \\
54191.60150  &     -1.9  &    6.1  \\
54192.60131  &      4.3  &    5.5  \\
54345.84686  &      0.1  &    8.3  \\
54377.85210  &    -14.5  &    6.6  \\
54402.79239  &     -3.6  &    5.9  \\
54460.80432  &     -5.7  &   10.3  \\
54496.58943  &     23.7  &    7.6  \\
54555.60469  &     21.5  &    6.7  \\
\enddata
\end{deluxetable}

\begin{deluxetable}{lrr}
\tabletypesize{\scriptsize}
\tablecolumns{3}
\tablewidth{0pt}
\tablecaption{HET Radial Velocities for HD 23596}
\tablehead{
\colhead{JD-2400000} & \colhead{Velocity (\ms)} & \colhead{Uncertainty
(\ms)}}
\startdata
\label{23596vels}
53581.96329  &     99.1  &    8.3  \\
53592.92364  &    102.1  &    8.2  \\
53593.94151  &    109.0  &    8.1  \\
53594.92991  &    103.8  &    8.0  \\
53605.89024  &     93.1  &    9.1  \\
53607.89908  &     94.0  &    8.1  \\
53608.88232  &     96.9  &    8.8  \\
53609.89877  &    105.4  &    8.7  \\
\enddata
\end{deluxetable}

\begin{deluxetable}{lrr}
\tabletypesize{\scriptsize}
\tablecolumns{3}
\tablewidth{0pt}
\tablecaption{2.7m Radial Velocities for HD 23596}
\tablehead{
\colhead{JD-2400000} & \colhead{Velocity (\ms)} & \colhead{Uncertainty
(\ms)}}
\startdata
\label{hjs23596}
53636.88979  &     73.6  &    7.1  \\
53692.83650  &     48.7  &    7.6  \\
53787.66076  &     11.9  &    6.0  \\
53808.63976  &     13.0  &    7.1  \\
54020.93260  &    -51.4  &    8.2  \\
54158.60541  &    -95.8  &    6.1  \\
\enddata
\end{deluxetable}

\begin{deluxetable}{lrr}
\tabletypesize{\scriptsize}
\tablecolumns{3}
\tablewidth{0pt}
\tablecaption{HET Radial Velocities for HD 28185}
\tablehead{
\colhead{JD-2400000} & \colhead{Velocity (\ms)} & \colhead{Uncertainty
(\ms)}}
\startdata
\label{28185vels}
53653.92112  &    -42.3  &    5.8  \\
53663.89377  &    -38.0  &    7.0  \\
53692.82400  &     24.8  &    6.3  \\
53695.80664  &     42.7  &    6.8  \\
53697.80395  &     34.2  &    6.8  \\
53701.79069  &     43.3  &    8.0  \\
53996.97502  &    -70.8  &    7.1  \\
53998.99146  &    -64.8  &    7.7  \\
54044.85384  &    -32.7  &    8.1  \\
54051.83487  &    -17.9  &    6.7  \\
54053.83800  &     -9.3  &    7.7  \\
54061.80571  &     14.6  &    7.0  \\
54063.79720  &      8.1  &    6.4  \\
54066.79838  &     17.8  &    7.3  \\
54068.78387  &     12.2  &    6.6  \\
54071.77898  &     25.8  &    6.6  \\
54073.77510  &     23.2  &    6.5  \\
54075.76724  &     34.4  &    7.9  \\
54105.68544  &    113.2  &    7.0  \\
54107.68662  &    115.1  &    7.7  \\
54110.67889  &    119.3  &    7.1  \\
54142.58869  &    209.6  &    6.7  \\
54368.95856  &    -53.6  &    6.8  \\
54370.96073  &    -49.5  &    6.8  \\
54374.94882  &    -50.1  &    6.8  \\
54376.93838  &    -59.9  &    7.0  \\
54390.90631  &    -43.2  &    6.1  \\
54396.89578  &    -66.0  &    5.8  \\
54400.87895  &    -53.6  &    6.7  \\
54402.86922  &    -53.2  &    6.4  \\
54404.86864  &    -44.3  &    6.3  \\
54418.82544  &    -38.2  &    7.3  \\
54425.80631  &    -38.5  &   12.6  \\
54433.79452  &    -12.4  &    7.3  \\
\enddata
\end{deluxetable}

\begin{deluxetable}{lrr}
\tabletypesize{\scriptsize}
\tablecolumns{3}
\tablewidth{0pt}
\tablecaption{HET Radial Velocities for HD 38529}
\tablehead{
\colhead{JD-2400000} & \colhead{Velocity (\ms)} & \colhead{Uncertainty
(\ms)}}
\startdata
\label{38529vels}
53341.50743  &   -186.9  &    3.2  \\
53355.84573  &   -167.2  &    5.3  \\
53357.85963  &   -181.3  &    5.4  \\
53358.72474  &   -157.0  &    5.1  \\
53359.72919  &   -144.0  &    5.9  \\
53360.84952  &   -127.9  &    5.4  \\
53365.81675  &    -82.4  &    4.6  \\
53367.81264  &    -88.1  &    5.3  \\
\enddata
\end{deluxetable}

\begin{deluxetable}{lrr}
\tabletypesize{\scriptsize}
\tablecolumns{3}
\tablewidth{0pt}
\tablecaption{2.7m Radial Velocities for HD 38529}
\tablehead{
\colhead{JD-2400000} & \colhead{Velocity (\ms)} & \colhead{Uncertainty
(\ms)}}
\startdata
\label{hjs38529}
53633.96726  &    -44.5  &    5.7  \\
53636.91593  &    -19.0  &    4.9  \\
53691.91356  &    -24.9  &    5.8  \\
53746.78728  &    -51.1  &    5.7  \\
53809.64793  &     47.7  &    5.5  \\
53984.94973  &     -1.4  &    5.1  \\
54020.90382  &     93.2  &    6.4  \\
\enddata
\end{deluxetable}

\begin{deluxetable}{lrr}
\tabletypesize{\scriptsize}
\tablecolumns{3}
\tablewidth{0pt}
\tablecaption{HET Radial Velocities for HD 40979}
\tablehead{
\colhead{JD-2400000} & \colhead{Velocity (\ms)} & \colhead{Uncertainty
(\ms)}}
\startdata
\label{40979vels}
53341.72130  &    -80.8  &   13.1  \\
53346.73432  &    -99.4  &   15.1  \\
53348.70934  &    -57.5  &   17.1  \\
53350.72298  &    -81.2  &   14.7  \\
53352.92178  &    -95.5  &   15.0  \\
53355.68899  &    -48.4  &   13.4  \\
53357.93997  &    -59.4  &   16.1  \\
53359.68105  &    -88.5  &   16.3  \\
\enddata
\end{deluxetable}

\begin{deluxetable}{lrr}
\tabletypesize{\scriptsize}
\tablecolumns{3}
\tablewidth{0pt}
\tablecaption{2.7m Radial Velocities for HD 40979}
\tablehead{
\colhead{JD-2400000} & \colhead{Velocity (\ms)} & \colhead{Uncertainty
(\ms)}}
\startdata
\label{hjs40979}
53636.00527  &     78.9  &    8.9  \\
53787.77595  &    -46.2  &    8.3  \\
53864.61096  &    -23.4  &    7.6  \\
54020.94453  &     -9.2  &    6.7  \\
\enddata
\end{deluxetable}

\begin{deluxetable}{lrr}
\tabletypesize{\scriptsize}
\tablecolumns{3}
\tablewidth{0pt}
\tablecaption{HET Radial Velocities for HD 72659}
\tablehead{
\colhead{JD-2400000} & \colhead{Velocity (\ms)} & \colhead{Uncertainty
(\ms)}}
\startdata
\label{72659vels}
53342.98059  &      8.4  &   11.7  \\
53346.89816  &     19.2  &   10.5  \\
53348.89792  &     17.7  &   11.3  \\
53351.88711  &     20.4  &    9.6  \\
53355.85531  &     21.5  &    9.2  \\
53357.86778  &     15.6  &    9.7  \\
53359.86322  &     12.3  &   11.8  \\
53366.92023  &      1.9  &   12.2  \\
\enddata
\end{deluxetable}

\begin{deluxetable}{lrr}
\tabletypesize{\scriptsize}
\tablecolumns{3}
\tablewidth{0pt}
\tablecaption{HET Radial Velocities for HD 74156}
\tablehead{
\colhead{JD-2400000} & \colhead{Velocity (\ms)} & \colhead{Uncertainty
(\ms)}}
\startdata
\label{74156vels}
53342.23249  &   -119.8  &    7.9  \\
53347.00129  &   -102.6  &    7.5  \\
53355.83378  &    -93.1  &    7.9  \\
53357.84523  &    -90.2  &    8.0  \\
53359.85043  &    -79.3  &   10.8  \\
53360.97444  &    -97.9  &    9.0  \\
53364.97638  &    -99.2  &   10.4  \\
53365.82597  &   -121.7  &    7.8  \\
\enddata
\end{deluxetable}

\begin{deluxetable}{lrr}
\tabletypesize{\scriptsize}
\tablecolumns{3}
\tablewidth{0pt}
\tablecaption{HET Radial Velocities for HD 80606}
\tablehead{
\colhead{JD-2400000} & \colhead{Velocity (\ms)} & \colhead{Uncertainty
(\ms)}}
\startdata
\label{80606vels}
53346.88103  &      8.1  &    8.6  \\
53358.02089  &    -16.0  &    8.1  \\
53359.82400  &    -26.7  &    8.8  \\
53361.02985  &    -28.9  &    7.7  \\
53365.03079  &    -46.8  &    8.0  \\
53373.98282  &    -63.3  &    9.8  \\
53377.80112  &    -62.8  &    9.6  \\
53379.75230  &    -62.1  &    9.3  \\
53389.74170  &    -95.6  &    7.8  \\
53391.74400  &    -97.6  &    8.0  \\
53395.72763  &   -104.1  &    8.6  \\
53399.72518  &   -115.4  &    9.4  \\
53401.72497  &   -123.7  &    9.1  \\
53414.67819  &   -186.7  &    9.1  \\
53421.85529  &    314.1  &    8.1  \\
53423.86650  &    374.2  &    7.9  \\
53424.85231  &    302.6  &    7.3  \\
53432.87120  &    132.0  &    7.6  \\
53433.60628  &    119.8  &    7.6  \\
53446.79322  &     55.8  &    8.2  \\
54161.85400  &    -62.4  &    8.9  \\
54166.83797  &    -80.2  &    7.3  \\
54186.76189  &   -134.1  &    7.4  \\
\enddata
\end{deluxetable}

\begin{deluxetable}{lrr}
\tabletypesize{\scriptsize}
\tablecolumns{3}
\tablewidth{0pt}
\tablecaption{HET Radial Velocities for HD 89744}
\tablehead{
\colhead{JD-2400000} & \colhead{Velocity (\ms)} & \colhead{Uncertainty
(\ms)}}
\startdata
\label{89744vels}
53709.89685  &     93.4  &   11.8  \\
53723.85367  &     57.9  &    8.7  \\
53727.84573  &     70.0  &    9.0  \\
53734.81973  &     64.1  &    9.1  \\
53736.82101  &     60.8  &    9.9  \\
53738.03441  &     76.6  &    9.8  \\
53738.81040  &     76.7  &    9.6  \\
53742.79299  &     66.5  &    9.0  \\
53746.81778  &     51.1  &    8.7  \\
53751.78379  &     53.7  &   10.7  \\
53753.78381  &     45.4  &    8.7  \\
53755.76218  &     47.2  &    9.2  \\
53757.77181  &     35.1  &    9.1  \\
53797.64834  &   -165.1  &   10.0  \\
53809.62700  &   -353.0  &    8.7  \\
53837.77709  &      1.2  &    7.9  \\
53866.70329  &     97.6  &    7.5  \\
53868.68562  &    100.3  &   13.0  \\
53875.66956  &    102.4  &   11.6  \\
53883.65837  &    114.7  &    6.5  \\
53890.63954  &    107.9  &    7.1  \\
53893.63138  &    125.0  &    7.2  \\
54050.96453  &   -137.3  &    7.5  \\
54052.96762  &   -160.3  &    6.9  \\
54056.94785  &   -197.9  &    7.6  \\
54063.93165  &   -327.2  &    7.9  \\
54073.91476  &   -438.1  &    9.1  \\
54122.01243  &    102.1  &    9.5  \\
54129.74491  &     94.8  &   10.1  \\
54160.66031  &     98.0  &    9.6  \\
54163.66643  &     89.0  &   10.0  \\
54165.88148  &     98.8  &   15.3  \\
54421.94999  &     75.7  &    8.1  \\
\enddata
\end{deluxetable}

\begin{deluxetable}{lrr}
\tabletypesize{\scriptsize}
\tablecolumns{3}
\tablewidth{0pt}
\tablecaption{2.7m Radial Velocities for HD 89744}
\tablehead{
\colhead{JD-2400000} & \colhead{Velocity (\ms)} & \colhead{Uncertainty
(\ms)}}
\startdata
\label{hjs89744}
53690.03080  &    262.2  &    7.7  \\
53805.87856  &   -129.3  &    7.1  \\
53806.73923  &   -130.0  &    8.9  \\
53807.83562  &   -188.4  &   10.8  \\
53809.79691  &   -198.1  &    7.8  \\
53840.78664  &    168.0  &    9.8  \\
53864.76543  &    203.8  &    8.9  \\
53911.61226  &    252.5  &    9.3  \\
54068.94214  &   -240.7  &    8.7  \\
\enddata
\end{deluxetable}

\begin{deluxetable}{lrr}
\tabletypesize{\scriptsize}
\tablecolumns{3}
\tablewidth{0pt}
\tablecaption{HET Radial Velocities for 47 UMa}
\tablehead{
\colhead{JD-2400000} & \colhead{Velocity (\ms)} & \colhead{Uncertainty
(\ms)}}
\startdata
\label{47umavels}
53313.33003  &     24.4  &    4.7  \\
53314.99108  &     31.8  &    4.6  \\
53317.99083  &     30.8  &    4.4  \\
53334.95254  &     26.0  &    4.3  \\
53335.94709  &     26.8  &    4.4  \\
53338.93789  &     20.1  &    3.1  \\
53340.91724  &     24.8  &    4.4  \\
53346.92207  &     17.7  &    4.3  \\
\enddata
\end{deluxetable}

\begin{deluxetable}{lrr}
\tabletypesize{\scriptsize}
\tablecolumns{3}
\tablewidth{0pt}
\tablecaption{2.7m Radial Velocities for 47 UMa}
\tablehead{
\colhead{JD-2400000} & \colhead{Velocity (\ms)} & \colhead{Uncertainty
(\ms)}}
\startdata
\label{hjs47uma}
51010.62898  &     52.2  &    6.2  \\
51212.97474  &      1.7  &    4.5  \\
51240.81250  &     -1.7  &    4.6  \\
51274.78993  &     -8.8  &    4.6  \\
51326.70558  &    -24.5  &    4.9  \\
51504.95996  &    -39.7  &    5.0  \\
51530.01978  &    -27.0  &    5.1  \\
51555.94972  &    -15.5  &    4.8  \\
51655.74023  &     -0.7  &    4.7  \\
51686.75156  &     -8.8  &    6.0  \\
51750.60418  &     13.1  &    5.1  \\
51861.01895  &     53.7  &    4.9  \\
51917.93086  &     48.2  &    4.6  \\
51987.85527  &     53.7  &    6.2  \\
52004.83235  &     60.5  &    4.8  \\
52039.77936  &     54.5  &    5.9  \\
52116.60554  &     55.7  &    6.6  \\
52249.00010  &     21.6  &    5.4  \\
52303.89238  &     -8.3  &    5.1  \\
52305.84757  &     -1.5  &    4.7  \\
52327.86285  &     21.2  &   13.1  \\
52353.85949  &      2.6  &    6.1  \\
52661.95399  &    -30.8  &    4.7  \\
53017.93695  &     75.2  &    6.4  \\
53069.76686  &     54.9  &    4.2  \\
53692.03243  &    -44.4  &    6.5  \\
53748.89147  &    -40.4  &    4.6  \\
53787.91198  &    -28.0  &    4.7  \\
53805.88756  &    -28.3  &    4.5  \\
53809.80777  &    -25.0  &    4.7  \\
53840.77154  &    -19.2  &    5.1  \\
53841.70627  &    -19.3  &    5.8  \\
53841.72168  &      0.2  &    6.5  \\
53861.74397  &    -19.8  &    5.4  \\
53909.61977  &      3.6  &    5.7  \\
54280.64401  &     38.2  &    6.7  \\
54280.64893  &     31.4  &    5.6  \\
54551.92162  &    -36.6  &    5.2  \\
54569.78354  &    -39.5  &    5.8  \\
54569.78813  &    -37.8  &    5.9  \\
54569.79271  &    -40.3  &    6.1  \\
54605.77977  &    -52.7  &    6.5  \\
54632.71638  &    -43.6  &    6.9  \\
\enddata
\end{deluxetable}

\begin{deluxetable}{lrr}
\tabletypesize{\scriptsize}
\tablecolumns{3}
\tablewidth{0pt}
\tablecaption{HET Radial Velocities for HD 106252}
\tablehead{
\colhead{JD-2400000} & \colhead{Velocity (\ms)} & \colhead{Uncertainty
(\ms)}}
\startdata
\label{106252vels}
53351.00010  &    -86.7  &   10.2  \\
53392.87552  &    -19.4  &   11.1  \\
53396.02801  &    -18.2  &   10.0  \\
53399.86570  &     -0.6  &   11.5  \\
53422.81038  &     35.5  &    8.9  \\
53423.95949  &     38.5  &    9.9  \\
53424.95405  &     30.2  &    8.4  \\
53429.93417  &     43.4  &    8.3  \\
53436.93141  &     41.8  &    8.3  \\
53439.93691  &     44.2  &    9.3  \\
53440.93473  &     46.8  &   10.0  \\
53448.72404  &     54.7  &    8.7  \\
53449.71116  &     60.8  &    9.4  \\
53450.72302  &     50.1  &    8.7  \\
53451.72210  &     49.4  &    8.3  \\
53452.88054  &     66.6  &    8.6  \\
53454.86997  &     56.3  &    8.8  \\
53455.86893  &     64.1  &    8.8  \\
53457.70319  &     60.4  &    8.6  \\
53480.62813  &     72.9  &   10.0  \\
53487.78674  &     69.1  &    9.0  \\
53498.76669  &     82.5  &    9.9  \\
53543.64221  &     92.8  &    8.7  \\
53736.92178  &    -13.3  &   10.6  \\
53736.92786  &     -9.9  &    9.9  \\
53743.93096  &      1.6  &   11.4  \\
53744.90904  &    -12.4  &    9.7  \\
53745.89603  &    -11.5  &    9.8  \\
53753.89366  &    -17.2  &    9.3  \\
53755.88492  &    -24.8  &    9.0  \\
53758.87460  &    -32.4  &    9.9  \\
53765.02009  &    -25.5  &   11.8  \\
53779.82267  &    -24.6  &   10.3  \\
53796.76996  &    -34.1  &    8.4  \\
53866.75798  &    -41.3  &   10.2  \\
53868.73636  &    -39.5  &    9.5  \\
53877.71039  &    -35.1  &    8.4  \\
53891.67012  &    -40.6  &    8.4  \\
54090.96158  &   -100.9  &    8.1  \\
54110.90994  &   -106.5  &    8.8  \\
54122.04519  &   -118.5  &    9.6  \\
54161.76867  &   -124.4  &    9.8  \\
54191.69138  &   -124.3  &    8.6  \\
\enddata
\end{deluxetable}

\begin{deluxetable}{lrr}
\tabletypesize{\scriptsize}
\tablecolumns{3}
\tablewidth{0pt}
\tablecaption{2.7m Radial Velocities for HD 106252}
\tablehead{
\colhead{JD-2400000} & \colhead{Velocity (\ms)} & \colhead{Uncertainty
(\ms)}}
\startdata
\label{hjs106252}
52116.61921  &     46.4  &   11.5  \\
52307.00335  &     16.9  &    9.8  \\
52328.89167  &    -32.1  &   11.8  \\
52357.80730  &    -40.7  &    9.6  \\
52743.85538  &   -129.6  &    9.1  \\
53465.74189  &     55.7  &   10.4  \\
53504.65833  &     73.1  &    9.8  \\
53564.63133  &     73.7  &   10.5  \\
53566.62685  &     84.6  &   10.8  \\
53808.85919  &    -23.7  &   10.3  \\
53842.77638  &    -24.6  &   10.9  \\
53861.78041  &    -35.8  &   10.2  \\
\enddata
\end{deluxetable}

\begin{deluxetable}{lrr}
\tabletypesize{\scriptsize}
\tablecolumns{3}
\tablewidth{0pt}
\tablecaption{HET Radial Velocities for HD 108874}
\tablehead{
\colhead{JD-2400000} & \colhead{Velocity (\ms)} & \colhead{Uncertainty
(\ms)}}
\startdata
\label{108874vels}
53370.93479  &     -4.1  &    7.6  \\
53377.90734  &      3.4  &    8.0  \\
53392.86111  &     -3.8  &    8.3  \\
53399.84978  &      4.9  &    7.1  \\
53423.00572  &      1.5  &    6.0  \\
53424.00026  &     -1.7  &    6.4  \\
53429.76827  &      9.2  &    5.6  \\
53446.93347  &      3.5  &    6.4  \\
53449.92196  &     -0.3  &    5.6  \\
53451.70786  &      4.7  &    5.5  \\
53452.91476  &     -0.6  &    5.0  \\
53454.90261  &    -10.7  &    6.1  \\
53455.91231  &     -2.2  &    5.6  \\
53457.89845  &     -9.3  &    5.8  \\
53460.88699  &     -5.9  &    6.0  \\
53708.98585  &    -35.6  &    7.5  \\
53723.95560  &    -17.0  &    7.7  \\
53730.92665  &    -15.4  &    7.5  \\
53751.89671  &     -3.0  &    7.1  \\
53753.88047  &     -3.1  &    7.8  \\
53755.87205  &      9.2  &    6.3  \\
53773.81280  &     11.5  &    6.7  \\
53806.96496  &      3.7  &    6.8  \\
53847.61993  &      5.4  &    6.1  \\
53866.78797  &     -3.3  &    6.2  \\
53868.77287  &     -6.4  &    5.8  \\
53880.76028  &     -3.9  &    6.5  \\
53895.71194  &     -5.6  &    5.3  \\
53897.70685  &     -3.1  &    5.7  \\
53912.66615  &    -12.8  &    5.3  \\
54080.96833  &    -24.1  &    7.0  \\
54084.96842  &    -25.8  &    7.1  \\
54109.89738  &     -1.9  &    7.0  \\
54127.85761  &      3.1  &    6.8  \\
54142.02401  &     17.2  &    6.6  \\
54158.76779  &     28.1  &    7.7  \\
54160.75928  &     23.2  &    6.7  \\
54162.75513  &     19.0  &    7.0  \\
54180.92025  &     23.0  &    5.8  \\
54190.88453  &     29.0  &    6.2  \\
\enddata
\end{deluxetable}

\begin{deluxetable}{lrr}
\tabletypesize{\scriptsize}
\tablecolumns{3}
\tablewidth{0pt}
\tablecaption{HET Radial Velocities for HD 114783}
\tablehead{
\colhead{JD-2400000} & \colhead{Velocity (\ms)} & \colhead{Uncertainty
(\ms)}}
\startdata
\label{114783vels}
53366.02839  &    -17.8  &    6.0  \\
53368.02525  &    -25.3  &    5.5  \\
53374.00882  &    -13.9  &    6.5  \\
53378.99750  &    -16.3  &    5.3  \\
53395.95201  &    -23.9  &    5.4  \\
53399.94906  &    -25.0  &    5.2  \\
53415.90535  &    -27.0  &    6.2  \\
53416.96080  &    -27.8  &    5.7  \\
53421.88527  &    -21.5  &    5.5  \\
53423.88025  &    -29.5  &    5.9  \\
53429.88531  &    -19.4  &    5.7  \\
53436.92048  &    -22.8  &    5.6  \\
53440.91863  &    -24.7  &    6.9  \\
53446.81643  &    -20.4  &    5.2  \\
53447.81352  &    -16.1  &    5.5  \\
53448.81555  &    -13.6  &    5.6  \\
53450.81211  &    -18.4  &    5.5  \\
53451.82012  &    -22.9  &    5.2  \\
53452.81032  &    -16.8  &    5.5  \\
53455.78717  &    -20.4  &    5.3  \\
53779.91198  &      6.9  &    5.7  \\
54097.02928  &     27.2  &    5.7  \\
54106.00717  &     32.2  &    6.0  \\
54121.96750  &     24.7  &    5.3  \\
54127.95124  &     28.9  &    5.3  \\
54130.94440  &     35.9  &    5.3  \\
54140.98702  &     26.0  &    5.7  \\
54143.92794  &     21.4  &    6.0  \\
54156.87064  &     35.6  &    5.8  \\
54158.87301  &     34.8  &    5.8  \\
54163.85683  &     38.2  &    5.8  \\
54168.83901  &     42.2  &    5.6  \\
54186.78882  &     32.9  &    5.0  \\
54191.77547  &     36.7  &    5.2  \\
\enddata
\end{deluxetable}

\begin{deluxetable}{lrr}
\tabletypesize{\scriptsize}
\tablecolumns{3}
\tablewidth{0pt}
\tablecaption{HET Radial Velocities for HD 128311}
\tablehead{
\colhead{JD-2400000} & \colhead{Velocity (\ms)} & \colhead{Uncertainty
(\ms)}}
\startdata
\label{128311vels}
53462.96527  &   -137.0  &    4.1  \\
53479.81567  &   -147.0  &    4.4  \\
53480.91902  &   -136.9  &    5.7  \\
53482.89103  &   -144.3  &    5.7  \\
53486.72671  &   -159.9  &    5.5  \\
53488.70965  &   -168.8  &    6.4  \\
53509.81664  &   -160.8  &    6.0  \\
53510.82254  &   -172.4  &    5.6  \\
\enddata
\end{deluxetable}

\begin{deluxetable}{lrr}
\tabletypesize{\scriptsize}
\tablecolumns{3}
\tablewidth{0pt}
\tablecaption{HET Radial Velocities for HD 130322}
\tablehead{
\colhead{JD-2400000} & \colhead{Velocity (\ms)} & \colhead{Uncertainty
(\ms)}}
\startdata
\label{130322vels}
53471.80558  &    -99.8  &    7.2  \\
53481.88526  &   -106.9  &    6.6  \\
53486.85864  &    105.1  &    6.4  \\
53488.75815  &     72.2  &    5.9  \\
53509.79117  &    101.3  &    6.1  \\
53512.78123  &    -65.6  &    5.1  \\
53527.74971  &     27.0  &    6.2  \\
53542.69985  &     55.4  &    5.6  \\
53543.70614  &     -4.9  &    6.1  \\
53550.70420  &    105.5  &    6.1  \\
53837.89677  &    -12.6  &    5.9  \\
53842.88880  &     27.6  &    6.3  \\
53868.80896  &    -78.8  &    5.7  \\
53882.78043  &     83.0  &    6.0  \\
53897.72683  &    -44.2  &    6.1  \\
53900.72079  &    -85.4  &    6.0  \\
53936.63557  &    110.4  &    6.6  \\
54122.01834  &    -12.7  &    6.8  \\
54128.00335  &     47.0  &    6.7  \\
54135.98084  &   -113.2  &    6.5  \\
54139.97029  &     89.9  &    7.2  \\
54140.96840  &     98.1  &    6.1  \\
54144.96962  &    -99.4  &    6.6  \\
54157.01611  &   -112.4  &    6.8  \\
54158.92425  &    -40.9  &    6.8  \\
54163.92465  &     26.3  &    6.7  \\
54168.90656  &    -71.6  &    6.5  \\
54173.98269  &     69.7  &    7.4  \\
54176.87914  &    -90.6  &    5.7  \\
54191.92631  &     20.5  &    6.2  \\
\enddata
\end{deluxetable}

\begin{deluxetable}{lrr}
\tabletypesize{\scriptsize}
\tablecolumns{3}
\tablewidth{0pt}
\tablecaption{2.7m Radial Velocities for HD 130322}
\tablehead{
\colhead{JD-2400000} & \colhead{Velocity (\ms)} & \colhead{Uncertainty
(\ms)}}
\startdata
\label{hjs130322}
53585.64900  &     83.7  &    7.5  \\
53843.89253  &    -18.0  &    7.5  \\
53863.78301  &     75.5  &    8.6  \\
53910.78043  &    -68.5  &    8.1  \\
54251.84318  &    -72.8  &    9.4  \\
\enddata
\end{deluxetable}

\begin{deluxetable}{lrr}
\tabletypesize{\scriptsize}
\tablecolumns{3}
\tablewidth{0pt}
\tablecaption{HET Radial Velocities for HD 136118}
\tablehead{
\colhead{JD-2400000} & \colhead{Velocity (\ms)} & \colhead{Uncertainty
(\ms)}}
\startdata
\label{136118vels}
53471.83434  &     37.0  &    7.8  \\
53480.89983  &     11.8  &   11.6  \\
53482.88067  &      8.1  &    9.0  \\
53486.86878  &     12.1  &   14.7  \\
53487.88314  &     17.0  &    9.0  \\
53509.80853  &     15.5  &   12.8  \\
53527.76309  &      5.7  &    9.4  \\
53544.72664  &      4.5  &    9.3  \\
\enddata
\end{deluxetable}

\begin{deluxetable}{lrr}
\tabletypesize{\scriptsize}
\tablecolumns{3}
\tablewidth{0pt}
\tablecaption{2.7m Radial Velocities for HD 136118}
\tablehead{
\colhead{JD-2400000} & \colhead{Velocity (\ms)} & \colhead{Uncertainty
(\ms)}}
\startdata
\label{hjs136118}
53585.66699  &     13.5  &   11.4  \\
53805.93741  &     -1.7  &    9.1  \\
53863.76953  &    -12.0  &   13.8  \\
53911.75165  &      0.2  &   10.0  \\
\enddata
\end{deluxetable}

\begin{deluxetable}{lrr}
\tabletypesize{\scriptsize}
\tablecolumns{3}
\tablewidth{0pt}
\tablecaption{HET Radial Velocities for HD 178911B}
\tablehead{
\colhead{JD-2400000} & \colhead{Velocity (\ms)} & \colhead{Uncertainty
(\ms)}}
\startdata
\label{178911vels}
53653.66528  &    -57.4  &    6.2  \\
53801.00302  &   -217.2  &    6.1  \\
53803.02412  &   -268.6  &    6.2  \\
53837.93570  &    257.0  &    6.7  \\
53846.89693  &    323.3  &    5.6  \\
53866.83649  &    -19.7  &    5.6  \\
53868.81760  &    -86.7  &    6.2  \\
53883.79325  &   -343.1  &    4.5  \\
53954.82320  &   -349.3  &    5.7  \\
53955.82606  &   -336.2  &    5.4  \\
53956.82027  &   -326.8  &    5.8  \\
53958.81704  &   -291.4  &    5.5  \\
53960.80356  &   -245.8  &    5.5  \\
53965.80165  &   -102.8  &    5.4  \\
53966.78694  &    -69.2  &    5.3  \\
53971.77135  &     72.4  &    5.6  \\
53976.76788  &    181.7  &    5.7  \\
53979.76859  &    242.2  &    6.0  \\
53988.72794  &    328.9  &    6.0  \\
53988.73584  &    330.3  &    5.8  \\
53993.71885  &    306.6  &    7.7  \\
54014.66291  &   -175.4  &    5.5  \\
54016.65503  &   -236.1  &    5.4  \\
54035.60988  &   -147.1  &    5.1  \\
54039.58877  &    -29.0  &    5.2  \\
54055.55127  &    302.5  &    5.6  \\
54063.54556  &    327.1  &    5.7  \\
54165.01752  &   -354.1  &    6.4  \\
54167.00332  &   -362.1  &    6.4  \\
54190.93933  &    182.1  &    6.4  \\
54251.77452  &   -106.8  &    5.5  \\
54323.81680  &    -86.4  &    5.5  \\
54332.79368  &    173.3  &    7.9  \\
54335.79347  &    228.2  &    6.0  \\
54338.78881  &    275.7  &    6.3  \\
54340.77677  &    301.3  &    6.1  \\
54344.77039  &    324.4  &    6.3  \\
54365.69757  &     22.9  &    6.3  \\
54396.61972  &    -44.4  &    5.5  \\
54400.59558  &     75.6  &    5.5  \\
\enddata
\end{deluxetable}

\begin{deluxetable}{lrr}
\tabletypesize{\scriptsize}
\tablecolumns{3}
\tablewidth{0pt}
\tablecaption{HET Radial Velocities for HD 190228}
\tablehead{
\colhead{JD-2400000} & \colhead{Velocity (\ms)} & \colhead{Uncertainty
(\ms)}}
\startdata
\label{190228vels}
53581.89473  &    -75.5  &    8.2  \\
53589.86259  &    -92.8  &    6.8  \\
53605.81074  &    -83.7  &    9.0  \\
53606.81111  &    -84.1  &    8.8  \\
53607.80141  &    -84.9  &    8.4  \\
53609.80786  &    -87.7  &    8.1  \\
53628.75754  &    -85.0  &    7.8  \\
53635.71968  &    -76.7  &    8.4  \\
53653.67654  &    -81.1  &    7.5  \\
53655.68368  &    -76.3  &    8.3  \\
53686.58579  &    -61.1  &    9.0  \\
53844.93852  &    -20.2  &    9.3  \\
53867.88496  &    -34.8  &    9.3  \\
53877.84874  &    -19.3  &    9.1  \\
53883.82777  &    -12.2  &    8.2  \\
53888.81145  &     -8.6  &    9.4  \\
53897.79760  &      1.5  &    8.9  \\
53935.90971  &      6.8  &    8.5  \\
53956.84836  &     -9.1  &    7.5  \\
53966.81728  &     -0.8  &    8.3  \\
53976.79348  &     -2.1  &    8.4  \\
53996.75947  &      5.3  &    9.0  \\
53998.73191  &      7.8  &    9.4  \\
53998.73697  &     -1.5  &    9.0  \\
54008.70583  &     10.4  &    7.8  \\
54013.70146  &      1.4  &    8.1  \\
54019.66975  &      7.1  &    9.0  \\
54032.62480  &     16.7  &    9.9  \\
54217.92567  &     49.1  &    8.9  \\
54265.79081  &     59.8  &    8.6  \\
54284.96126  &     62.5  &    9.1  \\
54326.62326  &     57.0  &    8.5  \\
54328.62216  &     60.8  &    7.0  \\
54331.61953  &     68.4  &    8.0  \\
54336.59244  &     67.8  &    9.0  \\
54344.79398  &     60.7  &    8.7  \\
54352.77160  &     68.1  &    9.1  \\
54368.71625  &     75.5  &    8.9  \\
54370.72492  &     81.4  &    8.4  \\
54377.68917  &     76.2  &    9.0  \\
54401.62808  &     74.4  &    8.3  \\
54428.57314  &     78.5  &    8.3  \\
\enddata
\end{deluxetable}

\begin{deluxetable}{lrr}
\tabletypesize{\scriptsize}
\tablecolumns{3}
\tablewidth{0pt}
\tablecaption{2.7m Radial Velocities for HD 190228}
\tablehead{
\colhead{JD-2400000} & \colhead{Velocity (\ms)} & \colhead{Uncertainty
(\ms)}}
\startdata
\label{hjs190228}
53584.82785  &    -47.3  &    6.2  \\
53585.80785  &    -27.8  &    8.9  \\
53635.70876  &    -42.5  &    6.9  \\
53636.74761  &    -41.8  &    5.5  \\
53691.63145  &    -16.2  &    7.9  \\
53862.94967  &     14.5  &    6.4  \\
53927.85541  &     43.2  &    6.8  \\
54403.65001  &    118.0  &    7.1  \\
\enddata
\end{deluxetable}

\end{document}